\documentclass[review,11pt,sort&compress]{elsarticle}
\usepackage[margin=1.18 in]{geometry}
\usepackage{array}
\usepackage{multirow}
\usepackage{siunitx}
\usepackage{notoccite}
\usepackage{alltt}
\usepackage{bm}
\usepackage[title]{appendix}
\usepackage{amsbsy,amsmath,amsthm,latexsym,amssymb}
\usepackage{graphicx,stmaryrd,sectsty}
\usepackage{setspace}
\usepackage[font=small,labelfont=bf]{caption}
\usepackage{verbatim}
\usepackage{graphicx}
\usepackage{caption}
\usepackage{subcaption}
\usepackage{float,color}
\usepackage{xcolor}
\usepackage{url}
\usepackage{textcomp}
\usepackage{booktabs}
\newcommand{\bc}{\begin{center}}
	\newcommand{\ec}{\end{center}}
\newcommand{\bfr}{\begin{flushright}}
	\newcommand{\efr}{\end{flushright}}

\newcommand{\no}{\noindent}
\newcommand{\be}{\begin{enumerate}}
	\newcommand{\ee}{\end{enumerate}}
\newcommand{\bi}{\begin{itemize}}
	\newcommand{\ei}{\end{itemize}}
\newcommand{\bd}{\begin{description}}
	\newcommand{\ed}{\end{description}}
\newcommand{\beq}{\begin{equation}}
\newcommand{\eeq}{\end{equation}}
\newcommand{\bea}{\begin{eqnarray}}
\newcommand{\eea}{\end{eqnarray}}

\newcommand{\bfi}{\begin{figure}}
	\newcommand{\efi}{\end{figure}}
\newcommand{\bay}{\begin{array}{l}}
	\newcommand{\eay}{\end{array}}

\newcommand{\cref}[1]{(\ref{#1})}   %to make cross reference easy.

\sectionfont{\normalsize}
\subsectionfont{\normalsize}
\subsubsectionfont{\normalsize}
\journal{Composites Part C: Open Access}

\begin{document}
\begin{frontmatter}
	\title{Elastic and Fracture Behavior of Three-Dimensional Ply-to-Ply Angle Interlock Woven Composites: Through-Thickness, Size Effect, and Multiaxial Tests}
	
	\author[label1]{Weixin Li}
	\author[label2]{Yao Qiao}
	\author[label1]{Joel Fenner}
	\author[label3]{Kyle Warren}
   \author[label2]{{Marco Salviato}\corref{cor1}}
   \ead{salviato@aa.washington.edu}
	\author[label1]{Zden\v ek P. Ba\v zant}
	\author[label1]{Gianluca Cusatis}
    
    \address[label1]{Department of Civil and Environmental Engineering, Northwestern University, Evanston, IL 60208, USA}
    \address[label2]{Department of Aeronautics and Astronautics, University of Washington, Seattle, WA 98195, USA}
    \address[label3]{Albany Engineered Composites, Rochester, NH 03867, USA}

	\cortext[cor1]{Corresponding Author}
	%Marco: my name appears as corresponding author and I was not able to fix it
	
	%\maketitle

	%\date{ }

	%\pagestyle{plain}\thispagestyle{empty}
	\begin{abstract}
\linespread{1}\selectfont
	
	{\small \no{ 
	%\color{blue} 
	This work presents a comprehensive investigation of the elastic and fracture behavior of ply-to-ply angle interlock three-dimensional woven composites. The research investigated novel splitting and wedge-driven out-of-plane fracture tests to shed light on the tensile fracture behavior in the thickness direction and to provide estimates of the out-of-plane tensile strength and fracture energy. In addition, size effect tests on geometrically-scaled Single Edge Notch Tension (SENT) specimens were performed to fully characterize the intra-laminar fracture energy of the material and to study the scaling of structural strength in this type of three-dimensional composites. The results confirmed that size effect in the structural strength of these materials is significant. 
%   Marco and Weixin: why more significant?	 % Rephrased
    In fact, even if the range of sizes investigated was broader than in any previous size effect study on traditional laminated composites and two-dimensional textile composites, all the experimental data fell in the transition zone between quasi-ductile and brittle behavior. This implies strong damage tolerance of the investigated three-dimensional composites. The analysis of the data via Ba\v zant's Type II Size Effect Law (SEL) enabled the objective characterization of the intra-laminar fracture energy of three-dimensional composites for the first time. Finally, Arcan rig tests combined with X-ray micro-computed tomography allowed unprecedented insights on the different damage mechanisms under multi-axial nominal loading conditions, particularly tension-dominated and shear-dominated conditions.
	}}
\end{abstract}
		%\end{abstract}
		
%	}
%	{\small \no {\bf   Keyword}: }

	%\newpage
	
%	\tableofcontents
	
	%\listoffigures
	
	%\listoftables
	
	%\newpage
\end{frontmatter}
\section{Introduction}
Thanks to their outstanding mechanical performance, textile composites were developed for various industrial applications including land, marine and air transportation, wind and tidal energy production, and blast protection of civil infrastructures and vehicles \cite{Tsaibook, Barbero}. Among them, three-dimensional (3D) woven composites provide extra reinforcement in the out-of-plane direction and enhanced fracture toughness and damage resistance compared to two-dimensional (2D) textile composites or conventional laminated composite systems \cite{Waas11,Soutis17,Waas18,Waas19}.  The mechanical characterization of composite materials  is essential not only to provide a database of their mechanical properties necessary for the design of structural components but also to advance the understanding of their deformation and failure mechanisms for the design of next generation material systems. However, comprehensive laboratory characterization of three dimensional (3D) woven composites is not straightforward and requires (1) measuring both in-plane and out-of-plane macroscopic elastic and strength properties considering material inhomogeneity associated with large periodic unit cells and complex microstructure; (2) recognizing and quantifying various damage mechanisms and failure events under different loading conditions; (3) characterizing the fracture properties and capturing the scaling of their mechanical response. %A large amount of work has been dedicated to investigating the elasticity, strength, and fracture properties of 3D woven composites in the in-plane and out-of-plane directions. 

The in-plane properties have been usually characterized by preforming uniaxial tension and compression tests, which can easily provide measures of Young's modulus, Poisson's ratio, and strength in both the longitudinal (warp) and transverse (weft) directions \cite{cox1992mechanisms,cox1994failure,cox1995macroscopic,Soutis17,Castaneda17,Lomov09,Warren15a,Warren15b,Behera15}. Although simple and straightforward, the means of strain measurement needs to be considered carefully to assure an adequate representation of the material with large unit cell size. % which was usually observed for textile composites.
The in-plane elastic and strength properties of 3D woven composites have been widely studied and were found to be strongly related to fabric architecture \cite{callus1999tensile,gu2002tensile,dai2015influence,mouritz2010mechanistic} and geometrical defects such as fiber crimp or tow waviness \cite{cox1995macroscopic,callus1999tensile,kuo2002failure,mouritz2010mechanistic,Warren15a}. A variety of damage mechanisms and failure modes have been documented by means of uniaxial tension and compression tests. For on-axis tension loading, transverse crack in fiber bundle, matrix cracking, tow/matrix debonding, and tow straightening were usually observed prior to failure \cite{cox1996tensile,callus1999tensile,osada2003initial,bogdanovich2013quasi,lomov2014monitoring,Warren15a} by means of techniques such as acoustic emission, optical microscopy, and computed tomography. These damage events contribute to the nonlinear stress-strain response frequently recorded in tension tests. Tow rupture and pullout were identified as the primary tensile failure modes. For on-axis compression loading, micro-buckling of fibers and growth of kink bands were found to be the key initiating event that ultimately led to failure \cite{cox1992mechanisms,cox1994failure,yang2000bending,kuo2000compressive,kuo2002failure,kuo2007effect,Warren15a}. Both fiber kinking at the microscopic level and kinking of fiber bundles at the miniscopic level were identified, and are strongly related to fiber misalignment and yarn imperfection (waviness) \cite{kuo2000compressive,kuo2002failure,kuo2007effect}. Off-axis tension/compression test with a bias loading direction ($\pm \ang{45}$) with respect to the material principal direction provides information on in-plane (intra-laminar) shear properties, and the dominant failure mode was recognized as matrix cracking \cite{gerlach2012plane,saleh2016characterising,visrolia2013multiscale}. In-plane shear behaviors were also frequently characterized by various direct shear methods such as Iosipescu \cite{pochiraju1999three} and rail shear \cite{Warren15a} tests. However, the applicability of the direct shear methods was questioned \cite{buchanan2012determination} because of the large periodic unit cells. 

Characterization of the out-of-plane properties of 3D textile composites remains a challenging problem, although the main purpose of 3D reinforcement is to increase the mechanical performance in the out-of-plane direction. Very limited literature data for out-of-plane tension is available. Conventional uniaxial tension test on dog bone or waisted specimens was adopted for thick 2D composites \cite{daniel2008three,abot2004through}. This method is not suitable for 3D composites because the contribution of 3D reinforcement usually vanishes at the surfaces of 3D composite plates, which poses a challenge for load transmission usually enabled by an adhesive bond \cite{gerlach2012plane}.  Various designs of ring or curved beam specimens have been proposed to address the difficulties arising from the inter-laminar tensile strength measurement of woven laminates \cite{cui1996interlaminar,olsson2011survey}. The L-shaped beam specimen design has been used to determine the out-of-plane tensile properties of textile composites \cite{hufenbach2011determination,hufenbach2013characterisation}. Unfortunately, the curved specimen design requires special treatment at the manufacturing stage and cannot apply to the characterization of conventional planar composite panels. A novel test with a cross specimen loaded in compression using U-shaped steel rings was proposed to generate failure of the out-of-plane reinforcement in tension, and achieved promising results \cite{gerlach2012plane}. However, the non-uniformity of the tensile stress field generated by tests based the curved specimen and cross specimen designs needs to be examined carefully. Compression experiment in the out-of-plane direction is conceptually simple, and has been widely used \cite{park2005through,song2014mechanical,gerlach2012plane}. One needs to pay attention to the design of specimen geometry to assure a sufficient aspect ratio and the specimen alignment during test setup to avoid invalid test results and failure patterns. For out-of-plane shear properties, short-beam (3-point bending) was commonly used. It was suggested that the short-beam test cannot provide a valid measure of the out-of-plane (or inter-laminar) shear strength of 3D textile composites because the test would not result in shear failure \cite{whitney1985short, cox1994failure,walter2010monotonic}, although it was shown to be a reliable method to characterize the out-of-plane shear modulus with the assistance of a full-field measurement \cite{gras2013identification}. Punch test based on a cutting mechanism with sharp corners \cite{kuo2003failure,kuo2002failure}, notched shear test based on asymmetrical loading \cite{gerlach2012plane}, Iosipescu shear test \cite{hufenbach2011analysing,hufenbach2013characterisation}, and V-notched rail shear test \cite{gude2015modified} were also adopted to determine the shear properties in the out-of-plane direction. Their applicability needs further examination in various aspects, especially in the representativeness of the generated stress states. 

Fracture tests have been widely accepted as a means to assess the fracture/damage resistance, damage tolerance, and ductility of composite materials. Both out-of-plane (inter-laminar) \cite{bascom1980interlaminar,alif1998effect,hufenbach2013influence,koh2011experimental,rys2010investigation,mouritz1999mode,Waas11} and in-plane (intra-laminar) fracture properties \cite{guenon1989toughness,liu2008fracture,blanco2014intralaminar,laffan2012translaminar} have been characterized in laboratory for various woven composite systems, and are useful for the design of structural components. It has been long recognized that the delamination fracture behavior of laminated and woven composites is nonlinear or of resistance type and can be described by equivalent linear fracture mechanics or quasi-brittle mechanics. Double cantilever beam (DCB) and wedge-driven delamination tests were frequently used to measure their fracture properties. On the other hand, composite in-plane fracture behaviors were frequently described by Linear Elastic Fracture Mechanics (LEFM), which, however, neglects the presence of a non-linear Fracture Process Zone (FPZ) with finite size. This is not accurate because the size of the FPZ occurring in the presence of a traction-free crack or notch is large compared to the typical size of laboratory specimens for carbon-polymer composites due to the complex mesostructure \cite{hughes2002fracture,green2007experimental}. Various damage mechanisms such as crack bridging, fiber-matrix debonding, and fiber pull-out occur and dissipate a large amount of energy within the FPZ, resulting in a nonuniform stress field which decreases with increasing crack opening. The size of the FPZ is even larger for textile composites than other quasibrittle materials (such as concrete, rock, ceramics, just to name a few) considering the coarse internal structure, the considerable unit cell size, and the complicated fiber-fiber and fiber-matrix interactions. Failing to capture the effects of a finite FPZ, one may obtain apparent measures of material fracture properties which are size dependent \cite{li2019size}.
% Weixin: cite here your paper on shale size-effect where you showed this very well. %Done
Although the complex fracturing process and the significant process zone size have been highlighted and investigated since the early 90s \cite{bao1992remarks,cox1994failure}, the application of nonlinear or quasi-brittle fracture mechanics in experimental characterization of carbon-polymer composites has fallen behind, presumably due to the difficulties of attaining a stable fracture test without exhibiting snap-back instability \cite{salviato2016direct}. A possible remedy is size effect testing based on a energetic size effect law, which not only has been applied to identify the fracture properties of various quasi-brittle materials \cite{cedolin2008identification,cusatis2009cohesive,wan2018age,cusatis-and-diluzio}
% Weixin: please cite here my papers on size effect: 11, 12, 47, 55. Thanks.
including carbon-polymer composites \cite{bazant1996size,bavzant1999size,catalanotti2014measurement,salviato2016experimental}, but also provide a new insight into the scaling characteristics of their mechanical properties. 

Another important, yet overlooked, aspect of three-dimensional woven composites is their behavior under multi-axial loading conditions. Multi-axial properties of composites were typically characterized by using tabular or cruciform specimens in the literature \cite{tube1,tube2,tube3,tube4,cruciform1,cruciform2,cruciform3,cruciform4}. The former one can provide various combinations of local multi-axial stresses (e.g. tension/shear, compression/shear, etc.) depending on the global loading condition whereas the latter one is generally used to investigate the tension/tension behavior of the materials. However, the drawbacks for the tabular specimens can be the difficulties in manufacturing specimens with this geometry and the possible buckling failure in the tests. The fabrication of the cruciform specimens is relatively easier compared to the aforementioned specimens but the experiments require an advanced testing system and the range of the multiaxiality ratio is also not sufficiently wide. Alternatively, to overcome the foregoing disadvantages, the Arcan rig test with different modifications was used to provide a variety of multi-axial stress states in the materials with less efforts \cite{arcandevice,Yaomulti1,Pearce,Tan,laux,liechti,Akhtar,Tan2,Alfonso,Zeinedini}. By leveraging these methods, the complete envelope of failure stresses and related damage mechanisms for the composites under multi-axial loading condition can be successfully achieved. On similar grounds, further experimental characterization on the fatigue multi-axial behavior of composites (e.g. fatigue lifetime, endurance limit, stiffness degradation, fatigue damage progression, etc.) can also be obtained \cite{cruciformfatigue1,cruciformfatigue2,cruciformfatigue3,Yaomulti1,Yaomulti2,tubefatigue,tubefatigue2,tubefatigue3,tubefatigue4,tubefatigue5}. These multi-axial properties are significantly important for the correct calibration of either continuum or discrete computational models used for the real and complex composite structures under various loading conditions.

%In this work, a comprehensive experimental investigation on a 3D woven composite was performed. A series of tests were preformed and summarized in Table \ref{tab:summary}. %11/17 Modified below MS

%{\color{blue} 
As a first step towards filling the several knowledge gaps discussed previously, this work presents a comprehensive investigation of the elastic and fracture behavior of ply-to-ply angle interlock three-dimensional woven composites.
%
%*****************
% Marco and Weixin: I find this type of blurbs pleonastic even though they are common in papers in our field. I would suggest removing it but I leave the final decision to you.
%I agree with you. I have never done it before. The main reason why I did it this time is that the manuscript is really very long. So I thought this would have helped. I am totally fine with removing it. MS
%In Sections 3.1 and 3.2, a complete characterization of the in-plane behavior in uniaxial tension and compression is provided for on-axis (warp and weft directions) and off-axis loading. Out-of-plane compression tests and tensile splitting tests are presented in Sections 3.3 and 3.4 for the characterization of the out-of-plane elastic behavior and strength. Section 4 discusses the results of a large size effect testing campaign which enabled the objective characterization of the intra-laminar fracture energy and the scaling of structural strength. To shed light on the out-of-plane fracture behavior, novel wedge-driven tests are presented in Section 5, where a lower-bound estimate of the through-thickness fracture energy is provided. Finally, the fracture tests are completed in Section 6 by multi-axial tests on notched specimens. By taking advantage of an Arcan rig to apply nominal multi-axial loading conditions and X-ray micro-computed tomography, the tests provided unprecedented insights on the different mechanisms of damage in tension-dominated and shear-dominated conditions. 
%*******************
A summary of all the tests performed is given in Table \ref{tab:summary} while a detailed description of all the tests and their results is provided next. 
%}

% Table generated by Excel2LaTeX from sheet 'Sheet5'
\begin{table}[htbp]
	\centering
	\caption{Summary of the in-plane and out-of-plane tests performed in this work.}
	\resizebox{\textwidth}{!}{
		\begin{tabular}{>{\raggedright\arraybackslash}p{4.8cm}cc>{\centering}p{3cm}c}
			\hline
			Type  & Specimen geometry & Dimensions [mm] & \multicolumn{1}{c}{Direction} & \multicolumn{1}{l}{Measured properties} \\
			\hline
			\multirow{2}[0]{4cm}{In-plane on-axis tension} & \multirow{2}[0]{*}{Coupon} & 250$\times$18.5$\times$7.2 & Warp  & $E_1$, $\nu_{12}$, $F_{1t}$ \\
			&       & 250$\times$35.6$\times$7.2 & Weft  & $E_2$, $\nu_{12}$, $F_{2t}$ \\
			\multirow{2}[0]{4cm}{In-plane on-axis compression} & \multirow{2}[0]{*}{Coupon} & 200$\times$18.5$\times$7.2 & Warp  & $E_1$, $\nu_{12}$, $F_{1c}$ \\
			&       & 200$\times$35.6$\times$7.2 & Weft  & $E_2$, $\nu_{12}$, $F_{2c}$ \\
			\multirow{2}[0]{4.5cm}{In-plane off-axis tension/compression} & \multirow{2}[0]{*}{Coupon} & 220$\times$30.5$\times$7.2 & $+\ang{45}$    & \multirow{2}[0]{*}{$G_{12}$, $F_{12}$} \\
			&       & 220$\times$30.5$\times$7.2 & $- \ang{45}$   &  \\
			{Out-of-plane compression} & {Prism} & 28.8$\times$17.5$\times$17.5 & Out-of-plane & $E_3$, $\nu_{31}$, $\nu_{32}$, $F_{3c}$ \\
			Tensile splitting test & Cube  & 28.8$\times$28.8$\times$28.8 &Out-of-plane & $F_{3t}$ \\
			\multirow{3}[0]{4cm}{In-plane fracture and size effect test} & \multicolumn{1}{c}{\multirow{3}[0]{2cm}{\centering SENT specimen}} & 300$\times$55$\times$7.2 & \multirow{3}[0]{*}{Warp} & \multirow{3}[0]{*}{$G_f$, $c_f$} \\
			&       & 246$\times$40.2$\times$7.2 &       &  \\
			&       & 200$\times$27.5$\times$7.2 &       &  \\
Wedge-driven out-of-plane fracture test & V-notch prism & 100$\times$28.8$\times$24 & Out-of-plane & $G_{F3}$\\
\multirow{2}[0]{4cm}{In-plane multi-axial fracture tests} & \multirow{2}[0]{*}{Center-notched specimen} & 250$\times$20$\times$7.2 & Warp  & \multirow{2}[0]{*}{$\sigma^{multi}_{Nc}$, $\tau^{multi}_{Nc}$} \\
			&       & 250$\times$40$\times$7.2 & Weft   &  \\
			\hline
		\end{tabular}}%
		\label{tab:summary}%
	\end{table}%
	
	\section{Materials}
\begin{figure}[htbp]
		\begin{center}
			\includegraphics[width = 0.5\textwidth]{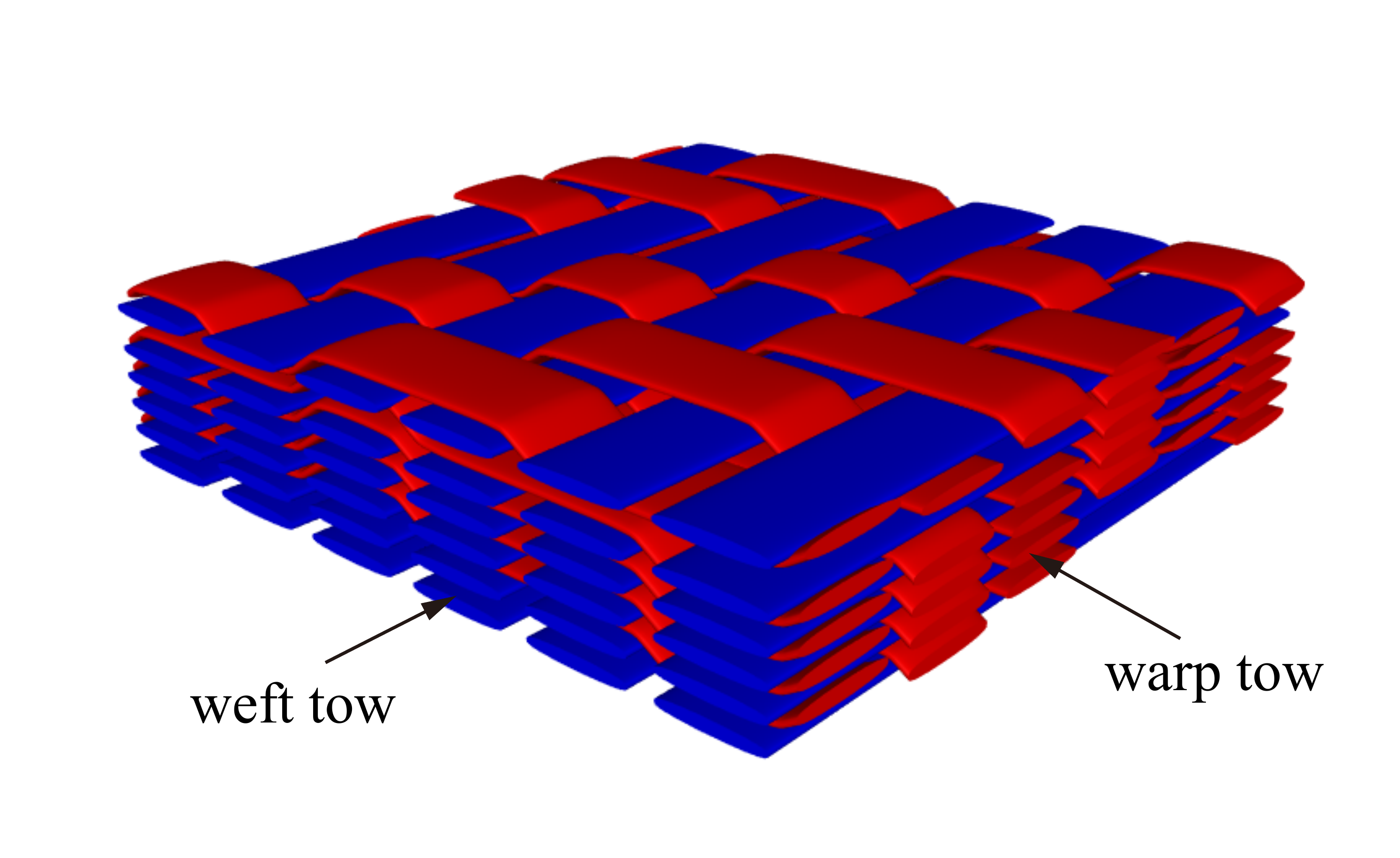}
			\caption{Geometric representation of a ply to ply angle interlock architecture of a 3D woven composite. 
			%{\color{blue} Weixin, this picture is very cool but, with soo many colors, it may be difficult to understand it. Would it be possible to have the weft in red and the warp in light blue? It should be easy to do if you have the file in Texgen. Another benefit is that, this way, the figure should be easy to understand also when printed in grayscale.MS. Good suggestion, Marco. I've modified the figure accordingly. WL.}
			}
			\label{fig: ARCH}
		\end{center}
	\end{figure}

\begin{figure}[htbp]
		\begin{center}
			\includegraphics[width = 0.8\textwidth]{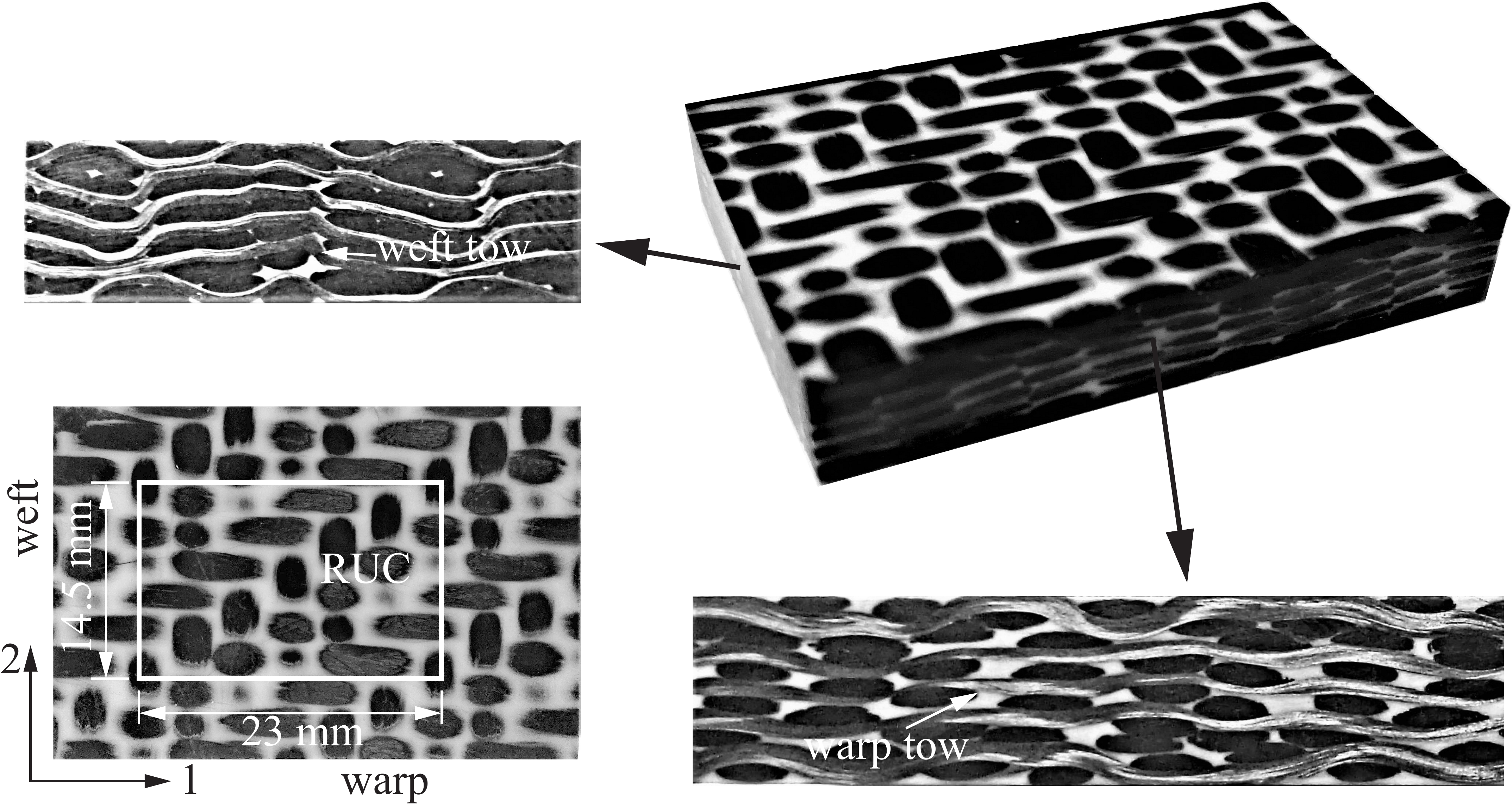}
			\caption{Images of a 3D woven composite sample and representation of unit cell.}
			\label{fig: MAT}
		\end{center}
	\end{figure}

	The material investigated in this work is a three-dimensional woven composite with a ply to ply angle interlock fiber architecture. The preforms were woven with 48K Hexcel IM7 carbon fiber, and Cytec PR520 toughened epoxy resin was injected using a resin transfer molding (RTM) process. A geometric representation of the three-dimensional woven architecture is illustrated in Fig. \ref{fig: ARCH}. The particular architecture has 6$\times$6 columns of warp and weft tows in each unit cell. Column spacing was adjusted to achieve a target warp content of 60\%. Warp tows also work as binders to provide through-thickness reinforcement. The overall fiber volume fraction is 0.58. Dimensional measurement were performed on the material surface and the repeated unit cell is approximately 23 mm in the warp direction and 14.5 mm in the weft direction, as illustrated in Fig. \ref{fig: MAT}. 
	
	Planar panels with various thicknesses were prepared by Albany Engineered Composites (AEC). The thinnest composite panels have 6 warp layers and the nominal thickness is 7.2 mm, whereas the thickest panels of 28.8 mm in thickness have 24 warp layers. The core warp layers are repeated through the thickness, and the in-plane unit cell dimensions remain the same throughout the configurations. The pictures of a sample lateral surface by sectioning the thin panels showed wavy warp and weft tows, as shown in Fig. \ref{fig: MAT}. It is difficult to quantify the tow waviness or the amount of crimp from the cross section because they are distinctive at different section planes. However, since waving over the thickness, warp tows span 23 mm horizontally whereas weft tows 14.5 mm, the overall waviness of weft tows is expected to be greater than that of warp tows. 
	
	Specimens were cut from the panels into designed geometries and nominal dimensions (Table \ref{tab:summary}) by using a sliding table tile saw with a water-cooled diamond blade. Surface layers damaged by cutting were removed by grinding and were further polished by fine grit sandpapers to ensure the high specimen quality. The actual dimensions of the prepared specimens were measured before each test was carried on. 
	
	\begin{table}[htbp]
		\centering
		\caption{Summary of measured elastic and strength properties}
		\begin{tabular}{llcr}
			\hline
			Description & Symbol [units] & \multicolumn{1}{l}{Measured value} & \multicolumn{1}{l}{Sections} \\
			\hline
			In-plane Young's modulus in direciton 1 & $E_1$ [GPa] & 71.4  & \ref{sec:onaxis} \\
			In-plane Young's modulus in direction 2 & $E_2$ [GPa] & 37.8  & \ref{sec:onaxis} \\
			In-plane Poisson's ratio & $\nu_{12}$ [-] & 0.218 & \ref{sec:onaxis} \\
			In-plane shear modulus & $G_{12}$ [GPa] & 5.6  & \ref{sec:offaxis} \\
			Out-of-plane Young's modulus & $E_3$ [GPa] & 8.4   & \ref{sec:TTComp} \\
			Out-of-plane Poisson's ratio in plane 3-1 & $\nu_{31}$ [-] & 0.056 & \ref{sec:TTComp} \\
			Out-of-plane Poisson's ratio in plane 3-2 & $\nu_{32}$ [-] & 0.396 & \ref{sec:TTComp} \\
			In-plane tensile strength in direction 1 & $F_{1t}$ [MPa] & 893   & \ref{sec:onaxis} \\
			In-plane tensile strength in direction 2 & $F_{2t}$ [MPa] & 271   & \ref{sec:onaxis} \\
			In-plane compressive strength in direction 1 & $F_{1c}$ [MPa] & 327   & \ref{sec:onaxis} \\
			In-plane compressive strength in direction 2 & $F_{2c}$ [MPa] & 196   & \ref{sec:onaxis} \\
			In-plane shear strength & $F_{12}$ [MPa] & 80    & \ref{sec:offaxis} \\
			Out-of-plane tensile strength & $F_{3t}$ [MPa] &  $>12$  & \ref{sec:TTBraz} \\
			Out-of-plane compressive strength & $F_{3c}$ [MPa] & 648   & \ref{sec:TTComp} \\
			\hline
		\end{tabular}%
		\label{tab:prop}%
	\end{table}%
	
	\section{Characterization of elasticity and strength}
	% Marco and Weixin: I typically do not use "I, we" in my papers but also the use of passive tense should be somewhat limited. Up to you to decide whether to rephrase the sentence below or other sentences using either "I, we"  or passive tense. % Rephrased a little bit, could improve in proof. WL
	This section discusses the tests for characterizing the elastic behaviors, strengths, and pre-peak responses of the investigated 3D woven composites. The on-axis and off-axis uniaxial tests were used to measure the composite in-plane behaviors, and the out-of-plane compression tests the elastic properties and compressive strength in the third direction through the panel thickness. As a first try, we proposed the tensile splitting test to estimate the out-of-plane tensile strength of the material. 
	
	For all the tests, the geometry of the specimens have to be determined in a way that the specimens are large enough compared to the unit cell size such that the measurements are representative while not exceeding the capacity of the loading systems and load cells. The specimen geometry specification for the conducted tests was reported in Table \ref{tab:summary}. The large unit cell size also poses a challenge for the measurement of strain since commercially available strain gauges were normally designed for materials with relatively small unit cells such as laminate composites and hence fail to provide a sufficient gauge length. To avoid this difficulty, a Digital Image Correlation (DIC) system (Correlated Solutions \cite{DIC}) was adopted to provide an in-situ full-field strain measurement on the specimen surfaces. This also enables one to correlate the non-linearity observed in the loading curves with the abnormal strain distribution detectable on the specimen surfaces. 
	
	The in-plane tension and compression tests were performed on a closed-loop servohydraulic Instron machine with a 222.4 kN (50 kip) load cell and hydraulic wedge grips, while the out-of-plane tests on a MTS machine with 978.6 kN (220 kip) loading capacity and proper fixtures as discussed latter. All of the tests were operated in a stroke-control mode, and the loading rate was adjusted such that the corresponding strain rate is within the quasi-static range. 
	
	\subsection{On-axis uniaxal test}\label{sec:onaxis}
\begin{figure}[htbp]
		\begin{center}
			\includegraphics[width = 1\textwidth]{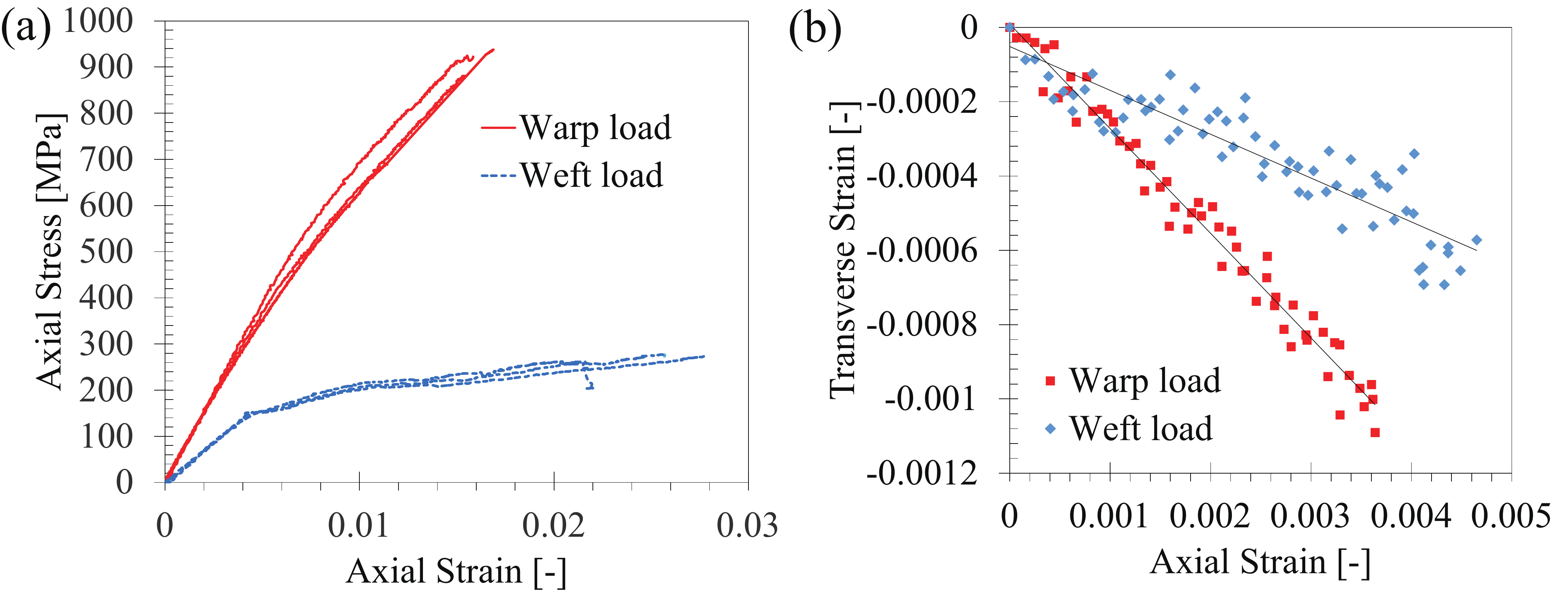}
			\caption{Mechanical responses of on-axis loaded specimens in tension tests: (a) axial stress versus axial strain curves; (b) typical relation between transverse and axial strains.}
			\label{fig: OnAxisTension}
		\end{center}
	\end{figure}

	Rectangular coupons cut along the warp direction were labeled $\ang{0}$, while the ones along the weft direction $\ang{90}$. The loading direction was aligned with long axis, and hence $\ang{0}$ specimens were warp loaded and \textbf{$\ang{90}$} specimens weft loaded. Typical mechanical responses of the warp loaded ($\ang{0}$) and weft loaded ($\ang{90}$) specimens in the on-axis uniaxial tension tests are shown in Fig. \ref{fig: OnAxisTension}. Figure \ref{fig: OnAxisTension}a plotted axial stress versus axial strain curves measured in the experiment, whereas Fig. \ref{fig: OnAxisTension}b plotted the typical relation between the transverse and the axial strains in the elastic region. The axial stress was approximated by a nominal stress calculated as the load applied on the specimens divided by the initial area of the specimen cross section. The axial and the transverse strains, on the other hand, were obtained by averaging the corresponding strain fields over the selected area of interest (AOI) centered at the middle of the specimen images. The AOI was selected properly such that it spans more than one repeated unit cell of the material, and the results with the AOI centered at various positions of different sizes were compared to ensure that the calculated average strains did not depend on the selection of AOI.
	
\begin{figure}[htbp]
		\begin{center}
			\includegraphics[width = 0.5\textwidth]{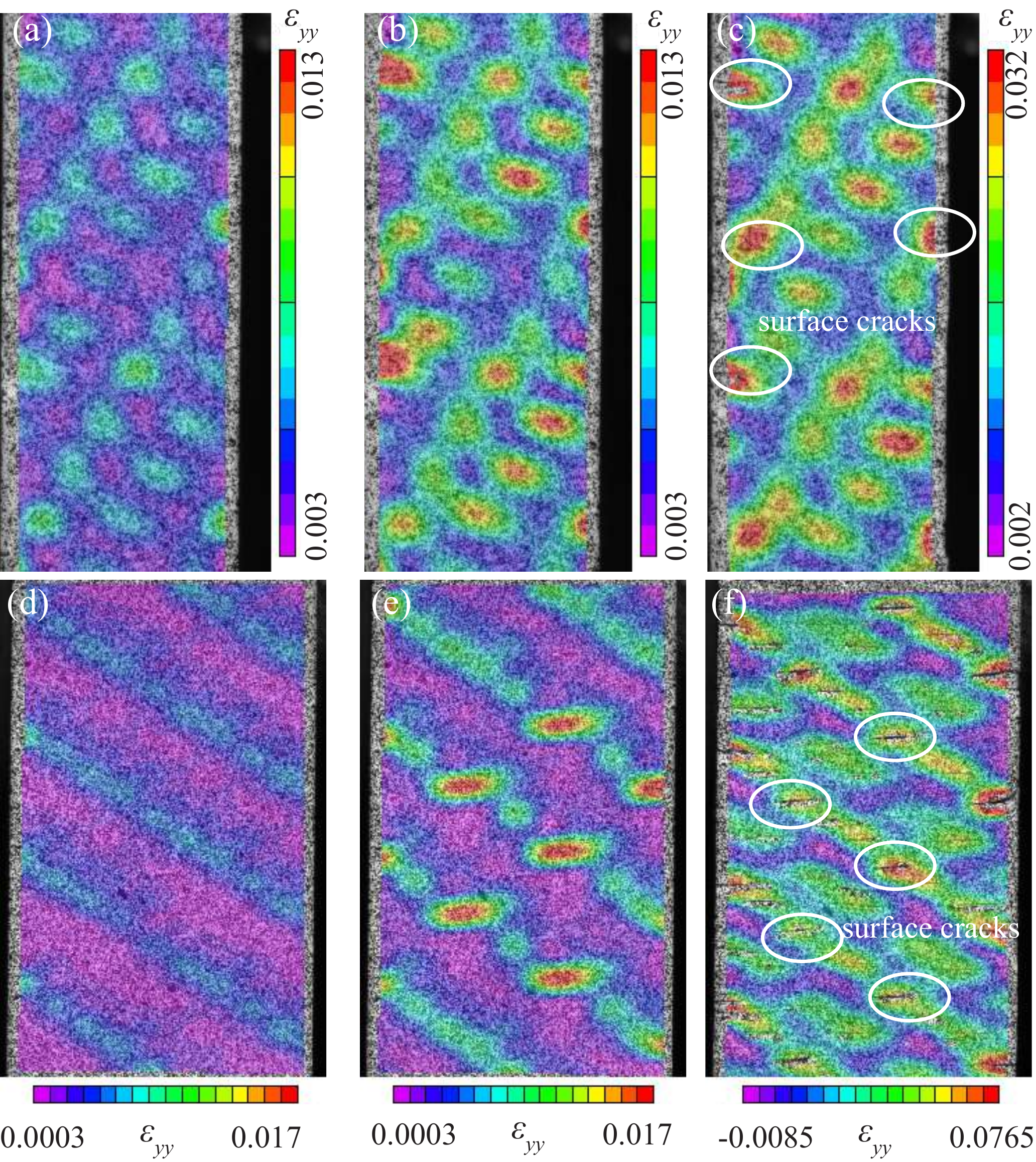}
			\caption[Typical surface strain fields in the direction of tension loading on on-axis loaded specimens.]{Typical surface strain fields in the direction of tension loading on (a)-(c) warp loaded specimen and (d)-(f) weft loaded specimen in which (a) and (d) represent a state in the elastic region, (b) and (c) represents a state right beyond the elastic region, and (c) and (f) represent a state prior to failure.  }
			\label{fig: OnAxisDIC}
		\end{center}
	\end{figure}

	The red solid lines and the blue dashed lines shown in Fig. \ref{fig: OnAxisTension}a represent the obtained stress and strain relations of the specimens loaded in the warp and weft directions, respectively. Each of the curve can be described by a bi-linear response, which is especially evident for the weft loaded specimens. The first linear segment is related to the elastic behavior, and the axial modulus was calculated as the slope of the segment. The results of the axial modulus with their mean values and Coefficient of Variance (COV) are listed in Table \ref{tab:OnAxis}. The axial modulus for the specimens loaded in the warp direction is larger than the one in the weft direction, which reflects the difference of fiber content evaluated in these two directions. The Poisson's ratio can be also estimated from Fig. \ref{fig: OnAxisTension}b as the slope of transverse strain versus axial strain curve. The scattered data were fitted by a linear function through the least square method to reduce the measurement error and uncertainty arising from the heterogeneous deformation in the investigated material. Initiation of the nonlinear response, or knee point in the stress-strain curves, was believed to be related to transverse crack in weft fiber bundle \cite{osada2003initial} or matrix cracking \cite{warren2015experimental}. Similar to the previous work by Warren et al.\cite{warren2015experimental}, one observed high surface strain concentrations after the knee point, as revealed by Fig. \ref{fig: OnAxisDIC}. Figure \ref{fig: OnAxisDIC}a, b, and c show the typical surface strain fields of the warp loaded specimen in the direction of loading prior to the knee point, after the knee point, and prior to failure, respectively. Similarly, Fig. \ref{fig: OnAxisDIC}d, e, and f show the results of the weft loaded specimens. The strain concentrations were prevalent after the knee points for both specimens, and it can be seen from Fig. \ref{fig: OnAxisDIC}c and f that the strain concentrations coincide with the locations where surface cracks occurred. As discussed in the following section, the surface cracks in the tension tests resulted from matrix cracking. Therefore, one may conclude that the presence of the knee points is strongly related to matrix cracking. It is also worth noting that the strain concentrations are notably evident for the weft loaded specimens, suggesting significant modulus reduction due to matrix cracking as shown in the corresponding stress-strain curves. The nonlinear segment of the stress-strain curves was believed to be related to tow straightening, which imparts strain on the surrounding matrix. As a consequence, since the higher waviness of weft tows allows more tow straightening, it may contribute to the greater strain to failure for the weft loaded specimens as one may note from Fig. \ref{fig: OnAxisTension}a. However, the tensile strength when evaluated in the weft direction is significantly lower compared to the strength in the warp direction, as shown in Fig. \ref{fig: OnAxisTension}a and Table \ref{tab:OnAxis}. This could be the result of lower fiber content in the weft direction than the one in the warp direction. 
	
	\begin{table}[htbp]
		\centering
		\caption{Results of tension and compression tests on on-axis loaded specimens}
		\begin{tabular}{lc>{\centering}m{4cm}>{\centering}m{3.5cm}>{\centering\arraybackslash}m{3cm}}
			\hline
			&  & Axial modulus (COV) [GPa] & Poisson's ratio (COV) [-] & Strength (COV) [MPa] \\
			\hline
			\multirow{2}[0]{*}{Tension} & warp  & 76.8 (2.06\%) & 0.226 (7.45\%) & 893 (3.02\%) \\
			& weft  & 35.2 (1.40\%) & 0.105 (2.04\%) & 271 (3.03\%) \\
			\multirow{2}[0]{*}{Compression} & warp  & 71.4 (3.55\%) & 0.210 (3.92\%) & 327 (2.43\%) \\
			& weft  & 40.3 (0.62\%) & 0.106 (3.96\%) & 196 (2.72\%) \\
			\hline
		\end{tabular}%
		\label{tab:OnAxis}%
	\end{table}%
	
\begin{figure}[htbp]
		\begin{center}
			\includegraphics[width = 0.7\textwidth]{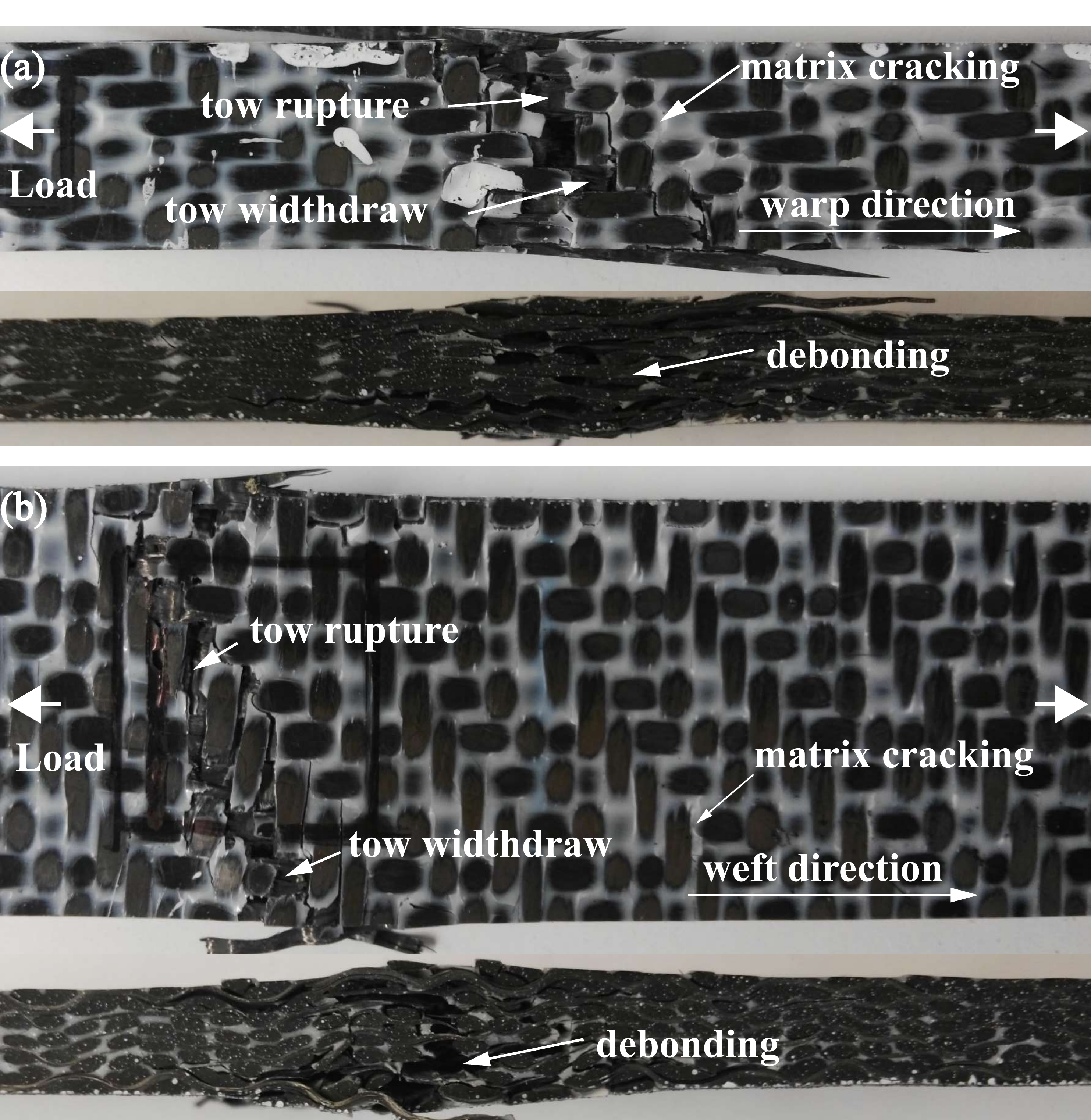}
			\caption{Typical failed specimens in on-axis tension test loaded in the (a) warp direction (b) weft direction. }
			\label{fig: TenFailure}
		\end{center}
	\end{figure}

	The typical failure mechanisms in the on-axis tension tests can be recognized from Fig. \ref{fig: TenFailure}, and coincide with a previous study by Warren et al.\cite{Warren15a}. Tow rupture and withdrawal from surrounding tow were the main failure mechanisms. Matrix cracking adjacent to tows can be also identified on the surface of the specimens, and is responsible for surface cracks one observed in Fig. \ref{fig: OnAxisDIC}c and f. The failure planes were mostly located between the tow/matrix interfaces. When evaluated in the weft direction, as shown in Fig. \ref{fig: TenFailure}b, the failure path was found to be inclined with respect to the weft tows, and the pattern is repeatable among all investigated specimens. On the other hand, evaluation of the specimens loaded in the warp direction shows both inclined failure path as demonstrated in Fig. \ref{fig: TenFailure}a, and the path normal to the loading direction similar to the one shown in Ref. \cite{warren2015experimental}. In addition, extensive debonding between the matrix and the surrounding tows was also observed on the lateral surfaces of the specimens loaded in the both warp and weft directions. The brooming type failure in the lateral view is the result of reinforcement failure in the through-thickness direction following tow rupture. 
	
\begin{figure}[htbp]
		\begin{center}
			\includegraphics[width = 1\textwidth]{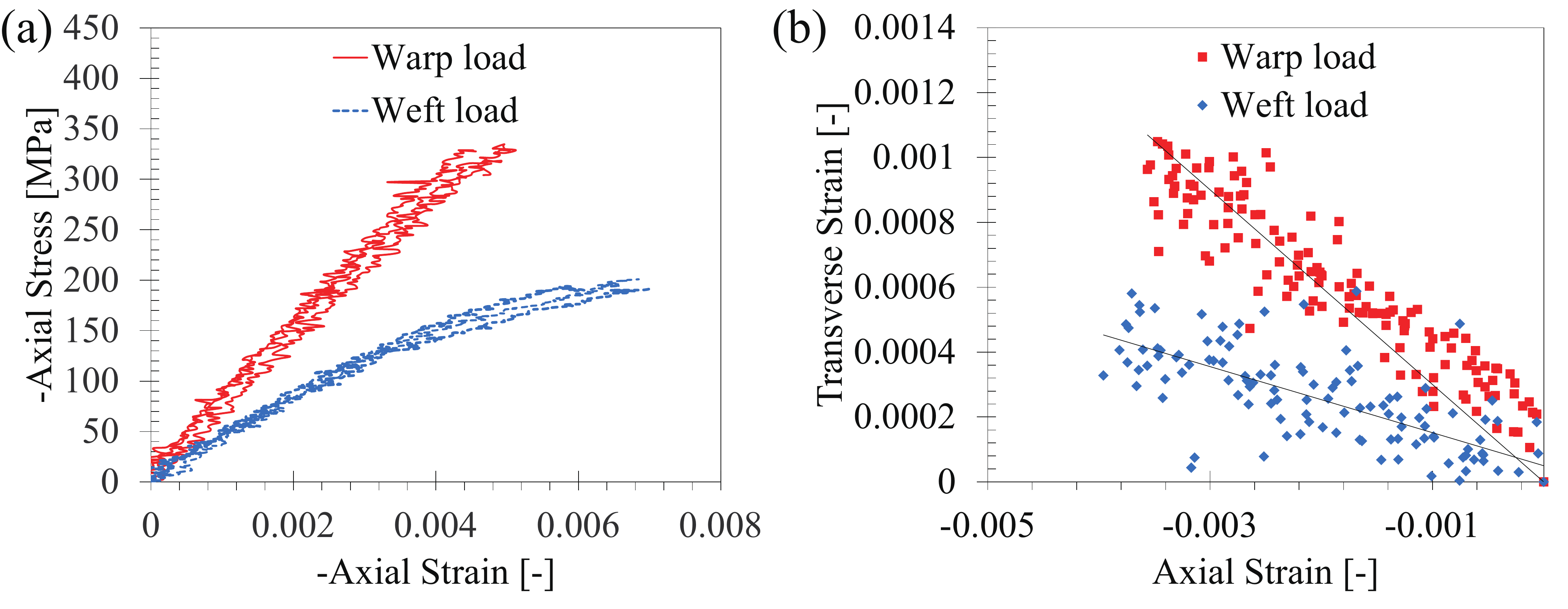}
			\caption{Mechanical responses of on-axis loaded specimens in compression tests: (a) axial stress versus axial strain curves; (b) typical relation between transverse and axial strains.}
			\label{fig: OnAxisComp}
		\end{center}
	\end{figure}

	The experimental results of the compression tests are shown in Fig. \ref{fig: OnAxisComp}a for the axial stress versus axial strain curves and Fig. \ref{fig: OnAxisComp}b for the typical relation between the transverse and the axial strains in the elastic region. Similar to the results of the tension tests, the axial moduli in compression and the Poisson's ratios of both warp and weft loaded specimens can be estimated from the initially linear portion of the stress-strain curves shown in Fig. \ref{fig: OnAxisComp}a and fitted curves in Fig. \ref{fig: OnAxisComp}b, respectively. They are also reported in Table \ref{tab:OnAxis}. The stress-strain curves were almost linear until failure for the specimens compressed in the warp direction, whereas a small portion of pre-peak nonlinearity can be observed for the weft loaded specimens. Similar to the tension results, the compressive strength evaluated in the warp direction is higher than the strength in the weft direction as reported in Table \ref{tab:OnAxis}, which, once again, could be attributed to the difference of fiber content and tow waviness in these two directions. 
	
	As one can see from Table \ref{tab:OnAxis}, the elastic properties measured from the compression test are consistent with the measurements from the tension tests. Accordingly, the axial moduli evaluated in the tension and compression tests were averaged, and reported in Table \ref{tab:prop} for $E_1$ representing in-plane modulus in the warp direction whereas $E_2$ in the weft direction. Similarly, the in-plane Poisson's ratio, $\nu_{12}$ was calculated from the results of the warp-loaded specimens, and reported in Table \ref{tab:prop}. One may also calculate $\nu_{21}$ from the weft-loaded specimens, however, it is related to $\nu_{12}$ through the relation $\nu_{12}/E_1 = \nu_{21}/E_2$, which can be examined by using the results reported in Table \ref{tab:OnAxis}.

\begin{figure}[htbp]
		\begin{center}
			\includegraphics[width = 0.7\textwidth]{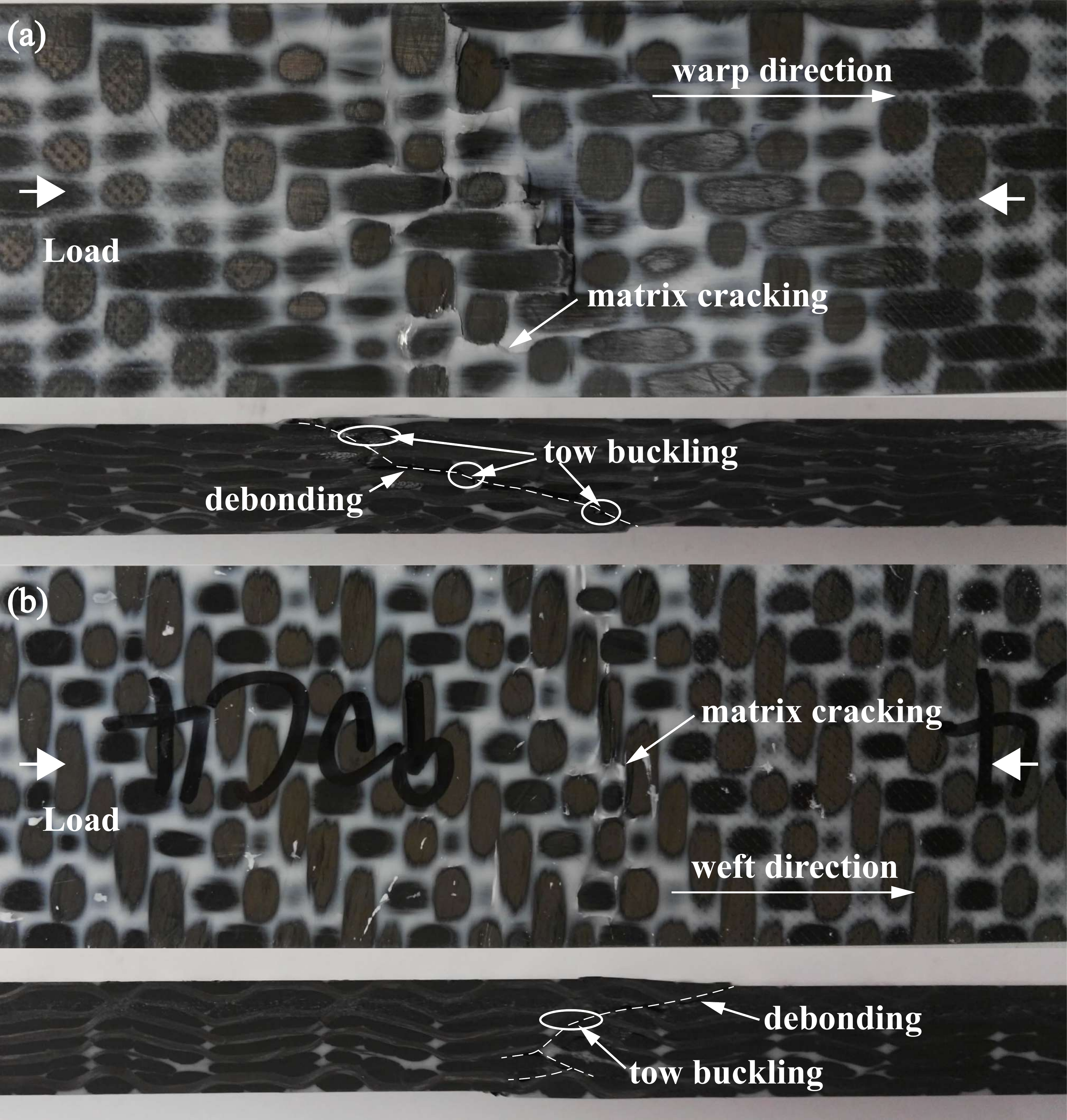}
			\caption{Typical failed specimens in on-axis compression test loaded in the (a) warp direction (b) weft direction.}
			\label{fig: CompFailure}
		\end{center}
	\end{figure}

	The dominant failure modes in the on-axis compression tests are tow microbuckling, formation of kink band, and matrix/tow debonding, as shown in Fig. \ref{fig: CompFailure}. Matrix cracking can be also observed in the front view of the failed specimens. No brooming type failure was observed for the investigated material, which was commonly observed in 3D woven with orthogonal fiber architecture, 2D textile, and unidirectional composites. This could be attributed to the high tow waviness of the preform and the absence of straight tows (stuffer) for the investigated material. As a consequence, microbuckling rather than macroscopic compressive failure of tows was observed, and a low angled failure path with respect to the loading direction can be found, which also leads to the relatively low compressive strength and the high ratio of tensile to compressive strengths of the material characterized in this work. 
	
	\subsection{Off-axis uniaxial test}\label{sec:offaxis}
\begin{figure}[htbp]
		\begin{center}
			\includegraphics[width = 1\textwidth]{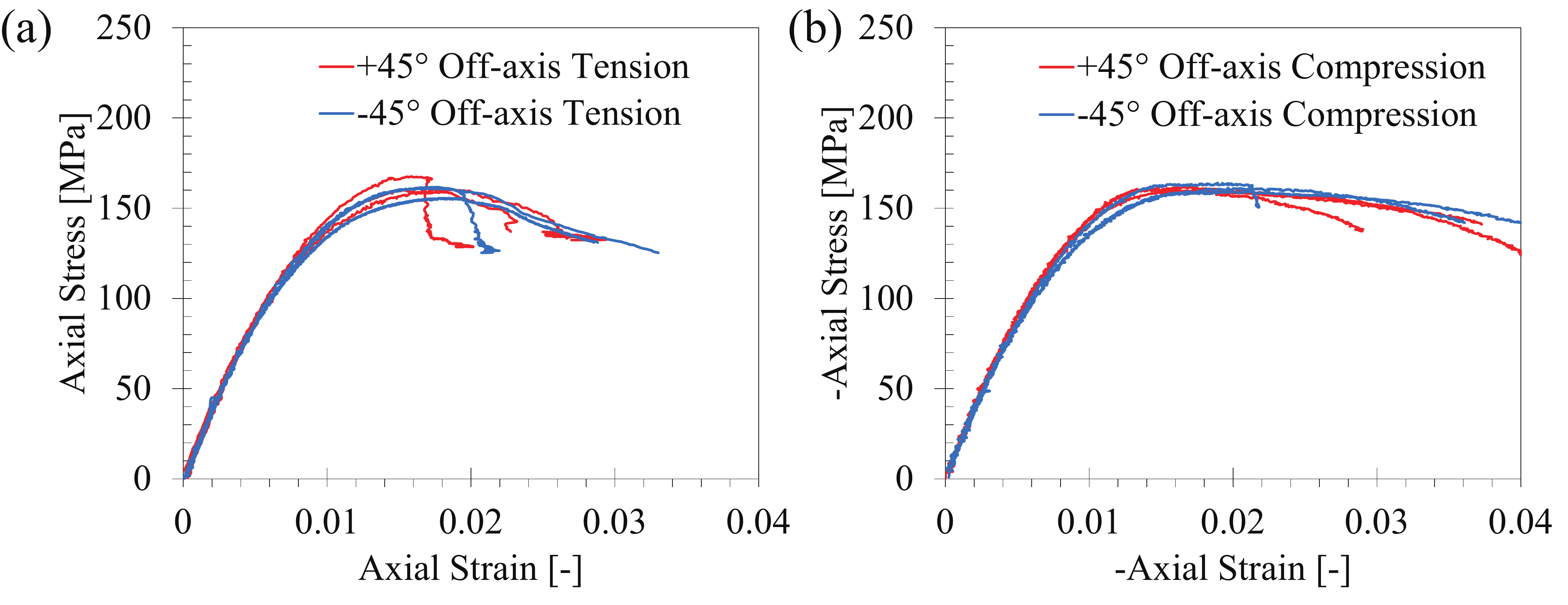}
			\caption{Axial stress versus axial strain responses for the specimens loaded in the $\pm \ang{45}$ off-axis direction under (a) tension and (b) compression.}
			\label{fig: OffAxis}
		\end{center}
	\end{figure}

\begin{figure}[htbp]
		\begin{center}
			\includegraphics[width = 0.7\textwidth]{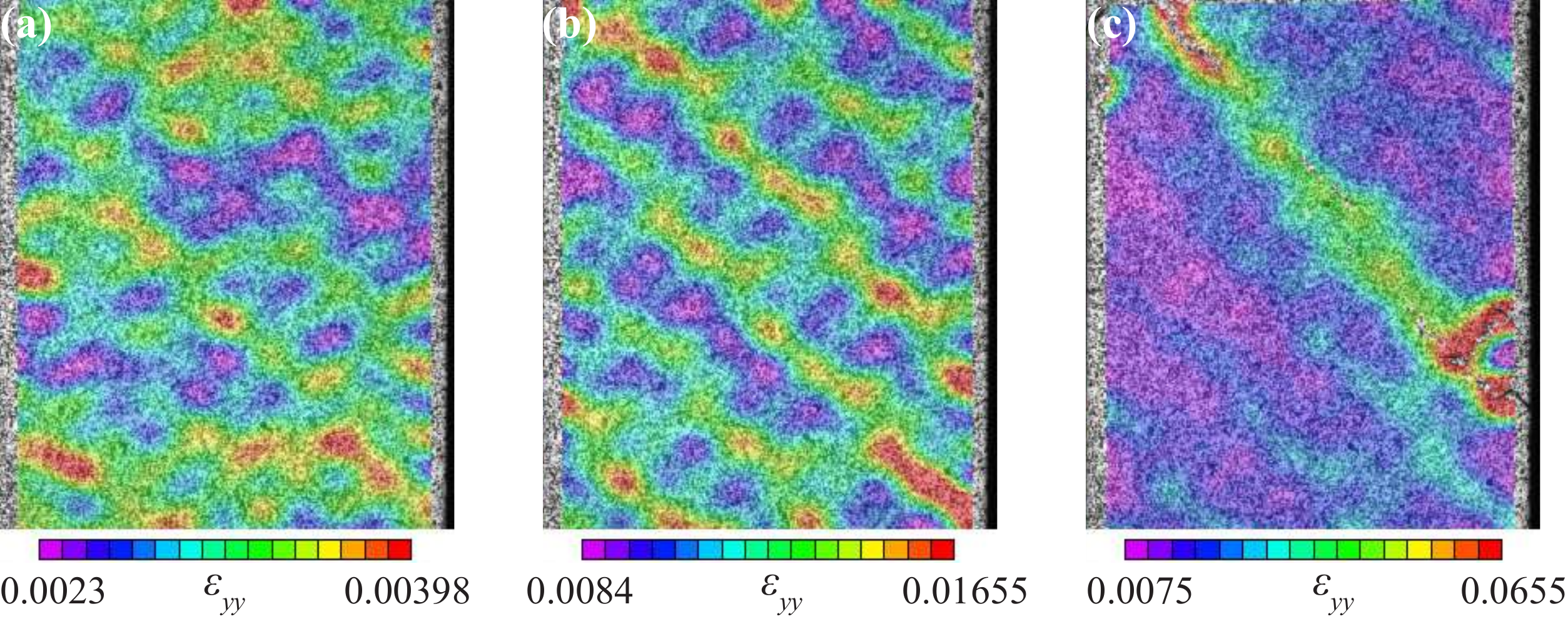}
			\caption[Typical strain fields in the direction of tension loading on a $+\ang{45}$ off-axis loaded specimen. ]{ Typical strain fields in the direction of tension loading on a $+\ang{45}$ off-axis loaded specimen, which represents a state (a) in the elastic region, (b) right beyond the elastic region, and (c) prior to failure.}
			\label{fig: OffAxisDIC}
		\end{center}
	\end{figure}

	Off-axis Specimens were cut and loaded in directions $+\ang{45}$ and $-\ang{45}$ with respect to the principal material axis (warp and weft direction) in order to measure in-plane shear properties. Axial stress versus axial strain curves of the $\pm \ang{45}$ off-axis specimens under tension and compression are shown in Fig. \ref{fig: OffAxis}a and b, respectively. The axial stress and strain were calculated in the same way as the ones for the case of on-axis loading. The stress-strain responses are consistent for all loading configurations, and are generally nonlinear. Three regions can be identified from the stress-strain curves, corresponding to three different stages of loading. Figures \ref{fig: OffAxisDIC}a-c show the strain fields in the loading direction at the three loading stages in a typical off-axis tension test. Fig. \ref{fig: OffAxisDIC}a represents the strain field in the elastic loading stage, and one can observe the inhomogeneous deformation as revealed by the presence of strain concentrations of which the location was randomly distributed. The location of the strain concentrations started following the surface weave pattern of the material when the inelastic region was attained, as shown in Fig. \ref{fig: OffAxisDIC}b. Strain localization with a band of localized strain inclined to the loading direction was observed in Fig. \ref{fig: OffAxisDIC}c as the post-peak region of the stress-strain curve was reached. 		
	
	\begin{table}[htbp]
		\centering
		\caption{Results of tension and compression tests on $\pm \ang{45}$ specimens}
		\resizebox{\textwidth}{!}{
			\begin{tabular}{lcccc}
				\hline
				& \multicolumn{2}{c}{Tension} & \multicolumn{2}{c}{Compression} \\
				\cline{2-5}
				& $+\ang{45}$    & $-\ang{45}$   & $+\ang{45}$    & $-\ang{45}$ \\
				\hline
				Axial modulus (COV) [GPa] & 19.1 (3.08\%) & 18.5 (3.80\%) & 18.1 (1.82\%) & 17.8 (0.95\%) \\
				Poisson's ratio  (COV) [-] & 0.645 (5.18\%) & 0.642 (0.95\%) & 0.671 (0.47\%) & 0.662 (3.52\%) \\
				Strength (COV) [MPa] & 160 (0.70\%) & 159 (0.21\%) & 161 (0.65\%) & 161 (1.50\%) \\
				\hline
			\end{tabular}%
		}
		\label{tab:OffAxis}%
	\end{table}%	
	
	The axial modulus and the Poisson's ratio can be determined from the initial linear segments of the measured responses for each case, and are reported in Table \ref{tab:OffAxis}. The obtained elastic properties are generally consistent with only small deviation. Great consistency of the pre-peak nonlinearity and the peak load can be also found for the investigated $+\ang{45}$ and $-\ang{45}$ off-axis specimens under both tension and compression, suggesting the same damage mechanism and deformation mode. The mechanical responses obtained from the tests with off-axis loading are related to the in-plane shear behaviors of the material. 	
	
	Given the axial modulus at an angle of $\theta = +\ang{45}$ or $-\ang{45}$ to the warp (or weft) direction of the material, denoted by $(E_y)_{\pm \ang{45}}$ in which the subscript $y$ represents the vertical loading direction in the tests, and the corresponding Poisson's ratio, denoted by $(\nu_{xy})_{\pm \ang{45}}$, one can estimate the in-plane shear modulus, $G_{12}$ of the material according to the theory of elasticity for orthogonal materials \cite{daniel2006engineering}, which read 
	\begin{equation}
		G_{12} = \frac{(E_y)_{\pm \ang{45}}}{2[1+(\nu_{xy})_{\pm \ang{45}}]}
	\end{equation}
	%G_{12} = (E_y)_{\pm \ang{45}}/2/(1+(\nu_{xy})_{\pm \ang{45}})$
	Substituting the values of the axial modulus $(E_y)_{\pm \ang{45}}$ and the Poisson's ratios $(\nu_{xy})_{\pm \ang{45}}$ reported in Table \ref{tab:OffAxis}, one can obtain the measure of $G_{12}$ with a mean of 5.6 GPa and a COV of 4.35 \%. The mean value of $G_{12}$ is also reported in Table \ref{tab:prop}.
	
	The in-plane shear strength of the material, $F_{12}$ could be also related to the off-axis tensile strengths, denoted by $(F_{yt})_{\pm \ang{45}}$, and the compressive strength, denoted by $(F_{yc})_{\pm \ang{45}}$, as listed in Table \ref{tab:OffAxis}. If one adopts the maximum stress theory as the failure criterion and realizes that the material failed in shear in the actual tests, the shear stress along the material axes could be considered to attain its maximum value (i.e. $F_{12}$) when the axial stress reaches the peak (i.e. $(F_{yt})_{\pm \ang{45}}$ or $(F_{yc})_{\pm \ang{45}}$). Note that the shear stress along the material axes $\tau_{12}$ is related to the axial stress $\sigma_y$ through the relation $\tau_{12} = -\sigma_y \sin\theta \cos \theta$. As a consequence, $F_{12}$ can be approximated as $F_{12} \approx (F_{yt})_{\pm \ang{45}}/2$ or $F_{12} \approx (F_{yc})_{\pm \ang{45}}/2$ given $\theta = \ang{45}$. Aware of the consistency of the off-axis tensile and compressive strengths, one can average $(F_{yt})_{\pm \ang{45}}$ and $(F_{yc})_{\pm \ang{45}}$, and thus obtain a measure of $F_{12}$ as 80 MPa. It is also possible to adopt more complex failure criteria considering the effect of stress interaction and the biaxial state of stress, such as Tsai-Hill and Tsai-Wu criterion, in the estimation of $F_{12}$. The accuracy of the estimation, however, is an open question in the absence of mechanical characterization under biaxial or multiaxial state of stress. It is worth noting that the measure of $F_{12}$ based on the maximum stress theory and the off-axis strengths is in the same order of magnitude as the in-plane shear strength data reported in Ref. \cite{warren2015experimental} for 3D woven composites with similar architectures and constituents. However, the shear properties were evaluated by using V-notched rail shear method in Ref. \cite{warren2015experimental}, and may not be comparable with the data presented in this work. In addition, as mentioned in Section 1, the accuracy of direct shear methods for 3D woven composites with large unit cells may be also questionable.

	%The accuracy of the shear properties obtained from the off-axis tests may be questioned. Especially, according to ASTM standard \cite{ASTM3518}, the highly inhomogeneous deformation and stress/strain distribution along the specimen cross section may impair the accuracy of the measured properties. However, the same problem also exists for Arcan, Iosipescu or other typical shear tests. Indeed, it is even worse if V-notched specimens, which are commonly adopted in Arcan and Iosipescu test, have to be used to introduce stress concentration. In this case, the location where the specimen is cut from the panel could impose a strong measurement bias considering the huge RUC and the acuity of the notch. In this work, the concern about the influence of material inhomogeneity may be alleviated by averaging the strain fields over a sufficiently large AOI. 
	
	%It was also suggested that the off-axis test tends to overestimate or underestimate the in-plane shear strength because of the presence of a normal tensile or compressive stress. Various interaction failure criteria, such as Tsai-Hill or Tsai-Wu criterion, were proposed to address this issue. 	
	
	\begin{figure}[htbp]
		\begin{center}
			\includegraphics[width = 0.7\textwidth]{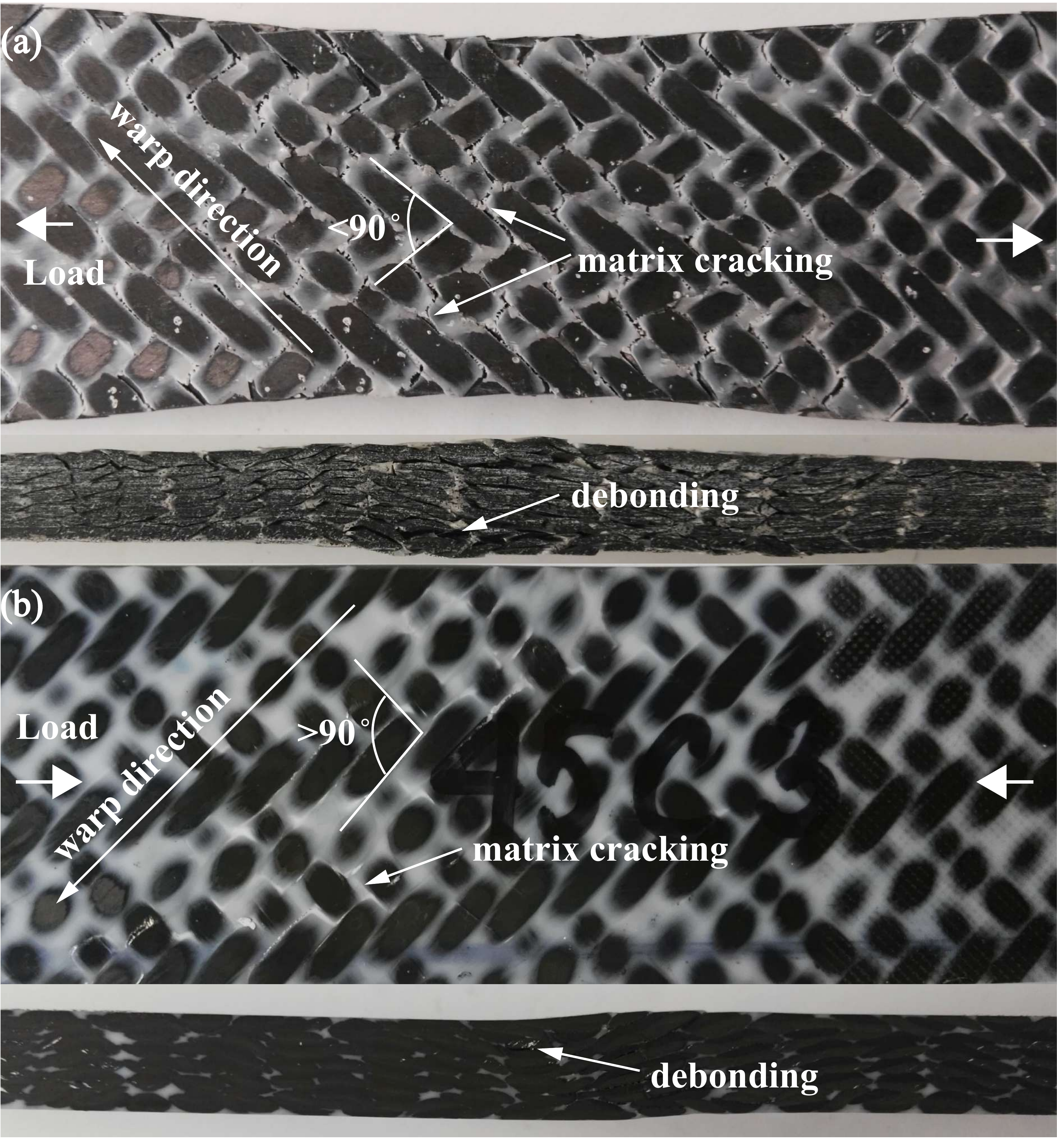}
			\caption{Typical $\pm \ang{45}$ off-axis specimens failed in (a) tension (b) compression. }
			\label{fig: ShearFailure}
		\end{center}
	\end{figure}					
	
	The failure modes of the typical $\pm \ang{45}$ off-axis specimens under tension and compression loads are shown in Fig. \ref{fig: ShearFailure}a and b, respectively. Extensive matrix cracking can be identified from the front view of the failed specimens, whereas matrix/tow debonding can be seen from the lateral view in both tension and compression tests. The failure modes observed in Fig. \ref{fig: CompFailure} were captured in the post-peak region in which matrix cracking already reached a state of crack saturation, and the behaviors were governed by fiber re-orienting, referred to as trellising \cite{hufner2009progressive}. Because no biased reinforcement was presented for the investigated material, the trellising behavior can be recognized from the observation that the angle between the originally orthogonal warp and weft tows changed from $\ang{90}$ (less than $\ang{90}$ in the tension tests and greater than $\ang{90}$ in the compression test). The trellising accompanying frictional behavior between tows consumes a larger amount of energy, and leads to larger strain measures as shown in Fig. \ref{fig: OffAxis}.

	\subsection{Out-of-plane compression}\label{sec:TTComp}~
	The uniaxial compression test was used to characterize the out-of-plane Young's modulus $E_3$, Poisson's ratios $\nu_{31}$ and $\nu_{32}$, and compressive strength $F_{3c}$. Prismatic specimens were cut in the third direction through the panel thickness of 28.8 mm. The nominal width and depth of the specimens are 17.5 mm to ensure a aspect ratio (height:width) of about 1.6. The prepared specimens were placed between two steel loading platens with Teflon sheets inserted to reduce the friction of the contact surfaces. Loading was along the out-of-plane direction, and two sets of tests were conducted with the images of the specimens captured on plane 1-3 and plane 2-3 respectively to enable DIC analysis on these two planar surfaces for strain measurements. 
	
\begin{figure}[htbp]
		\begin{center}
			\includegraphics[width = 1\textwidth]{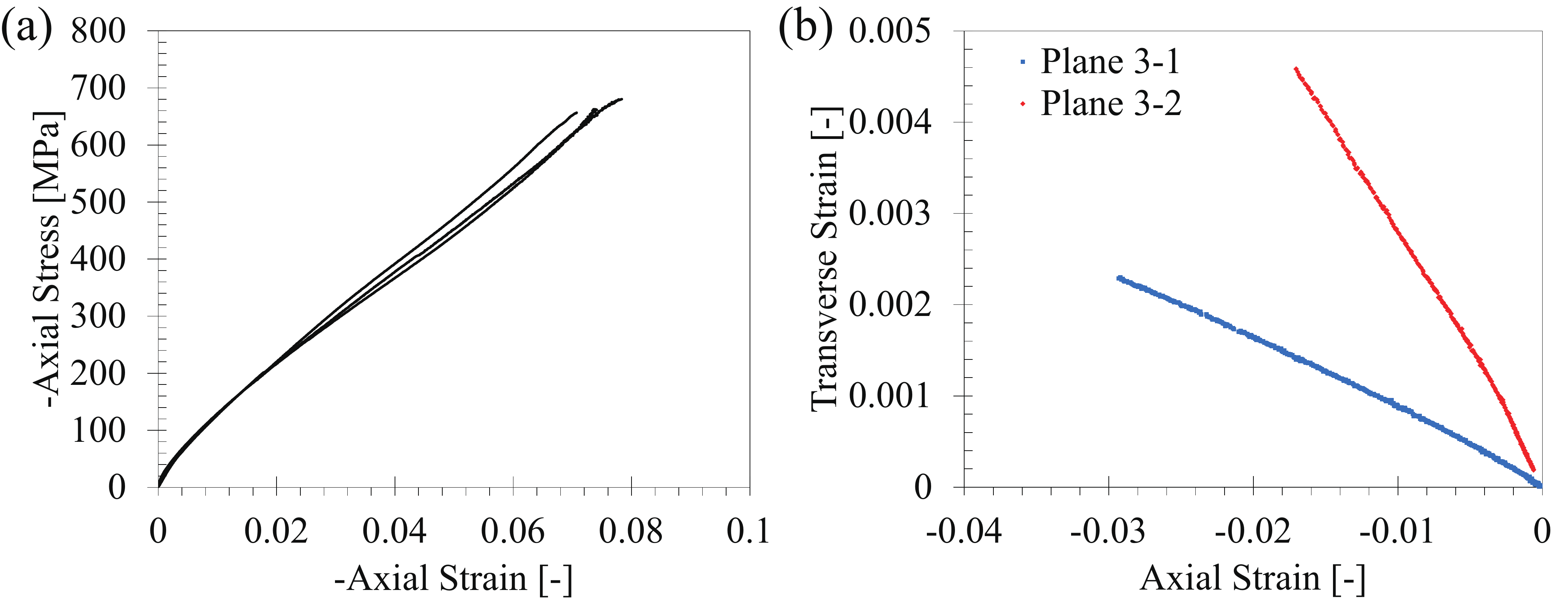}
			\caption{Mechanical responses of specimens under out-of-plane uniaxial compression: (a) axial stress versus axial strain curves; (b) typical relation between transverse and axial strains.}
			\label{fig: TTComp}
		\end{center}
	\end{figure}

    Fig. \ref{fig: TTComp}a shows the stress-strain responses of the specimens under out-of-plane compression, and Fig. \ref{fig: TTComp}b plots the typical transverse strain response versus axial strain measurements in the elastic region. The stress and strain measurements were calculated by using the same method adopted in the in-plane tests as described above. The stress-strain curves are mostly linear until failure, and $E_3$ and $F_{3c}$ were calculated as the slope and peak of the curves, respectively. Similarly, Poisson's ratios can be calculated from Fig. \ref{fig: TTComp}b. Note that the strain measurements conducted on plane 3-1 and plane 3-2 with two sets of test enable one to independently obtain two constants, $\nu_{31}$ and $\nu_{32}$. The obtained properties are also summarized in Table \ref{tab:prop}. It is also worth noting that the obtained out-of-plane compressive strength is about three time larger than the in-plane compressive strengths measured in the warp or weft directions. 
	
\begin{figure}[htbp]
		\begin{center}
			\includegraphics[width = 0.55\textwidth]{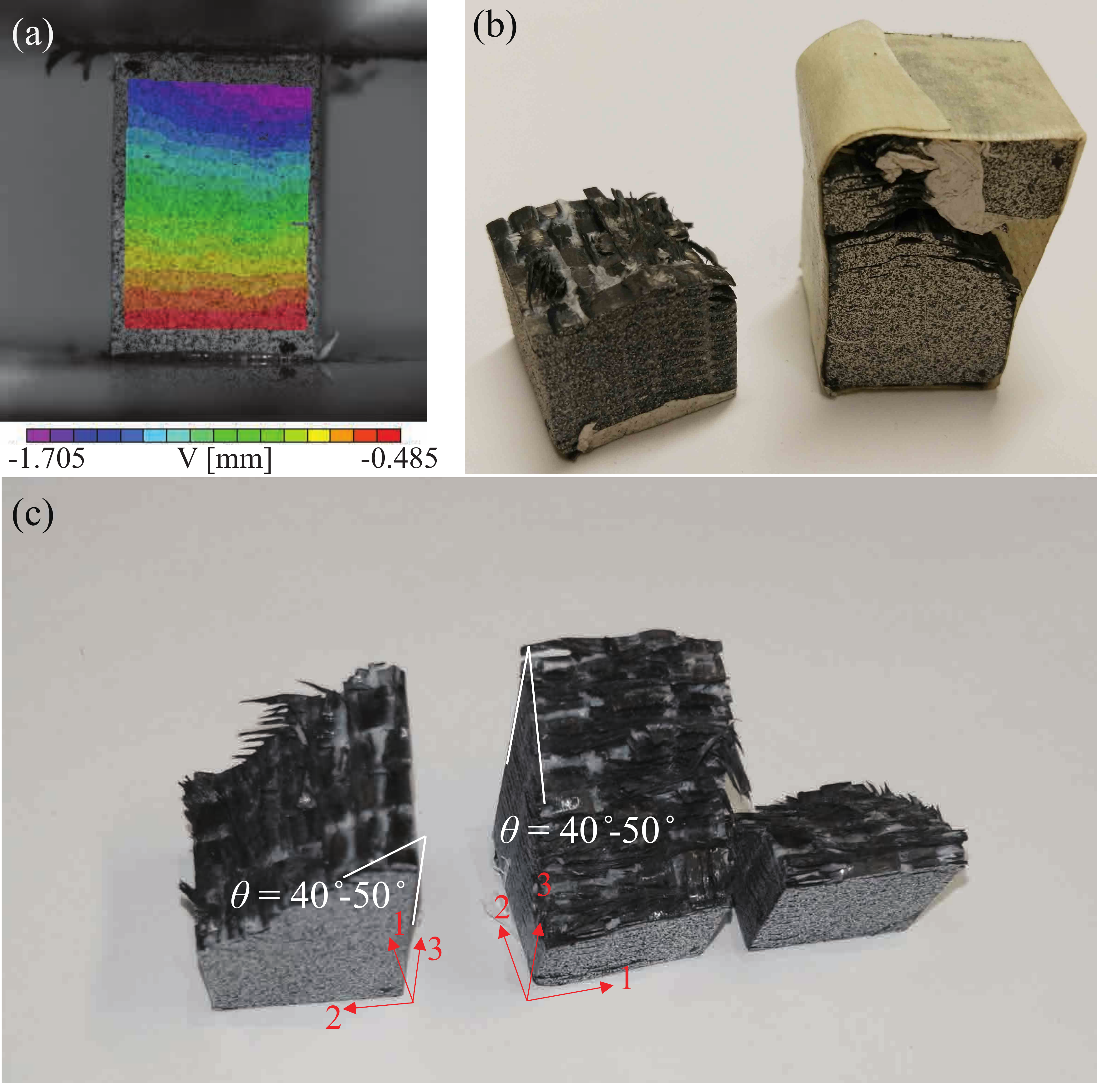}
			\caption{(a) Typical vertical displacement field obtain from DIC analysis in out-of-plane compression test. (b)-(c) Photograph of failed specimens under out-of-plane compression.}
			\label{fig: TTCompFailure}
		\end{center}
	\end{figure}

	Fig. \ref{fig: TTCompFailure}a shows the typical vertical displacement field of the specimen loaded in the out-of-plane direction prior to failure. The surface deformation revealed by the DIC system is generally uniform up to failure. This observation is consistent with the nearly linear stress-strain response plotted in Fig. \ref{fig: TTComp}a. All specimen failed in a catastrophic way right after the load reached its peak. The explosive ejection of fragments perpendicular to the loading direction was observed (see Fig. \ref{fig: TTCompFailure}b). A similar failure mode has been also reported in Ref. \cite{gerlach2012plane}. As shown in Fig. \ref{fig: TTCompFailure}c, some fragments exhibited a failure plane across plane 2-3 with a angle of about $\ang{45}$ to the loading direction, which coincides with the plane of maximum shear stress under uniaxial compression. No extensive fiber microbuckling and kinking were noticed.  
	
	\subsection{Tensile splitting test}\label{sec:TTBraz}~\\
	Tensile splitting test provides an indirect method for measurement of tensile strength of materials and has been widely used in laboratory for mechanical characterization of quasi-brittle and brittle materials such as rock, concrete, and ceramic. Given the difficulties of performing direct uniaxial tension tests on 3D woven composites in the out-of-plane direction associated with the high inter-laminar tensile strength featured by 3D reinforcement, the possibility of applying tensile splitting test for 3D woven composites are explored. 
	
	Cuboid specimens with an edge length of 28.8 mm, which is equal to the thickness of the thick composite panels, were prepared for the tensile splitting tests. Two steel pins as well as the attached bearing blocks were used to transmit the load provided by the machine and to apply a line load on the specimens placed between them. To reduce high stress concentration at the line of contact, two strips of plywood cushion were placed between the loading pins and the specimens. The specimens were positioned carefully to ensure that the line of vertical loading is perpendicular to the out-of-plane direction. Either warp or weft directions of the specimens could be positioned vertically (parallel to the loading direction). A continuously increasing compressive load was provided by the MTS machine until failure occurred.

\begin{figure}[htbp]
		\begin{center}
			\includegraphics[width = 0.5\textwidth]{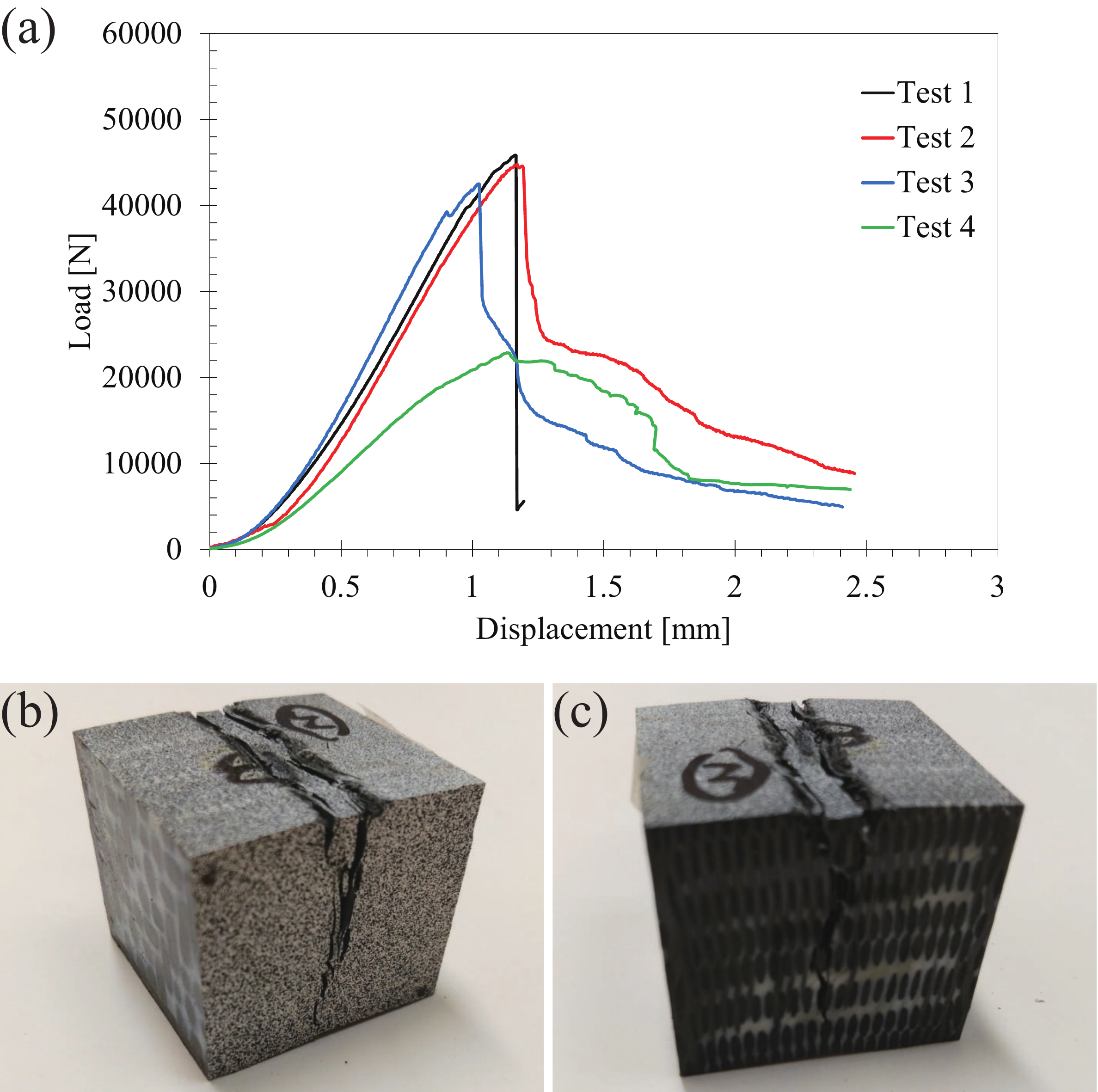}
			\caption{Experimental results of tensile splitting tests: (a) load-displacement responses, (b)-(c) photograph of a failed specimen.}
			\label{fig: BTSLD}
		\end{center}
	\end{figure}

	Figure \ref{fig: BTSLD} shows the load-displacement responses of the investigated specimens in the tensile splitting tests. Four tests were recorded and can be categorized into two groups: test 1, 2, and 3 were performed on the specimens loaded in the warp direction, whereas test 4 in the weft direction. All curves can be described by an initial stage with a gentle slope followed by a segment with linear growth of load and a small portion of pre-peak nonlinearity. The gentle slope at the initial stage can be attributed to adjusting contact between the specimens and the loading pins. The peak load with a mean of 44.4 kN and a COV of 3.84\% was reported for the warp loaded specimens, whereas 22.9 kN for the weft loaded one. Peak load in tensile splitting test was believed to be related to the tensile strength of material, and in this work, out-of-plane tensile strength $F_{3t}$ of the investigated 3D woven composite through the formula: $\sigma_t = k P_{max}/{Dt}$ in which the parameter $k$ depends on specimen geometry and loading condition. For anisotropic materials, the parameter $k$ also depends on the elastic constants. In this work, finite element analysis was performed to interpret the experimental results as shown in Appendix \ref{Sec:App1}. The parameter $k$ was estimated as 0.213 and 0.356 for specimens loaded in warp and weft directions, respectively. Accordingly, one obtains $F_{3t}$ of mean 12.5 MPa and COV 5.14\% from test 1, 2, and 3, whereas $F_{3t} \approx 11.0$ MPa from test 4. The values of $F_{3t}$ obtained from these two groups of tests are in a good agreement. 
	
	\begin{figure}[htbp]
		\begin{center}
			\includegraphics[width = 0.7\textwidth]{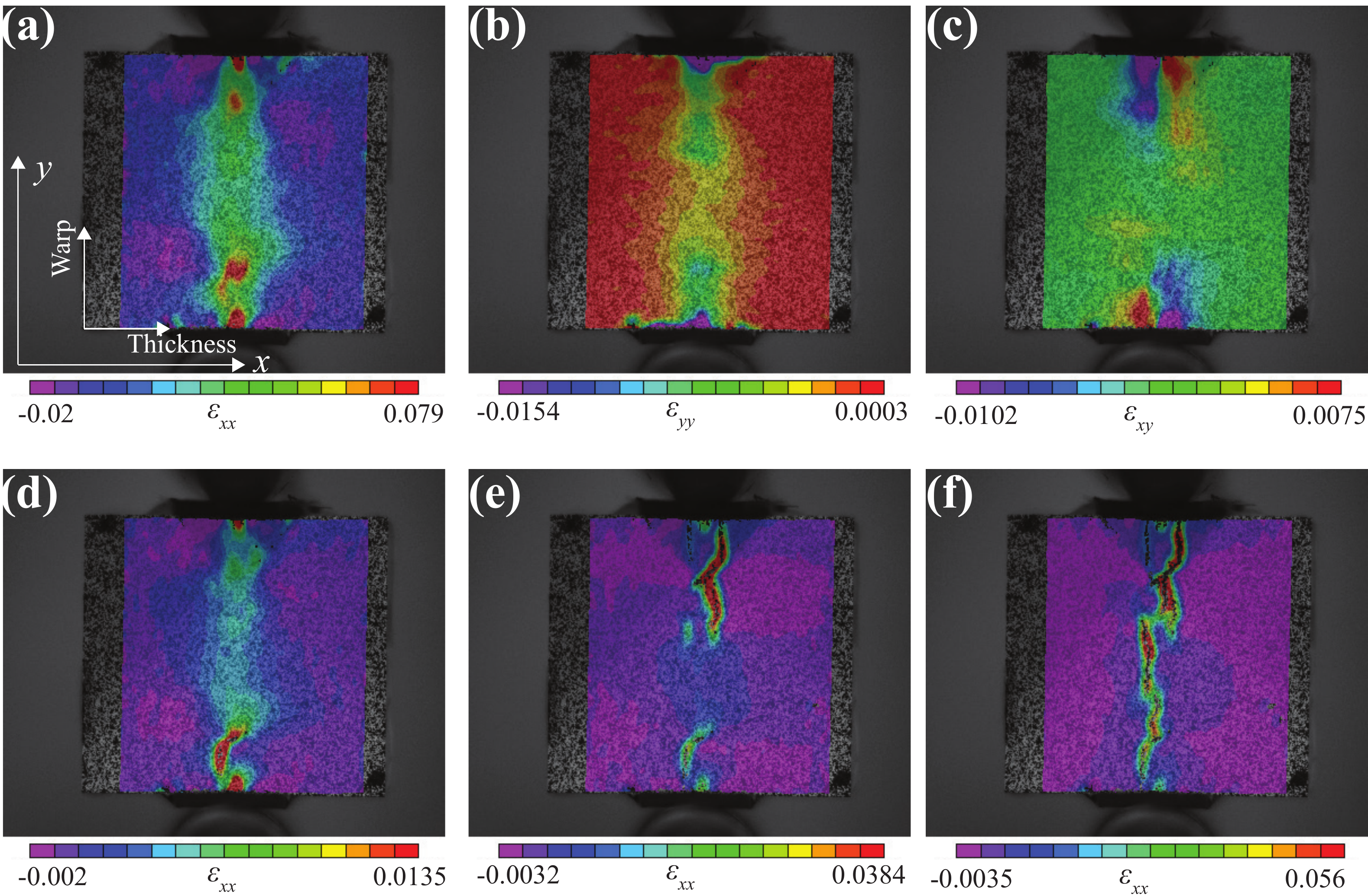}
			\caption[Strain fields obtained from DIC analysis for specimens during tensile splitting tests. ]{Strain fields obtained from DIC analysis: (a) $\varepsilon_{xx}$, (b) $\varepsilon_{yy}$, and (c) $\varepsilon_{xy}$ prior to failure, and (d)-(f) $\varepsilon_{xx}$ strain field prior to, upon, and after peak load.}
			\label{fig: BTSDIC}
		\end{center}
	\end{figure}
	
	The photographs of a typical failed specimen is shown in Fig. \ref{fig: BTSLD}b and c. A splitting failure mode is usually expected for tensile splitting tests. However, in addition to the splitting failure developed near the center of the specimen, one may also notice from Fig. \ref{fig: BTSLD} that a wedge was formed from the upper half of the specimen. The formation of wedges was commonly observed in tensile splitting test on concrete or rocks, and was believed to result from compression-shear failure in the vicinity of the load application zone due to high stress concentrations \cite{li2013brazilian}. Some researchers reported that the formation of wedges occurred after the development of main cracks driven by the tensile stress in the center of a specimen \cite{lopez2008meso,bazant1991size}, whereas some reported that failure initiation was driven by compression-shear rather than tensile stresses \cite{hobbs1964tensile,steen2005observed}. %The latter scenario may jeopardize the applicability of tensile splitting test in identifying the tensile strength of materials. 
	Thanks to the DIC technique, one can examine the two possibilities by keeping track of the occurrence of failure associated the evolution of the surface strain fields during a test. 
	
	Figure \ref{fig: BTSDIC}a, b, and c show the horizontal normal strain ($\varepsilon_{xx}$), the vertical normal strain ($\varepsilon_{yy}$), and the shear strain ($\varepsilon_{xy}$) obtained from the DIC analysis for test 3 prior to failure, respectively. It can be seen that significant strain concentrations for both $\varepsilon_{xx}$, $\varepsilon_{yy}$, and $\varepsilon_{xy}$ were developed in the vicinity of the top and bottom load application zones. Fig. \ref{fig: BTSDIC}d, e, and f show the snapshots of the $\varepsilon_{xx}$ strain field load obtained prior to, upon, and after the load reached its peak. As one can see, surface cracks initiated from the bottom (Fig. \ref{fig: BTSDIC}d) and top (Fig. \ref{fig: BTSDIC}e) load application zones, and propagated to develop a failure path through the specimen (Fig. \ref{fig: BTSDIC}f) along the loading direction. This observation is also echoed by the numerical analysis presented in Appendix \ref{Sec:App1_1} which showed large stress concentration in the load application zones. For this reason, the tensile strength calculated based on the measured peak load should be regarded as a lower bound of $F_{3t}$ of the material. 
	%One may also notice that the calculated strength value is lower compared to the existing literature data \cite{stig2009assessment,gerlach2012plane}. 	
	Possible ways to improve the test setup for achieving a more reliable strength measure is outside the scope of this work but were offered in Appendix \ref{Sec:App1_2} for the interest of readers.

	\section{Characterization of in-plane fracture properties through size effect testing}
	Size effect testing was employed in this work for measuring the in-plane fracture properties of the investigated 3D woven composite. Recognizing the occurrence of the non-negligible FPZ in the presence of a large stress-free crack for quasi-brittle materials, the size effect method was developed based on the energetic size effect law associated with a given structural geometry \cite{bazant1997fracture}. It is easy to implement given that only peak load measurements are needed and provides an alternative method of characterizing the nonlinear fracture properties of quasi-brittle materials which cannot be easily estimated by measuring the specimen post-peak responses in a fracture test. The size effect method has been successfully applied to obtain the intra-laminar fracture properties of laminated and two-dimensional textile composites \cite{bazant1996size,bavzant1999size,salviato2016experimental} and to characterize the inter-laminar fracturing behaviors \cite{salviato2019mode}. In this work, the in-plane fracture energy and a measure of the FPZ length of the investigated material were obtained through size effect testing. 
	
	\subsection{Test description}
\begin{figure}[htbp]
		\begin{center}
			\includegraphics[width = 0.6\textwidth]{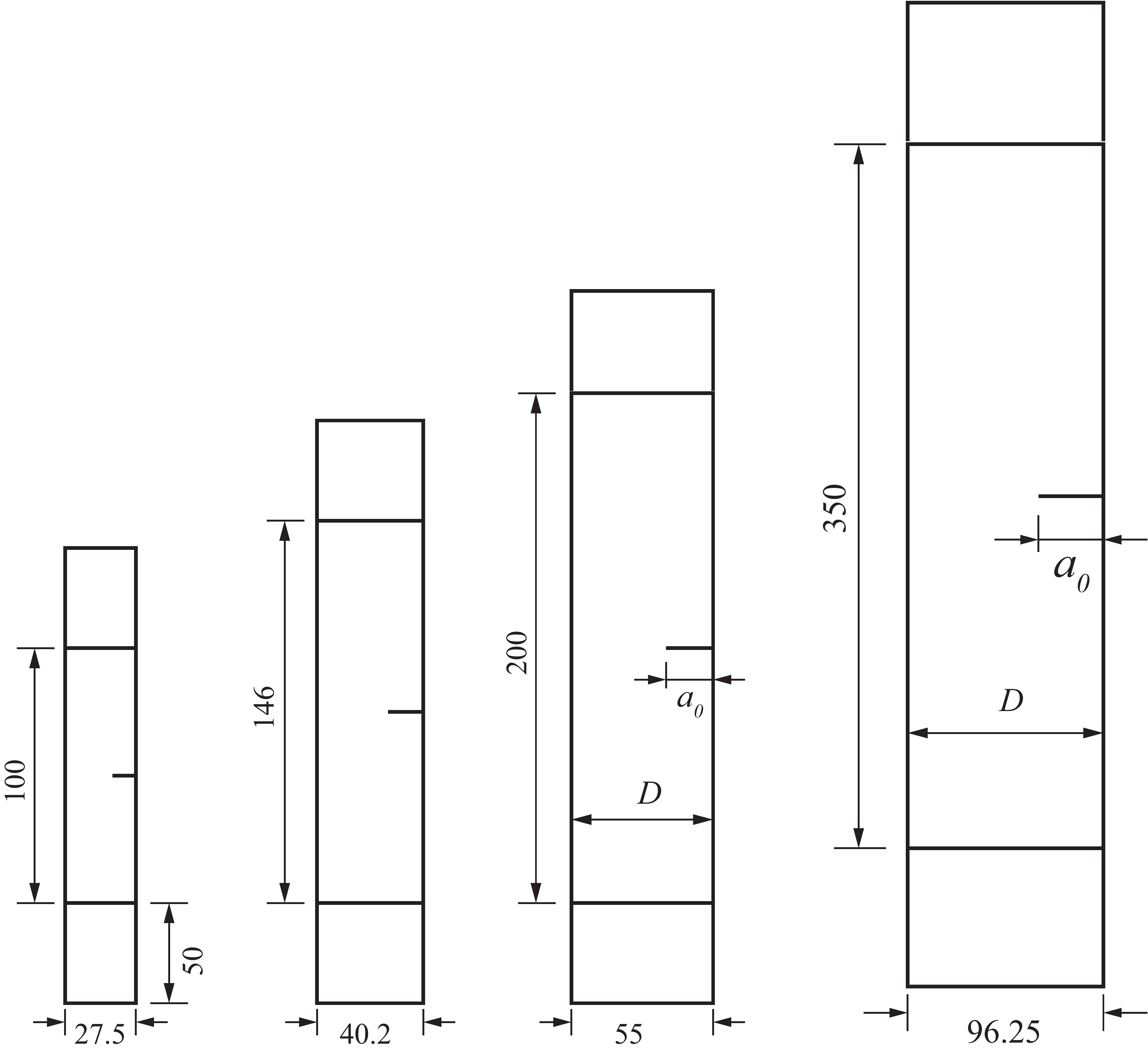}
			\caption{Geometry and dimensions of the SENT specimens. Units: mm}
			\label{fig: FracSP}
		\end{center}
	\end{figure}

	The size effect tests were performed on Single Edge Notched Tension (SENT) specimens. The SENT specimen design was adopted because it offers the simplest specimen geometry compared to other typical fracture tests such as Double Edge Notched Tension (DENT) and Compact Tension. In addition, single edge notch is preferred to double edge notch because the scenario that only one of the cracks can propagate was frequently observed \cite{bazant1992bifurcation} and the double edge notch was also associated with asymmetric response. As illustrated in Fig. \ref{fig: FracSP}, geometrically similar SENT specimens of four increasing sizes, with a size ratio of 1:1.5:2:3.5, were tested to investigate the size dependence of the measured fracture properties. The geometric specifications of the specimens are reported in Table \ref{tab:SE_Size}. Note that two-dimensional geometric similarity with scaled planar dimensions and a constant thickness (7.2 mm) was considered in this work. The notches were made by means of a hand saw which provided a width of roughly 1 mm. Although only blunt notches were prepared, trial tests confirmed that there is no significant difference of results obtained with blunt and sharp (notch tip radius $\approx 20 \si{\um}$) notches, which could be attributed to the fact that the fracturing behavior of the material can be similar if the notch radius is sufficiently small compared to the Irwin's characteristic length and the size of the Repetitive Unit Cell (RUC) \cite{cusatis-and-diluzio, salviatonotch,gomeznotch} and this length is definitely large for the investigated three-dimensional woven composites. % changed by Yao_11/10/2020

	%quasi-brittle nature of the material for the investigated sizes and the notch insensitivity featured by 3D reinforcement \cite{cox1994failure}. 
	
	%The tests on the first three sizes (size ratios of 1, 1.5, and 2) were conducted at Northwestern University, and the tests on the largest specimens (size ratio of 3.5) were conducted at University of Washington. 

    % Table generated by Excel2LaTeX from sheet 'Fracture SENT'
    \begin{table}[htbp]
      \centering
      \caption{Geometric specifications of the SENT specimens.}
        \begin{tabular}{lcccc}
			\hline
			Size  & \multicolumn{1}{>{\centering\arraybackslash}m{2cm}}{Width, $D$ [mm]} & \multicolumn{1}{>{\centering\arraybackslash}m{2.6cm}}{Notch length, $a_0$ [mm]} & \multicolumn{1}{>{\centering\arraybackslash}m{2.6cm}}{Gauge length, $L$ [mm]} & \multicolumn{1}{>{\centering\arraybackslash}m{2.0cm}}{Thickness, $t$ [mm]} \\
			\hline
        Size 1 & 27.5  & 14.6  & 100   & 7.2 \\
        Size 2 & 40.2  & 10.7  & 146   & 7.2 \\
        Size 3 & 55    & 7.3   & 200   & 7.2 \\
        Size 4 & 96.25 & 51.1  & 350   & 7.2 \\
        \hline
        \end{tabular}%
      \label{tab:SE_Size}%
    \end{table}%
	
	The tests were conducted on an Instron machine, and hydraulic wedge grips were used for gripping specimens. Note that although the grip length was not scaled, the size effect data should not be affected because gripping has no appreciable effect on the stored energy and fracture always occurred away from the grips \cite{salviato2016experimental}. Similar to the in-plane tension and compression tests, the fracture tests were carried out under a stroke control mode in a quasi-static setting. 	
	
	\subsection{Experimental results} \label{sec: fracturedata}
\begin{figure}[htbp]
		\begin{center}
			\includegraphics[width = 1.0\textwidth]{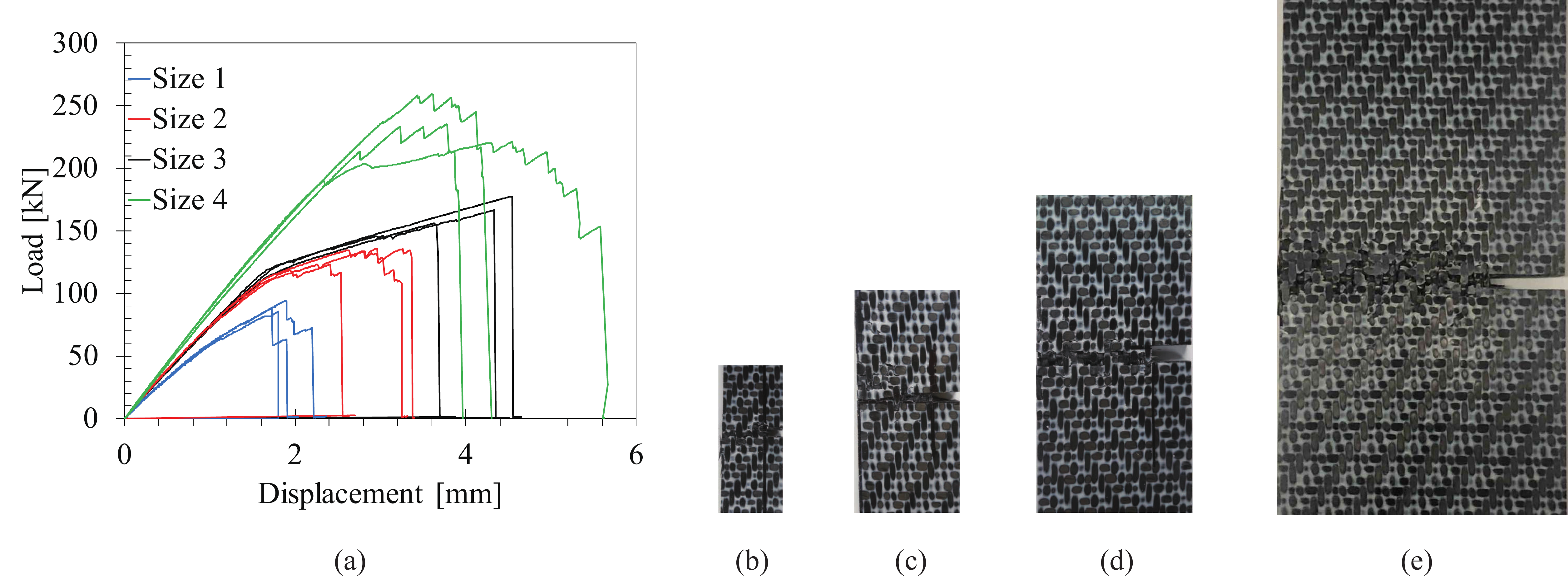}
			\caption[Load-displacement curves of the SENT specimens of various sizes and typical failure pattern. ]{(a) Load-displacement curves of the SENT specimens of various sizes. Typical failure patterns of SENT specimens of width (b) $D = 27.5$ mm (c) $D = 40.2$ mm (d) $D = 55$ mm (e) $D = 96.25$ mm. }
			\label{fig: SENT}
		\end{center}
	\end{figure}

	The load-displacement responses of the SENT specimens during the fracture tests are shown in Fig. \ref{fig: SENT}a. For the large and medium specimens, the response can be described by a bilinear curve, whereas for the small specimens, the load-displacement curves are mostly linear with a small portion of pre-peak nonlinearity. The bilinear or pre-peak nonlinear curves may be an indication of hardening inelastic behaviors or ductility. Nevertheless, specimens exhibited snap-back instability for all investigated sizes. Although a few small and medium sized specimens regained partial load bearing capacity after reaching the peak load, a sudden drop of load was always observed for all of the specimens. None of these tests yield a complete load-displacement curve with a gradual post-peak response, and all of the specimens failed in a catastrophic (dynamic) way shortly after the peak load. Figure \ref{fig: SENT}b-e shows the typical failure patterns of the investigated specimens of various sizes. Extensive tow rupture and pull-out can be noticed. The test results are summarized in Table \ref{tab:SizeEffect}. The failure paths, although mostly found to be along the extension of the notches, are zigzag and have the tendency to following the weaving pattern of the material. 
	
    % Table generated by Excel2LaTeX from sheet 'Fracture SENT'
    \begin{table}[htbp]
      \centering
		\caption{Experimental results of fracture tests on SENT specimens. }
		\begin{tabular}{m{1.2cm}>{\centering}m{3.5cm}cccl}
			\hline
			Size & Nominal strength, $\sigma_{Nu}$ [MPa] & \multicolumn{2}{>{\centering\arraybackslash}m{5cm}}{Apparent fracture toughness, $K_{IC}$  [MPa$\sqrt{\text{m}}$]} & \multicolumn{2}{>{\centering\arraybackslash}m{4cm}}{Apparent fracture energy, $G_f$ [N/mm]} \\
			\hline 
        \multirow{3}[0]{*}{Size 1} & 435.6 & 72.5  & \multirow{3}[0]{*}{76.1} & 140.9 & \multirow{3}[0]{*}{157.4} \\
              & 480.5 & 80.6  &       & 177.5 &  \\
              & 445.4 & 75.0  &       & 153.8 &  \\
        \hline
        \multirow{3}[0]{*}{Size 2} & 474.5 & 94.0  & \multirow{3}[0]{*}{91.7} & 229.9 & \multirow{3}[0]{*}{223.7} \\
              & 465.7 & 93.8  &       & 237.0 &  \\
              & 434.4 & 87.3  &       & 204.3 &  \\
        \hline
        \multirow{3}[0]{*}{Size 3} & 424.5 & 98.4  & \multirow{3}[0]{*}{98.5} & 252.0 & \multirow{3}[0]{*}{253.1} \\
              & 397.5 & 91.9  &       & 218.1 &  \\
              & 451.1 & 105.1 &       & 289.3 &  \\
        \hline
        \multirow{3}[0]{*}{Size 4} & 325.9 & 101.5 & \multirow{3}[0]{*}{107.6} & 270.9 & \multirow{3}[0]{*}{305.5} \\
              & 368.5 & 115.2 &       & 349.2 &  \\
              & 340.3 & 106.2 &       & 296.6 &  \\
        \hline
        \end{tabular}%
      \label{tab:SizeEffect}%
    \end{table}%
	
	The specimen nominal strength, $\sigma_{Nu}$, defined as the average stress at failure distributed on the unnotched cross section, $\sigma_{Nu} = P_{max}/Dt$, is reported in Table \ref{tab:SizeEffect}. One may notice that $\sigma_{Nu}$ varies with the specimens size. Indeed, according to strength-based criteria, the nominal strength is assumed to be size independent. On the other hand, classic Linear Elastic Fracture Mechanics (LEFM) predicts a decrease of $\sigma_N$ proportional to $D^{-1/2}$, which is also not the case as one can note from Table \ref{tab:SizeEffect}. As a consequence, one can conclude that neither strength-based criteria nor LEFM can capture the variation of the nominal strength with the structural size. 
	
	\subsection{Analysis of in-plane fracture tests by size effect law}
	For analyzing the fracture test data, LEFM is widely used by assuming an infinitesimal FPZ compared to the characteristic size of the investigated specimens. However, as one may notice from Table \ref{tab:SizeEffect}, the apparent fracture toughness and fracture energy calculated from the measured peak load according to LEFM are size-dependent. In addition, LEFM predicts that the nominal strength of a structure should be inversely proportional to the square root of the structural size, which is not the case as indicated in Section \ref{sec: fracturedata}. These two observations imply that LEFM is not applicable for the investigated material. Indeed, researchers have found that typical carbon-epoxy composites have a remarkable FPZ size \cite{hughes2002fracture,green2007experimental,bazant1996size,salviato2016experimental} indicating that quasi-brittleness has to be considered. 
	
	To gain an insight into the quasi-brittleness of the composite system under study, the size effect data of the fracture tests were analyzed by means of type II Size Effect Law (SEL) , which can be derived from an equivalent linear elastic fracture mechanics approach. The type II size effect occurs when a large notch or traction-free crack exists at maximum load, and the resulting SEL bridges the region between strength-based criteria and classic LEFM theory. The derivation of SEL is summarized in Appendix \ref{Sec:App_SEL}. The dimensionless form can be written as
	\begin{equation}\label{Eq:CSEL2}
		\frac{\sigma_{Nu}}{\sigma_0} = \frac{1}{\sqrt{1 + D/D_0}}
	\end{equation}
	where $\sigma_0 = \sqrt{E^*G_f/(c_fg'(\alpha_0))}$ and $D_0 = c_fg'(\alpha_0)/g(\alpha_0)$; $E^*$ is the effective elastic modulus for orthotropic materials; $G_f$ and $c_f$) represent the initial fracture energy and effective FPZ length of the material; $g(\alpha_0)$ and $g'(\alpha_0)$ denote the dimensionless energy release rate and its derivative, and $\alpha_0 = a/D$ the dimensionless initial crack length. Note that $\sigma_0$ has the dimension of stress, and $D_0$ has the dimension of length. It is key that the SEL is endowed with the characteristic length $D_0$ to describe the transition from ductile to brittle behavior with increasing structure size. 	
	
\begin{figure}[htbp]
		\begin{center}
			\includegraphics[width = 1.0\textwidth]{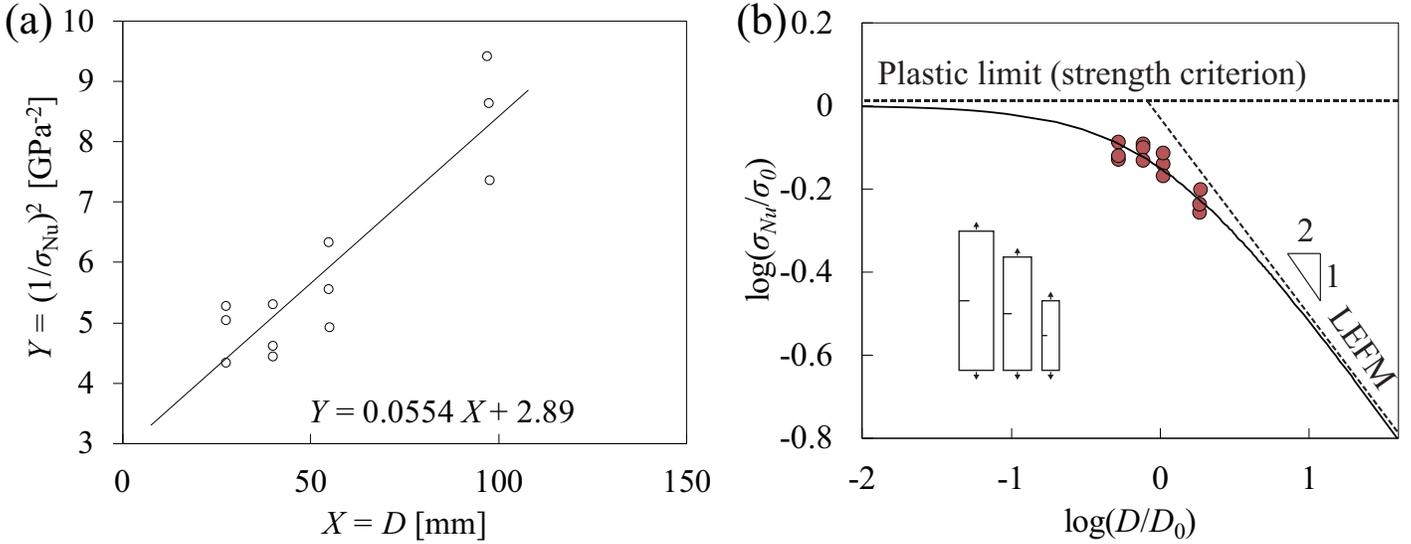}
			\caption{ Analysis of size effect data: (a) linear regression fitting of experimental data; (b) plot of size effect data for 3D woven composite.}
			\label{fig: SizeEffect}
		\end{center}
	\end{figure}

	%\subsubsection{Fitting of experimental data by SEL}~\\
	The parameters $\sigma_0$ and $D_0$ are constants given that the investigated specimens are geometrically similar, and can be determined by either linear or nonlinear regression analysis of the experimental data. For simplicity, a linear regression analysis is adopted in this work, which is enabled by reformulating Eq. \ref{Eq:CSEL2} as
	\begin{equation}\label{Eq:CLR}
		Y = AX + C
	\end{equation}
	in which $X = D$, $Y = \sigma_{Nu}^{-2}$, $A = {C}/{D_0}$, and $C = {\sigma_0^{-2}}$.
% Weixin: I would put these equations inline, they are not important % Done
	%\begin{subequations}
	%	\begin{align}
	%		X& = D, &  Y& = \sigma_{Nu}^{-2} \label{Eq:CXYP}\\
	%		A& = \frac{C}{D_0}, & C& = \frac{1}{\sigma_0^2}  \label{Eq:CACP}
	%	\end{align}
	%\end{subequations} 	
	The fitting of the experimental data through the linear regression analysis based on Eq. \ref{Eq:CLR} was conducted. The results are shown in Fig. \ref{fig: SizeEffect}a, and provided the estimates of the following parameters: $A = 0.0554$ GPa$^{-2}$mm$^{-1}$ and $C = 2.89$ GPa$^{-2}$. Correspondingly, $D_0$ and $\sigma_0$ can be estimated from the coefficients, and they are $D_0 = 52.14$ mm and $\sigma_0 = 589$ MPa. 
	
	The fracture properties of the material can be estimated from the fitting of the size effect data. The initial fracture energy $G_f$ and the effective FPZ length $c_f$ can be determined from $D_0$ and $\sigma_0$ supposing that $g(\alpha_0)$ and $g'(\alpha_0)$ are given (both can be calculated numerically as demonstrated in Ref. \cite{salviato2016experimental}). They were estimated as $G_f = 466$ N/mm and $c_f = 13.6$ mm. 
	
	Figure \ref{fig: SizeEffect}b plots the normalized strength $\sigma_{Nu}/\sigma_0$ as a function of the normalized characteristic size $D/D_0$ in double logarithmic scale, as a result of the fitting of the experimental data by SEL. The red circles represent the experimental data, and the black solid line represents SEL as shown in Eq. \ref{Eq:CSEL2}. It can be seen that SEL depicts a smooth transition from the strength criterion characterized by a horizontal asymptote (plastic limit) when $D \to 0$, in which no size effect on structural strength is expected, to LEFM by an inclined asymptote of slope $-1/2$ when $D \to \infty$, which represents the strongest size effect possible. The intersection of the two asymptotes corresponds to $D =D_0$, referred to as the transitional size. The experimental data fall within the transition zone of the SEL curve. As a consequence, one may conclude that the failure of the investigated 3D woven composites containing traction-free cracks exhibit a significant size effect.  
	
	%\subsubsection{Calculation of $g(\alpha)$}
	%Note that the functions $g(\alpha)$ in Eq. \ref{Eq:CG} and $k(\alpha)$ in Eq. \ref{Eq:CK} can be either obtained from close-form solutions or calculated numerically. For the specimen geometry adopted in the work, Bao et al. provided a solution of $k(\alpha)$, which, however, was  

	It is of vital importance to realize the size dependence of the measured mechanical properties in various situations such as material design, component manufacturing, and safety evaluation, especially, in the presence of a traction-free crack. In particular, these conclusions ought to be incorporated into a design process to enable extrapolation from small scale laboratory tests to real size structures. %In addition, future work on the development of a sophisticated fine scale model is planned to interpret the size effect data of the fracture tests in this work. 
	\subsection{Comparison of size effect experimental data for 3D and 2D composites}~\\
	The size effect data of the 3D woven composite investigated in this work are also compared to the data of a two-dimensional textile composite. The latter ones were reported in Ref. \cite{salviato2016experimental}, and the material consisted of DGEBA based epoxy resin matrix reinforced by a twill 2 $\times$ 2 fabric made of carbon fibers. The fiber volume fraction was reported to be $0.54$, which is close to the one of the 3D woven material investigated in this work. However, their elastic and strength properties are significantly different, most likely due to the contrasting fabric architectures. Note that the unit cell size of the twill 2 $\times$ 2 composite is about 7 mm, which is more than two times smaller than that of the 3D woven composite. The same as in the current work, the size effect data were obtained from the fracture tests on geometrically similar SENT specimens with increasing size, which were made of $[\ang{0}]_8$ twill 2 $\times$ 2 laminates. The investigated specimen width $D$ ranged from 20 to 80 mm. 
	
\begin{figure}[htbp]
		\begin{center}
			\includegraphics[width = 1.0\textwidth]{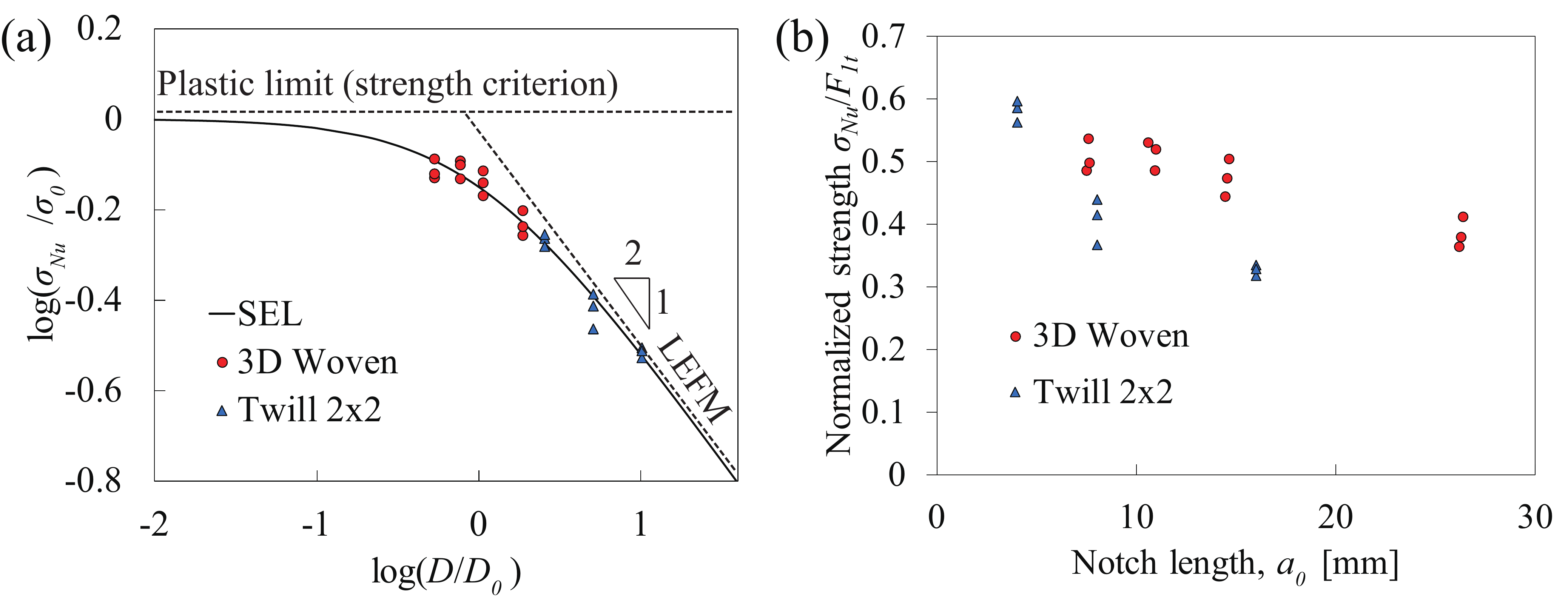}
			\caption[Comparison of size effect data of 3D woven and twill 2$\times$2 composites. ]{ Comparison of size effect data of 3D woven and twill 2$\times$2 composites: (a) plot of normalized nominal strength versus normalized specimen characteristic size; (b) relation between normalized strength and notch length. }
			\label{fig: SizeEffectComp}
		\end{center}
	\end{figure}

	The comparison of the size effect data is shown in Fig. \ref{fig: SizeEffectComp}. Fig. \ref{fig: SizeEffectComp}a depicts the normalized strength $\sigma_{Nu}/\sigma_0$ as a function of the normalized characteristic size $D/D_0$ in double logarithmic scale with $\sigma_0$ and $D_0$ obtained by the SEL fitting of each set of data. Comparing with the SEL curve, one can observe that the 3D woven data are close to the plastic limit while the twill 2 $\times$ 2 are near the LEFM asymptote. Considering that the investigated specimen size ranges are in the same order of magnitude, it is fair to conclude that the twill 2 $\times$ 2 is more brittle than the investigated 3D woven composite. In other words, the 3D woven composite exhibit higher ductility.
	
	Thanks to the high ductility, the 3D woven composite is presumably endowed with higher damage tolerance compared to the 2D textile composite. As illustrated in Fig. \ref{fig: SizeEffectComp}b, the nominal strength of the damaged specimen $\sigma_{Nu}$ was normalized by the strength of the undamaged one, i.e. tensile strength in the warp direction $F_{1t}$, and was plotted against the degree of damage characterized by the notch length $a_0$. It can be seen that within a comparable range of $a_0$, a more severe reduction of the normalized strength was observed for the twill 2 $\times$ 2 composite than for the 3D woven composite. Consequently, one may conclude that the 3D woven composite is less vulnerable to the presence of a large traction-free crack. This conclusion agrees with early findings that certain 3D polymer composites offer reduced notch sensitivity and superior damage tolerance far beyond those of 2D laminates \cite{chou1992effect,cox1994failure,walter2010monotonic}, which provides various benefits including extended allowable damage region, reduced maintenance cost, and potentially longer service life. 
	
	\section{Wedge-driven out-of-plane fracture test}
	The study of delamination is of particular importance for 3D woven composites which feature enhanced resistance to out-of-plane fracture, and is presented in this section. Strictly speaking, the term \textit{delamination} is somewhat confusing for 3D textile composites since there is not a clear laminated structure. Nevertheless, the word \textit{out-of-plane fracture} is used interchangeably with \textit{delamination} in this work for the sake of convenience. 
	
	Conventional delamination testing using the double cantilever beam specimens \cite{D5528} was originally designed to determine the inter-laminar fracture properties of laminated composites and may be also applicable to 2D fabric composites. However, the conventional method may not be suitable to determine the out-of-plane fracture properties of 3D woven composites for the following reasons. (1) The DCB specimens frequently suffer from bending failure prior to delamination given the large delamination resistance and the low in-plane compressive strength of 3D woven composites. To avoid the undesired bending failure, tabbed DCB specimens were frequently used \cite{tanzawa1999interlaminar,tamuzs2003delamination}. Although it seems effective, the effect of tabs has to be analyzed, and methods of preventing delamination between tabs and specimens are needed. (2) Special gripping systems \cite{tamuzs2001progressive,tamuzs2003delamination} should be used to avoid the failure of adhesive bonding specimens and loading devices when a conventional piano hinge or end block loading system is used. In particular, the gripping system has to be designed in such a way that it can withstand strong gripping force given that scaled-up specimens have to be used considering the materials' large RUC. (3) The applicability of LEFM to describing the mode I inter-laminar fracture behavior of 3D woven composites has not been examined yet. This would require a thorough fracture scaling analysis such as the size effect study completed by \cite{salviato2019mode} for laminated composites. A possible way to overcome the aforementioned difficulties is a wedge testing method. It was adopted in this work to investigate the out-of-plane fracture properties of the investigated 3D woven composite. 
	
	\subsection{Test description}
\begin{figure}[htbp]
		\begin{center}
			\includegraphics[width = 0.55\textwidth]{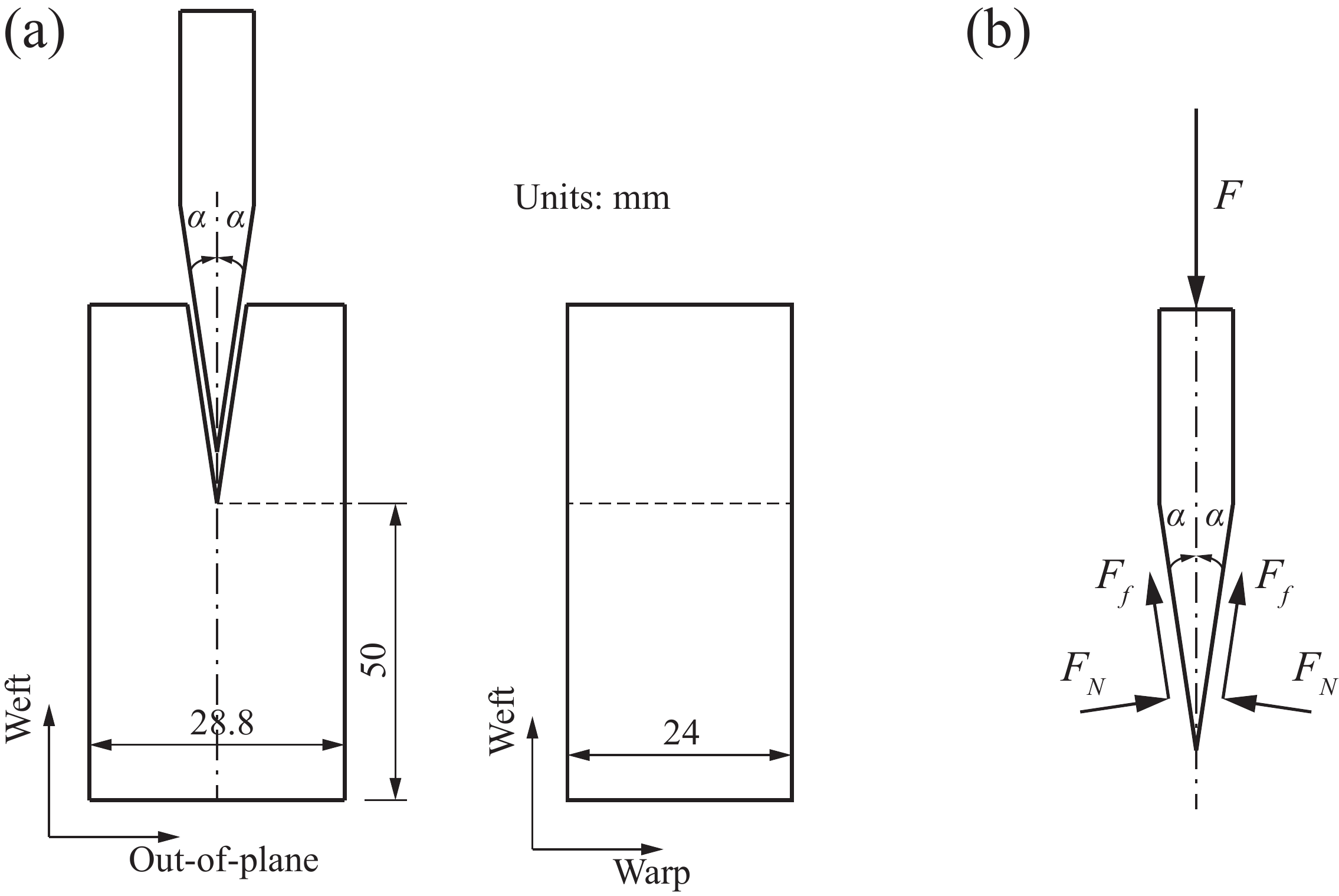}
			\caption[Schematic diagram of wedge test setup. ]{(a) Schematic diagram of wedge test setup and geometry of specimen. (b) Free-body diagram demonstrating forces applied on the wedge. }
			\label{fig: WedgeIllus}
		\end{center}
	\end{figure}

	The experimental setup of the wedge tests performed in this work consists of a prismatic specimen with a V-shaped notch, a steel wedge, and a servo-controlled testing machine. 
	The steel wedge was driven against the stationary specimen and to drive the propagation of crack initiated from the tip of the notch.   
	As depicted in Fig. \ref{fig: WedgeIllus}a, the out-of-plane direction of the material was placed perpendicular to the potential crack path emanated from the notch tip, and the weft direction was parallel to it. The V-shaped notch was cut through the warp direction, and the notch half-angle is the same as the wedge half-angle $\alpha$ which measures $7^{\circ}$ in this work. The specimen dimension is also depicted in Fig. \ref{fig: WedgeIllus}a. 
	
	Efforts were made to minimize the effect of friction resulting from sliding between the wedge and the notch surface. First of all, the wedge surface was polished to ensure a smooth contact. In addition, a stack of plastic sheets with smooth surfaces (made of grade G-10 glass fabric/epoxy) was bonded to the notch surfaces to reduce the contact area. Friction was further reduced by applying a dry film graphite lubricant to the contact surfaces. Ultimately, the coefficient of kinetic friction between the wedge surfaces and the plastic sheets, $\mu$, measures 0.15. 
	
	The tests were performed under the stroke control mode with a constant crosshead speed of about 1.5 mm/min. The load and displacement data were recorded during the tests. 
	
	\subsection{Experimental results}
\begin{figure}[htbp]
		\begin{center}
			\includegraphics[width = 0.8\textwidth]{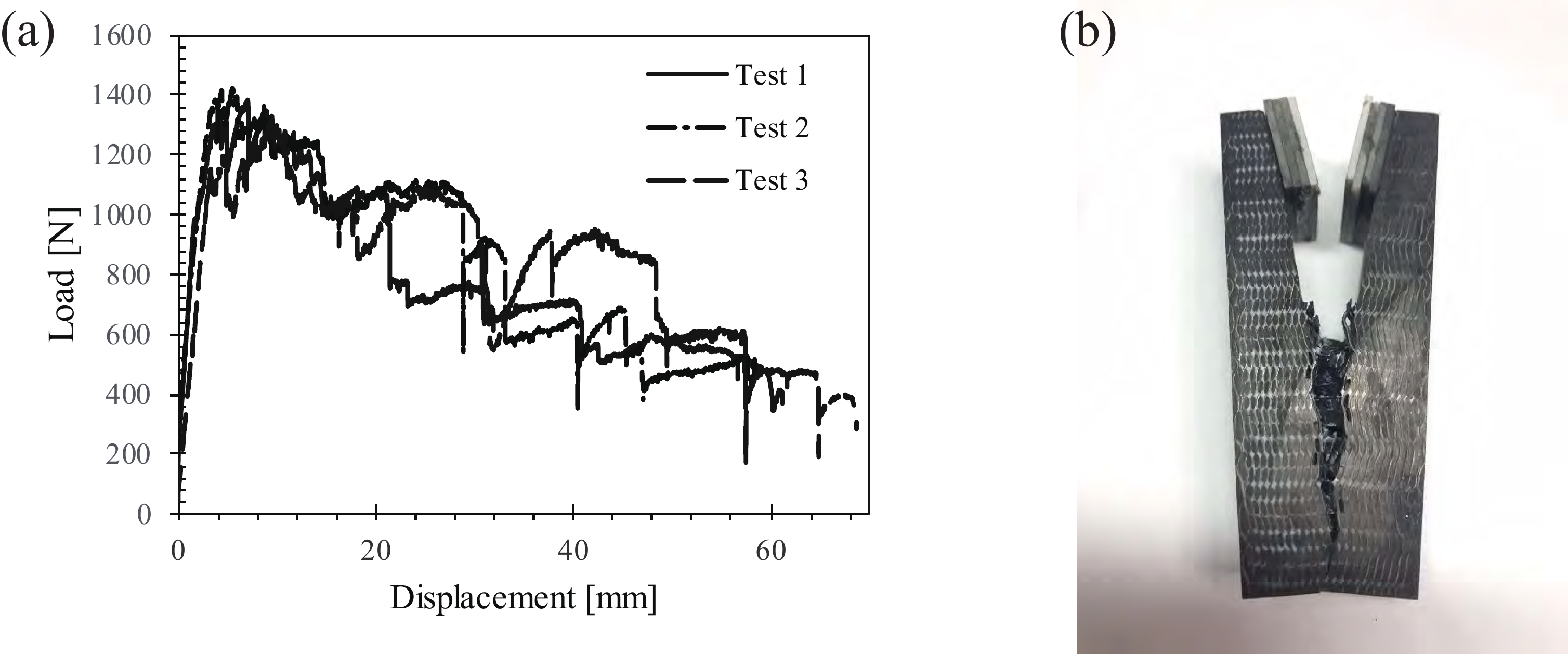}
			\caption[Results of the wedge test. ]{(a) Load-displacement curves of the wedge tests. (b) Typical failed specimen. }
			\label{fig: WedgeResults}
		\end{center}
	\end{figure}

	The load-displacement curves of three independent tests are plotted in Fig. \ref{fig: WedgeResults}a, which feature a rapid increase of load in the pre-peak region and a slow reduction of load in the post-peak region. All of the specimens exhibited saw-tooth post-peak response which is characterized by a stick-slip behavior corresponding to the observation that the crack was repeatedly arrested until sufficient energy was accumulated to enable further crack propagation. Such a stick-slip behavior has been commonly observed in woven composites \cite{alif1997mode,alif1998effect,mouritz1999mode,tamuzs2001progressive} and is related to the complex microstructure with a large RUC. 
	
	A typical failed specimen after the test is shown in Fig. \ref{fig: WedgeResults}b. One can find that the crack path is wavy, partially following the weaving pattern of the weft tows. In particular, a strip of damage zone instead of a single throughgoing crack characterizes the main failure pattern of the specimens. A large amount of broken fibers bridging the failure surfaces is noticeable. Furthermore, pullout of warp tows can be also observed. Consequently, one may highlight that the fracturing process during the tests involved breaking the interweaving fiber bundle structure in all three dimensions and thus created a large fracture process zone. This particular failure pattern resembles the one observed in the in-plane fracture tests discussed in the above section, and thus differentiates the out-of-plane fracture behavior of 3D woven composites from the typical delamination failure of laminated composites which is featured by a straight line crack separating the originally bonded interfaces of laminates. Hence, one has a clue that the out-of-plane fracture behavior of 3D woven composites should be also of the quasibrittle type, resembling the in-plane fracture behavior, although it has to be confirmed by further investigation such as size effect tests. 
	
	Thanks to the use of wedge setup, the tests were mechanically stable, and one can estimate the fracture energy using the work-of-fracture method given a complete load-displacement curve. Provided that the energy dissipation due to inelastic deformation other than fracturing is negligible, the fracture energy $G_F$ can be calculated as $G_F = W/Lb$ where $W = $ the area under the complete stable load-displacement curve representing the total amount of the energy dissipation during the entire test, $L = $ length of the broken ligament, and $b = $ specimen thickness. Note that although a residual load was observed at the end of the tests, it is relatively small compared to the peak, and it is expected that the associated remaining energy has negligible influence on the calculation of the fracture energy. The out-of-plane fracture energy calculated in this way has a mean of 42.8 N/mm and a COV of 7\%. 
	
	\subsection{Correction of friction effect}
	The merit of the wedge-type test is that it provides a solution to the fracture test in the face of strict requirements for post-peak stability \cite{bruhwiler1990wedge}, and is free from the complex setup and the difficulties of avoiding the unwanted failure experienced by a DCB test \cite{brunner2008status}. Accompanying the wedge test, the work-of-fracture method provides a straightforward solution to the estimation of the fracture property using the energy approach without assuming the applicability of LEFM theory and the need for seeking closed-form expression of the energy release rate or stress intensity factor. Nevertheless, the question of friction contributions can become a concern for a wedge-type test, which is probably why the wedge-type test method has never been widely used despite its obvious advantages \cite{glessner1989mode,brunner2000experimental,brunner2008status}. It is necessary to quantify the effect of friction. 
	
	Given the measured friction coefficient $\mu$ and the wedge half-angle $\alpha$, one can subtract the contribution of energy dissipation due to frictional sliding. According to the static equilibrium condition for the wedge (see Fig. \ref{fig: WedgeIllus}b for the free-body diagram), the driving force $F$ is balanced by the normal force $F_N$ applied on the wedge surfaces and the frictional force $F_f$, i.e., $F = 2(F_N \sin\alpha + F_f \cos\alpha)$. By assuming that the kinetic friction condition was met throughout the tests, one obtains $F_f = \mu F_N$. Accordingly, the frictional force can be estimated as $F_f = \mu F/2(\sin\alpha + \mu\cos\alpha)$. In addition, neglecting the deformation of the steel wedge and the machine, one can also approximate the amount of sliding as $u_s = u\cos\alpha$, where $u$ is the recorded crosshead displacement. Accordingly, the energy dissipated by friction can be written as $W_f = F_f u_s = k_f W$, where $k_f = \mu\cos\alpha/2(\sin\alpha+\mu\cos\alpha)$. Given this simple calculation and neglecting the other possible energy dissipation mechanisms, one can correct the calculation of fracture energy discussed above by a factor of $1-k_f$. The corrected fracture energy measures 31 N/mm. It can be seen that the frictional sliding accounts for $27\%$ of total dissipated energy. Attempts to further reduce the effect of friction will be made in the future work. 
	
	The DCB test method has been adopted by several researchers to determine the delamination resistance of 3D woven composites \cite{tanzawa1999interlaminar,mouritz1999mode,tamuzs2001progressive,tamuzs2003delamination,siddique2019finite}. Most of the work reported the out-of-plane fracture energy as the energy release rate for steady state crack propagation, which was calculated using either Modified Beam Theory (MBT) or Modified Compliance Calibration (MCC) method according to the ASTM standard \cite{D5528}. The reported values varied greatly, and depends strongly on the fiber architecture and the constituent materials. In addition, the validity of these measurements needs further investigation, and some of the tests failed to attain the steady-state crack propagation. Nevertheless, the out-of-plane fracture energy obtained in this work is comparable with some measurements reported in literature. For instance, \cite{tamuzs2003delamination} reported that the delamination resistance of a 3D carbon fiber woven material produced by 3TEX could reach 20-30 N/mm; \cite{siddique2019finite} measured the out-of-plane energy release rate of a composite material with a 3D angle-interlock woven fabric and reported that the maximum energy release rate could be larger than 20 N/mm. It is obvious that the out-of-plane fracture energy of 3D woven composites is significantly higher than the one of laminated composites with typical value of less than 1 N/mm. This confirms that the 3D architecture does feature a great enhancement of the delamination resistance. 
	
	\section{In-plane multi-axial fracture test}
	The multi-axial fracturing behavior of polymers and their fiber-reinforced laminates was recently investigated in the literature \cite{Yaomulti1,Pearce,Tan,laux,liechti,Akhtar,Tan2,Alfonso,Zeinedini}. However, to the best of authors' knowledge, both notch-free and notched three-dimensional woven composites under global multi-axial stress states were rarely reported so far in the open literature. The lack of experimental data definitely has hindered the understanding of the mechanical behavior of three-dimensional woven composites under complex loading scenarios. In this section, the effects of an intra-laminar central notch on the three-dimensional woven composites under multi-axial quasi-static load were studied. Two different geometries were prepared for the in-plane multi-axial tests including the rectangular specimens cut along the warp and weft directions. The intra-laminar central notch was manufactured by firstly drilling a hole using a 0.8 mm tungsten carbide drill bit and then completing the notch with a 0.8 mm wide diamond-coated saw. The dimensions of the specimens are illustrated in Figure \ref{fig:arcanrig}.

    \subsection{Testing description}
   
\begin{figure}[htbp]
		\begin{center}
			\includegraphics[width = 0.9\textwidth]{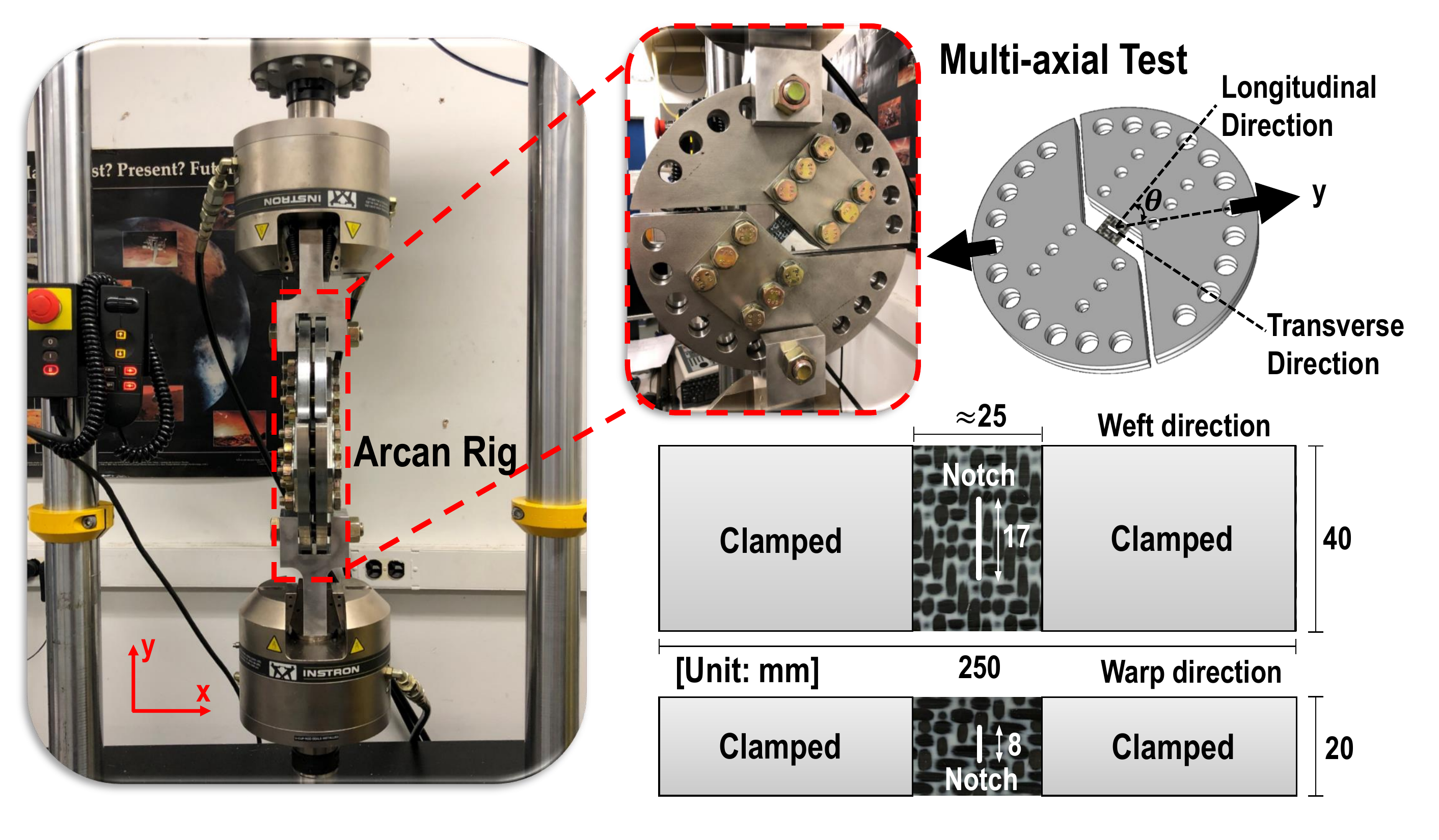}
			\caption{Test setup and Arcan rig used for the multi-axial tests on notched three-dimensional woven composites. Both warp and weft directions oriented towards the longitudinal direction of the specimen were investigated in this work.}
			\label{fig:arcanrig}
		\end{center}
	\end{figure}

    The multi-axial behavior of notched three-dimensional woven composites was evaluated by leveraging a modified Arcan rig as illustrated in Figure \ref{fig:arcanrig} which comprises four identical 17-4 PH stainless steel plates (two fronts and two backs) to clamp the specimens by friction through twelve M14 high-strength bolts. The global multi-axial loads were applied by rotating the modified Arcan rig with the angle $\theta$ between the loading direction and the longitudinal direction of the specimen. As illustrated in Figure \ref{fig:arcanrig}, a mixed tension and shear can be achieved by an intermediate angle between $0^{\circ}$ and $90^{\circ}$ whereas two extreme angles $\theta=0^{\circ}$ and $\theta=90^{\circ}$ represent the uni-axial tension and pure shear loading conditions respectively. To have a better description on the loading configuration, multiaxiality ratio can be defined as $\lambda=$ arctan$(\tau_{N}/\sigma_{N})$ where $\sigma_{N} = P$cos$\theta/(tD)$ is the nominal normal stress and $\tau_{N} = P$sin$\theta/(tD)$ is the nominal shear stress. The information on the multi-axial behavior of notched three-dimensional woven composites was achieved by investigating six sets of multiaxiality ratios with $\lambda= 0, 0.262, 0.524, 0.785, 1.047$ and 1.571 in this study. The foregoing multi-axial tests were performed on a servo-hydraulic 8801 Instron machine with closed-loop control and the displacement rate is 0.02 mm/s for all the specimens.
    
    \subsection{Damage detection and quantification}
    Speckled specimens were used for the acquisition of the entire strain field on the material surface by means of an open source Digital Image Correlation system programmed in MATLAB software and developed at Georgia Tech  \cite{Ncorr,Blader}. The detailed three-dimensional damage morphology at about the peak load was detected by means of a NSI X5000 X-ray micro-tomography scanning system \cite{northstar} with a X-ray tube setting of 110 kV in voltage and 160 $\mu$A in current. It is worth mentioning here that the sub-critical damage was better visualized by utilizing a dye penetrant \cite{Yaomulti2,shiyaocompact} in all the scans and the related solutions can be found in \cite{dye1,dye2}. After scanning, the damage characteristics were further analyzed through the reconstructed specimens via the software ParaView \cite{paraview}.
    
    About 100 reconstructed images were sliced throughout the thickness of the specimen leading to about 2 $\times$ $10^9$ pixels for the gauge volume of the specimen and the colors of all the pixels were converted into the grey scale values in order to quantify the sub-critical damage, By setting a range of the grey scale threshold, the crack volume can be measured by leveraging the pixels with the values higher than this threshold divided by the total pixels in the gauge volume of the specimen. This range was selected about 145-155 and carefully taken on the location of the specimen where no significant damage was observed. 
    
    \subsection{Experimental results and analysis}
\begin{figure}[htbp]
		\begin{center}
			\includegraphics[width = 0.9\textwidth]{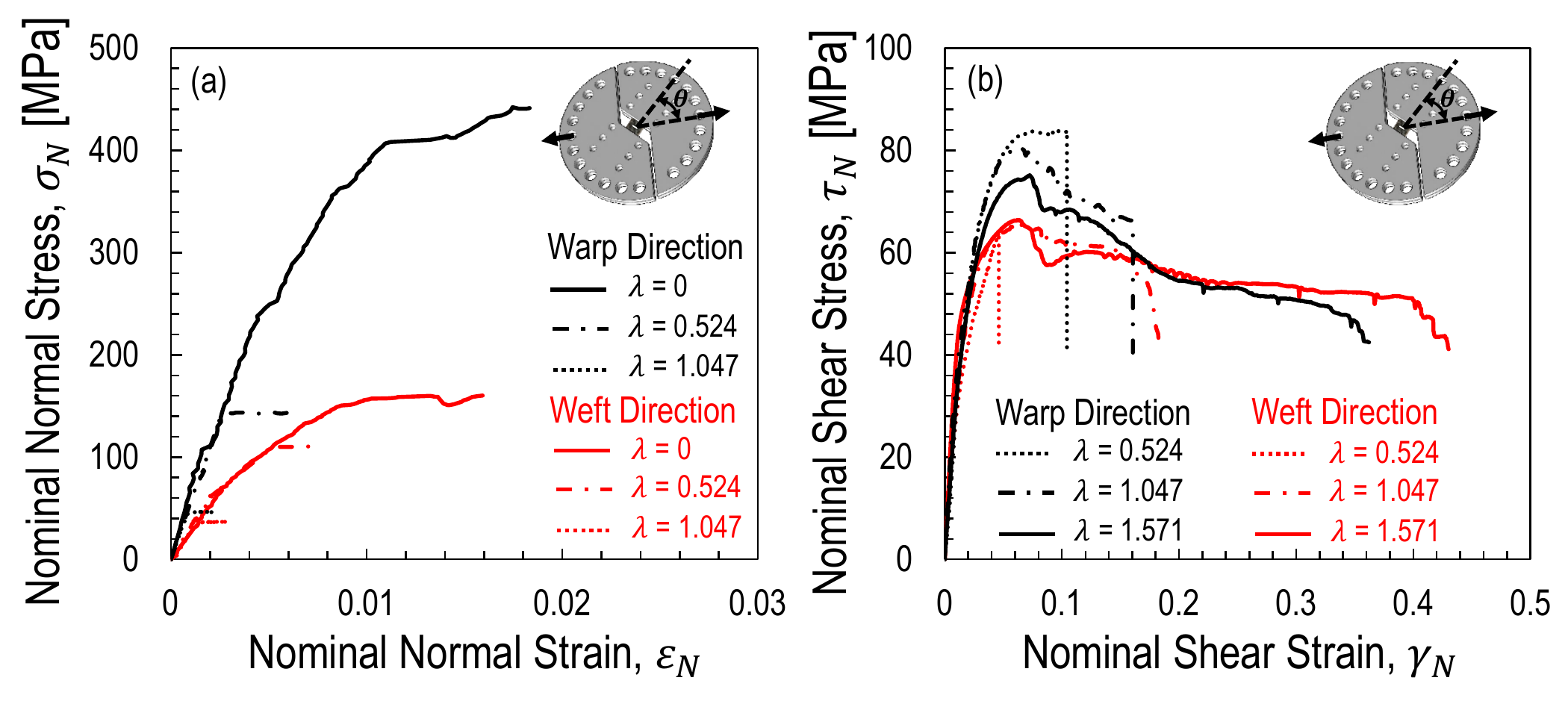}
			\caption{Nominal stress vs. strain curves in (a) longitudinal direction and (b) transverse direction obtained from the multi-axial tests on the notched three-dimensional woven composites with the longitudinal direction oriented in both warp and weft directions. Note that the nominal strain was measured based on the difference in the displacements close to the two ends of the gauge area of the specimen by means of the Digital Imaging Correlation (DIC).}
			\label{fig:multiaxialstressstrain}
		\end{center}
	\end{figure}

    The representative nominal stress and strain curves for the three-dimensional woven composites in presence of an intra-laminar central notch under multi-axial loads were plotted in Figure \ref{fig:multiaxialstressstrain}. For both warp and weft scenarios, the mechanical response of the materials in the longitudinal direction of the specimen exhibits less non-linearity up to the peak load with the curves departure from the linear part at least after 35\% of the strain at the peak load. However, the emergence of the material quasi-ductility in the transverse direction of the specimen was observed with increasing proportions of shear to tension since departure from linear behavior occurs only after 20\% of the strain at the peak load and it is followed by stable post-peak softening.
    %linear part only after 20\% of the strain at the peak load and followed by a stable post-peak behavior. %11/17: Changed by MS
    Similar phenomenon was also reported in other related studies \cite{Yaomulti1,Pearce,Tan,laux,liechti}. The non-linearity of the materials under shear-dominated loads can be explained mainly due to the formation of the significant matrix and debonding cracks for the energy dissipation in this case rather than the plasticity of the material as it will be shown through the morphological study in the next section. On the other hand, it is interesting to notice from the figure that the evolution of the nominal stress and strain curves can be dependent on the global multi-axial stress states. This implies the significant effects of the local stress states on the plastic deformation and fracture behavior of the materials which was also reported in other studies for various materials \cite{stressconrete,stresscomposite1,stresscomposite2,stressmetal1}.
    
\begin{figure}[htbp]
		\begin{center}
			\includegraphics[width = 0.9\textwidth]{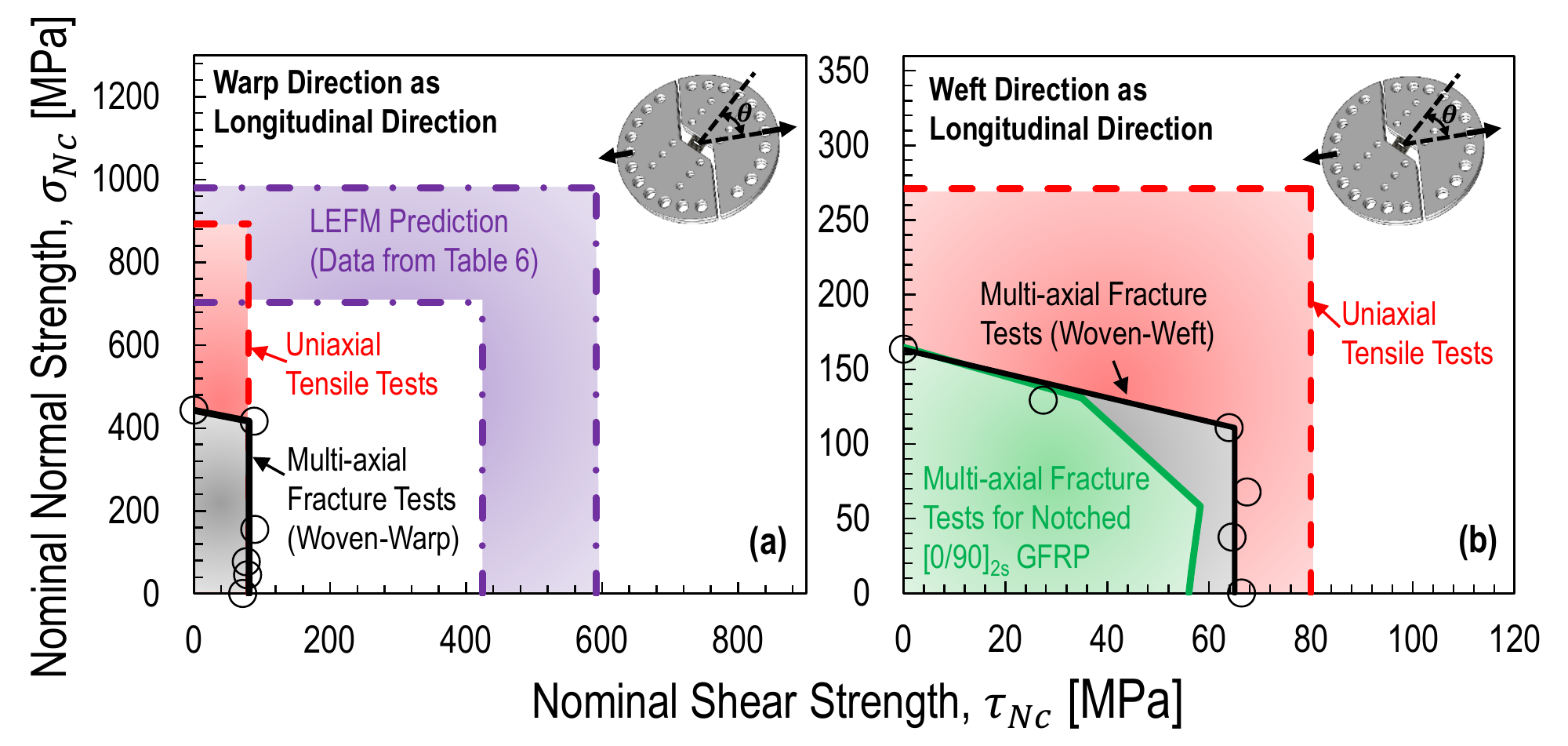}
			\caption{Failure envelops for the notched three-dimensional woven composites with the longitudinal direction of the specimen towards the (a) warp tows and (b) weft tows. Figure (a) compares the failure envelop measured from the multi-axial tests with the associated results obtained from the uni-axial tensile tests and predicted through the LEFM analysis. Figure (b) has a similar comparison but includes the previous results measured from the multi-axial tests on the notched $[0/90]_{2s}$ GFRP laminates.}
			\label{fig:failurecontour}
		\end{center}
	\end{figure}

    The correlation between the nominal normal and shear strength as a function of the multiaxiality ratio for the notched three-dimensional woven composites was further investigated and plotted in Figure \ref{fig:failurecontour}. As illustrated in Figure \ref{fig:failurecontour}a for the warp tows oriented towards the longitudinal direction of the specimen, the failure envelope features two distinguishable segments which represent both tension dominated ($\lambda=$ 0 to 0.262) and shear-dominated ($\lambda=$ 0.262 to 1.571) loading conditions. This is an indication of two different damage mechanisms with respect to the multiaxiality ratio as it will be clarified in the next section. On the other hand, the forgoing failure envelop is located in the area featuring the boundary with the in-plane tensile strength ($F_{1t}$) and shear strength ($F_{12}$) measured from the previous uni-axial tensile tests on the notch-free specimens. This aspect is not surprising but mainly attributed to the significant reduction on the nominal normal strength of the specimen since the nominal shear strength does not exhibit sensitive degradation for the existence of an intra-laminar central notch in the specimen compared to the one estimated from the uni-axial tensile tests on the previous off-axis specimens. In addition, the experimental results obtained from the previous Mode I fracture tests were further used through the LEFM analysis to predict the nominal normal and shear strength of the geometry in this case with the warp tows parallel to the longitudinal direction of the specimen. As can be noted from Figure \ref{fig:failurecontour}a, the prediction by means of the LEFM analysis can lead to a remarkable over-estimation for both nominal normal and shear strength and this mismatch confirms the quasi-brittleness of the investigated three-dimensional woven composites which cannot be simply analyzed by leveraging the LEFM approach. It is worth mentioning here that the LEFM prediction on the nominal shear strength of the specimen in Figure \ref{fig:failurecontour}a is based on the assumption that the apparent Mode II fracture energy $G^{II}_{f}$ of the material with the warp tows oriented toward the longitudinal direction of the specimen is equivalent to the apparent Mode I values as summarized in Table \ref{tab:SizeEffect} due to the lack of data on this aspect in the open literature. In fact, the larger Mode II fracture energy of the laminates compared to the Mode I counterparts was typically reported in the literature \cite{salviato2019mode} and this can lead to even more pronounced over-prediction on the forgoing nominal shear strength of the specimen.
    
    Similar analysis was further performed on the notched three-dimensional woven composites with the weft tows in the longitudinal direction of the specimen. As it can be noted from Figure \ref{fig:failurecontour}b, the failure envelope of the foregoing specimens under multi-axial loads is also composed of two noticeable parts representing the transition of the damage mechanism from tension to shear. Moreover, this aforementioned failure envelop was reasonably compared with the one in the previous study \cite{Yaomulti1} for the notched cross-ply $[0/90]_{2s}$ laminates made of glass fiber and thermoset polymer (GFRP) in the consideration of similar in-plane dimensions. As illustrated in Figure \ref{fig:failurecontour}b, the mechanical behavior of the foregoing cross-ply laminates is similar to the one of the notched woven composites plotted in this figure under tension-dominated loading conditions. However, this does not happen in the shear-dominated scenarios since this notched woven composites exhibit higher nominal shear strength indicating the better shear resistance of the materials than the two-dimensional GFRP laminates with similar fiber orientations. On the other hand, the investigated intra-laminar central notch leads to the reduction on both nominal tensile and shear strength of the notched woven composites with the weft tows in the longitudinal direction of the specimen and this aspect differs from the warp case as plotted in Figure \ref{fig:failurecontour}a showing almost no degradation on the nominal shear strength of the material with the non-negligible intra-laminar central notch. This can be explained by either the contribution of the warp tows to the better shear performance or the geometrical effects considering the different specimen sizes for both warp and weft cases. This requires further computational studies which will be covered in the future publications.

	\subsection{Morphological Characteristics and Damage Mechanisms}
	\subsubsection{Warp tow as longitudinal direction}  
\begin{figure}[htbp]
		\begin{center}
			\includegraphics[width = 0.9\textwidth]{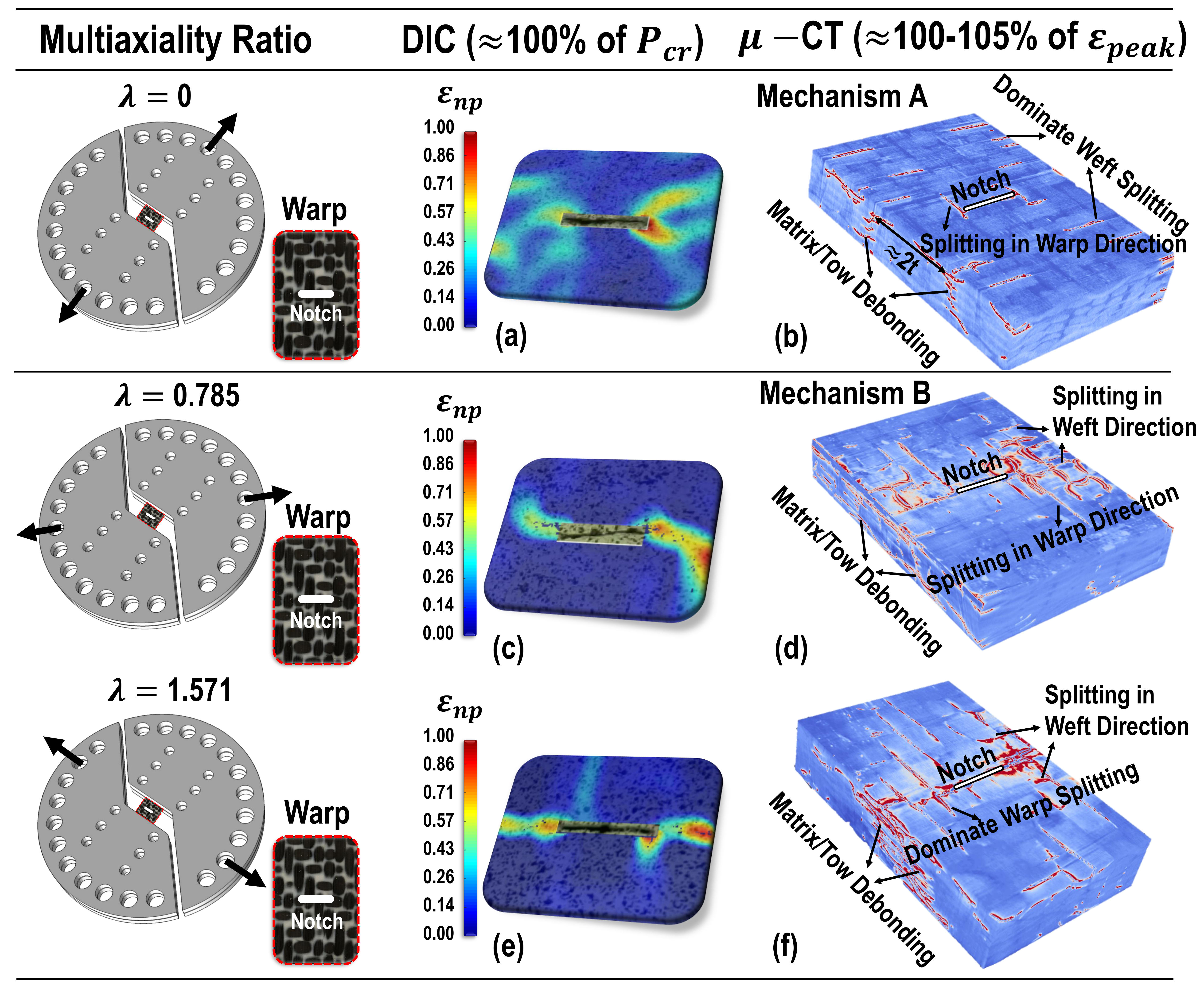}
			\caption{Analyses of quasi-static damage mechanisms of notched three-dimensional woven composites with the warp tows oriented toward the longitudinal direction of the specimens under multi-axial load by Digital Imaging Correlation (DIC) and micro-computed tomography ($\mu-$CT). These figures illustrate the detailed damage characteristics almost at the peak load for the forgoing composites weakened by a 8mm central notch under the multiaxiality ratio $\lambda$ = 0, 0.78 and 1.571. Note that the symbol $\varepsilon_{np}$ in the DIC analysis represents the normalized maximum principle strain. Also note that the warp direction is the longitudinal direction whereas the weft direction is the transverse direction in this case. %{\color{blue} Yao, maybe it is just my Overleaf but I keep having problems with this figure. Figure f does not show up. Can you take a look?} \color{blue} Yao, 11/18/20 - Sometimes there is a visualization problem in the OVERLEAF. I can see this figure through my laptop. If the picture is not shown up, this issue can be fixed by zooming in or out. In any cases, this figure can be seen in the downloaded PDF version.
			}
			\label{fig: warpmulti}
		\end{center}
	\end{figure}

    In the case that the warp tows are oriented towards the longitudinal direction of the specimens and these notched three-dimensional woven composites are subjected to multi-axial load,  the damage mechanisms can be reasonably categorized into two different scenarios. The damage characteristics belong to \emph{Mechanism A} when the tension-dominated loading conditions ($\lambda=0,0.262$) are applied on the specimens. In this mechanism, the fracturing features almost at the peak load are characterized by a remarkable distributed region of high deformation at both sides of the notch as illustrated in Figure \ref{fig: warpmulti}a through the two-dimensional DIC analysis. This deformed region is mainly formed by the significant splitting in the transverse direction distributed in the gauge area of the specimen and the noticeable splitting in the longitudinal direction at the notch tip as can be clearly seen from Figure \ref{fig: warpmulti}b by leveraging micro-computed tomography analysis. Thanks to this three-dimensional damage inspection technique, the significant damage through the entire thickness of the specimen almost at the peak load was also observed as illustrated in Figure \ref{fig: warpmulti}b showing the multiple debonding cracks spreading along the interfaces between tows and matrix as similarly reported in other related studies \cite{Yu2016}. Interestingly, these debonding cracks are characterized by a distance about two times the thickness of the specimen and this important feature is similar to the experimental morphology of the cross-ply laminates under uni-axial tension exhibiting individual transverse matrix crack at a distance of roughly one to three times the thickness of the lamina \cite{talrejadamagebook,geubelletransverse,herraez,mortell}. Based on the foregoing description, the damage evolution of \emph{Mechanism A} can be summarized into the following: 
    \begin{enumerate}
    
    \item splitting initiates in the transverse direction and has a distributed development (Figure \ref{fig: warpmulti}b);
    
    \item splitting in the longitudinal direction initiates and develops near the notch tip (Figure \ref{fig: warpmulti}b);
    
    \item matrix and debonding cracks initiate and develop through the specimen thickness (Figure \ref{fig: warpmulti}b);
    
    \item fibers break mainly in the longitudinal direction together with the further splitting propagation mainly in the transverse direction near the notch tip.
    \end{enumerate}
    
     On the other hand, the specimens exhibit completely different damage characteristics compared to the previous ones following \emph{Mechanism A} when the shear load component is involved ($\lambda=0.524,0.785,1.047$ and $1.571$). This significant difference was exemplified in Figures \ref{fig: warpmulti}c-f for the DIC and micro-computed tomography analyses on the specimens with the multiaxiality ratio $\lambda=0.7857$ and $1.571$. In these scenarios, the damage characteristics follow \emph{Mechanism B} which features localized region of high deformation at both sides of the notch almost at the peak load as shown in Figures \ref{fig: warpmulti}c and \ref{fig: warpmulti}e. The detailed damage morphology with respect to the forgoing localized damage region was illustrated in Figures \ref{fig: warpmulti}d and \ref{fig: warpmulti}f showing that the in-plane damage band is the combination of the localized splitting in both transverse and longitudinal directions whereas the minor matrix cracks and the pronounced debonding cracks between tows and matrix lead to the out-of-plane damage band which was not obtained from the two-dimensional DIC analysis. In addition to this, the splitting in the longitudinal direction is more pronounced in terms of both length and quantity as the multiaxiality ratio increases representing the transition from uni-axial tension to pure shear condition as illustrated in Figure \ref{fig: warpmulti}. Based on the foregoing discussion, the damage evolution of \emph{Mechanism B} consists of the following phases:
	
     \begin{enumerate}
    
     \item splitting initiates in the transverse direction and has a localized development near the notch (Figures \ref{fig: warpmulti}d,e);
    
     \item matrix and debonding cracks initiate and develop through the specimen thickness (Figures \ref{fig: warpmulti}d,e);
     
     \item splitting in the longitudinal direction initiates and develops in the specimen with the noticeable longer length for the pure shear loading condition (Figures \ref{fig: warpmulti}e);
     
     \item fiber breaking leads to the final separation of the specimen.
     \end{enumerate} 
     
				\begin{figure}[htbp]
		\begin{center}
			\includegraphics[width = 0.9\textwidth]{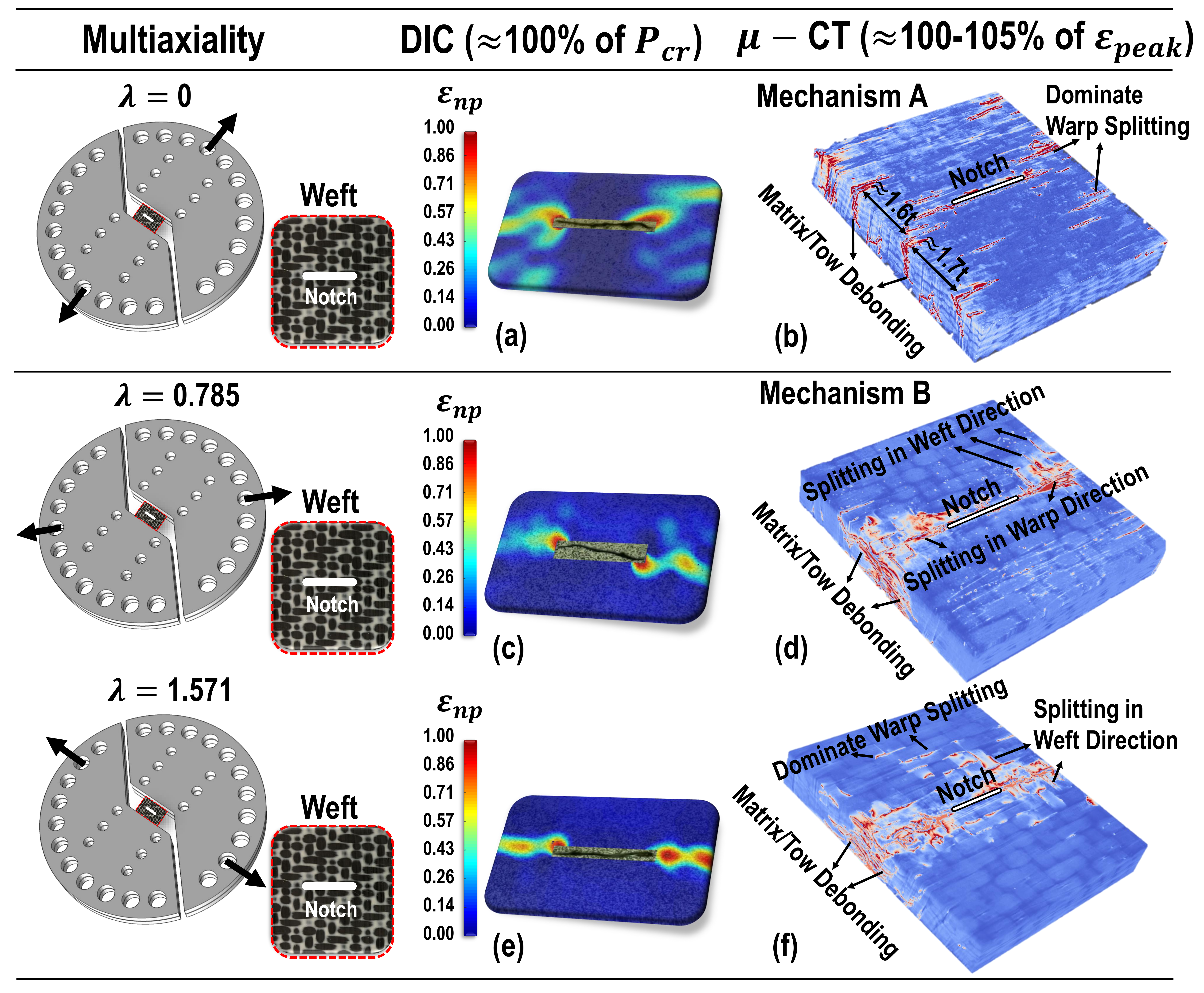}
			\caption{Analyses of quasi-static damage mechanisms of notched three-dimensional woven composites with the weft tows oriented toward the longitudinal direction of the specimens under multi-axial load by Digital Imaging Correlation (DIC) and micro-computed tomography ($\mu-$CT). These figures illustrate the detailed damage characteristics almost at the peak load for the forgoing composites weakened by a 8mm central notch under the multiaxiality ratio $\lambda$ = 0, 0.78 and 1.571. Note that the symbol $\varepsilon_{np}$ in the DIC analysis represents the normalized maximum principle strain. Also note that the weft direction is the longitudinal direction whereas the warp direction is the transverse direction in this case.}
			\label{fig: weftmulti}
		\end{center}
	\end{figure}
     
	\subsubsection{Weft tow as longitudinal direction}
    On similar grounds, the damage mechanisms can also be reasonably categorized into two different situations when the  notched three-dimensional woven composites are subjected to multi-axial load but the weft tows are oriented towards the longitudinal direction of the specimens. In this case, the application of tension-dominated loading conditions ($\lambda=0,0.262$ and $0.524$) on the specimens leads to the damage characteristics following \emph{Mechanism A}. As illustrated in Figure \ref{fig: weftmulti}a, a distributed region of high deformation at both sides of the notch was also observed almost at the peak load through the DIC analysis and the forgoing region is still the consequence of the distributed splitting in the transverse direction as shown in Figure \ref{fig: weftmulti}b through the micro-computed tomography analysis. These aspects are almost similar to the morphological features as discussed in the previous section for the warp tows parallel to the longitudinal direction of the specimen but the difference lies in the disappearance of the noticeable splitting in the longitudinal direction near the notch tip as clearly shown in Figure \ref{fig: weftmulti}b. On the other hand, the matrix and debonding cracks with an approximate distance of roughly two times the thickness of the specimen were also observed as can be noted in Figure \ref{fig: weftmulti}b. In summary, the damage evolution of \emph{Mechanism A} has the following steps:
    
     \begin{enumerate}
     \item splitting initiates in the transverse direction and has a distributed development (Figure \ref{fig: weftmulti}b);
    
     \item matrix and debonding cracks initiate and develop through the specimen thickness (Figure \ref{fig: weftmulti}b);
    
     \item fibers break mainly in the longitudinal direction together with the further splitting propagation in the transverse direction near the notch tip.
    \end{enumerate}
      
    When the specimens are subjected to shear-dominated loading conditions ($\lambda=0.785,1.047$ and $1.571$), the damage characteristics can be categorized into \emph{Mechanism B} which has similar features compared to the previous ones for the materials with the warp tows parallel to the longitudinal direction of the specimen under the same global multi-axial loads. As it can be noted in Figures \ref{fig: weftmulti}c-f for both DIC and $\mu$-CT analyses almost at the peak load, the specimens exhibit localized three-dimensional damage bands at both sides of the notch due to the mixed splitting in both longitudinal and transverse directions near the notch and the combination of matrix and debonding cracks through the entire thickness of the specimen. However, the splitting in the longitudinal direction of the specimen was significantly reduced in these scenarios which is substantially different from the previous \emph{Mechanism B} showing remarkable splitting in the longitudinal direction of the specimen. Based on these interesting morphological studies, the damage evolution of this \emph{Mechanism B} can be summarized in the following:
    
     \begin{enumerate}
     \item splitting initiates in the transverse direction and has a localized development near the notch (Figures \ref{fig: weftmulti}d,e);

     \item matrix and debonding cracks initiate and develop through the specimen thickness (Figures \ref{fig: weftmulti}d,e);
     
     \item minor splitting initiates and develops in the longitudinal direction  (Figures \ref{fig: weftmulti}d,e);
     
     \item fiber breaking leads to the final separation of the specimen.
     \end{enumerate} 
     
	             	\begin{figure}[htbp]
		\begin{center}
			\includegraphics[width = 0.5\textwidth]{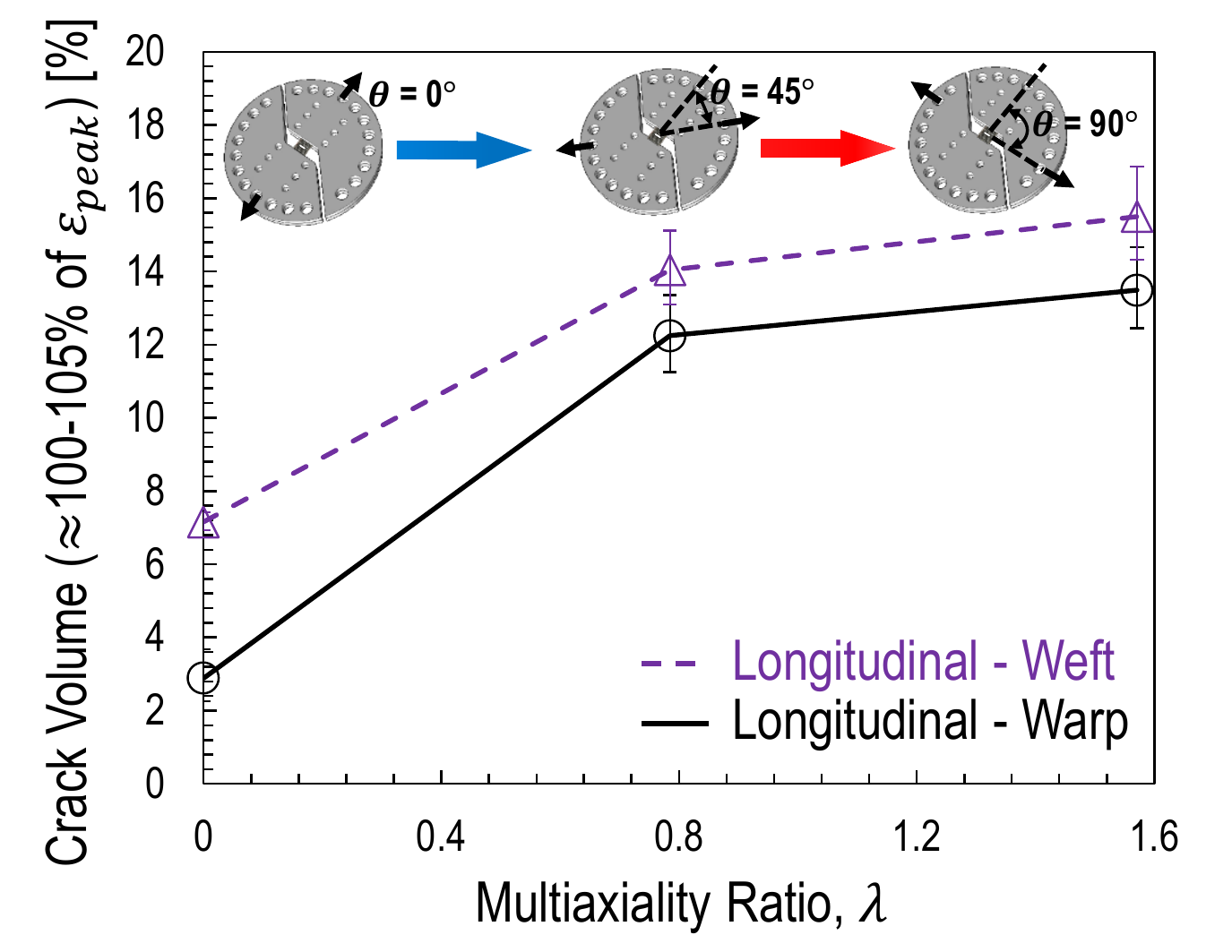}
			\caption{Crack volume as a function of the multiaxiality ratio almost at the peak load for the notched three-dimensional woven composites with the longitudinal direction in both warp and weft directions.}
			\label{fig:crackvolume}
		\end{center}
	\end{figure}
	
	 \subsubsection{Crack volume vs. multiaxiality ratio}
	The crack volume in the notched three-dimensional woven composites as a function of the multiaxiality ratio almost at the peak load was further estimated as illustrated in Figure \ref{fig:crackvolume}. As can be noted from this figure for both warp and weft cases, the crack volume increases in a non-linear way as the multiaxiality ratio increases. This is a confirmation that the material exhibits more quasi-brittleness when the loading condition transits from tension to shear. On the other hand, the foregoing phenomenon is also consistent with the previous nominal stress-strain curves of the notched three-dimensional woven composites under multi-axial loads since the less crack volume in the material for the tension-dominated loads leads to the less non-linearity of the curves whereas the energy dissipation due to the large crack volume can be the main reason for the significant non-linearity of the curves under shear-dominated loads.  
	
	\section{Conclusions}
	A comprehensive experimental study was performed in order to characterize various mechanical properties of a 3D woven carbon-epoxy composite with ply to ply angle interlock architecture. The following conclusions can be made based on the results provided in this work: 
	
	1) A database of the elastic, strength, and fracture properties of the 3D woven composite was provided through a series of quasi-static tests conducted in the warp, weft, and out-of-plane directions. The elastic properties were measured based on the theory of elasticity for orthotropic materials. 
	
	2) The nonlinear stress-strain responses were observed for the specimens under on-axis tension loading and $\pm \ang{45}$ off-axis tension and compression loading. The results of DIC analysis revealed that the failure initiation was strongly related to matrix cracking. In addition, material non-linearity was more evident for the specimens stretched in the weft direction, which also showed greater strain to failure although lower modulus and strength were detected. These observations could be attributed to the higher waviness of weft tows and lower fiber content in the weft direction. 
	
	3) The nonlinear mechanical responses of the specimens under $\pm \ang{45}$ tension and compression loading showed negligible difference, indicating that they share the same damage mechanism and deformation mode. Shear failure was recognized for both cases, and shear modulus and strength were determined from the off-axis specimens. 
	
	4) Various failure mechanisms and ultimate failure modes were identified from the in-plane tension and compression tests on both on-axis and off-axis specimens. 
	
	5) The out-of-plane elastic properties and compressive strengths were determined from the experimental results of the compression tests conducted in the out-of-plane direction. The explosive ejection of fragments perpendicular to the loading direction led to the ultimate failure of the specimens, and both horizontal and inclined failure planes were observed. 
	
	6) Tensile splitting test was proposed to estimate the out-of-plane tensile strength of the material. The relation between the tensile strength and the recorded peak load was determined through a finite element simulation. In addition, the DIC analysis revealed that the failure was initiated in the vicinity of the load application zones due to significant strain/stress concentrations, and both vertical crack and wedge formation were observed from the failed specimens. An improved tensile splitting test was suggested to reduce the degree of strain/stress concentrations, which was validated by finite element simulations. 
	
	7) The experimental results of the fracture tests on geometrically-scaled notched specimens showed a remarkable size effect. The strength-based failure criterion failed to capture the size effect whereas LFEM was also unsuccessful because the fracture properties determined from the measured peak load were size dependent. As a consequence, a nonlinear (quasibrittle) fracture mechanics theory with the capability of accounting for the finiteness of the FPZ was needed to interpret the fracture test results and to provide correct scaling analysis of mechanical properties, which is essential for damage tolerant design of large composite structures. 
	
	8) Comparison of the size effect data of the typical 3D and 2D woven composites suggested that the 3D woven composite provides higher quasi-ductility and superior damage tolerance which are beneficial in many aspects. In fact, the Mode I intra-laminar fracture energy, $G_f$ was estimated to the be $466$ N/mm or higher. This value is, on average, about seven times larger then the fracture energy of typical two-dimensional textile composites made of similar materials \cite{salviato2016experimental}.
	
	(9) The nominal stress-strain curves for the investigated center-notched three-dimensional woven composites exhibit significant non-linearity and relatively stable post-peak behavior as the loading multiaxiality ratio increases representing the involvement of shear stresses. This non-linearity is most likely due to the high volume of damage for energy dissipation and the plasticity of the material may have less contribution.
	
	(10) The failure envelop constructed by using the critical nominal normal and shear strength for the investigated center-notched three-dimensional woven composites cannot be simply predicted through the LEFM method due to the lack of characterizing the non-linear damage zone ahead of the notch in the materials. The forgoing multi-axial fracture behavior must be described by leveraging the quasi-brittle fracture mechanics and related non-linear computational modeling.
	
	(11) Regarding the damage mechanisms of the investigated center-notched three-dimensional woven composites under multi-axial scenario, a distributed damage with the formation of splitting in the transverse direction and the fiber/matrix debonding cracks throughout the specimen thickness mainly characterizes the material under tension-dominated loading whereas a localized damage region features the material under shear-dominated loading. This region is typically formed by various damage morphologies including matrix/fiber debonding, matrix cracking, and splitting in both longitudinal and transverse directions. The foregoing damage characteristics were observed for the material with either weft or warp tows oriented towards the longitudinal direction of the specimen.
	
	\section*{Acknowledgements}
	The work was partially supported under NSF (Funder ID: 10.13039/100000001) grant No. CMMI-1435923 to Northwestern University. In-kind contribution by Albany Composites is gratefully acknowledged. 
	%Thanks also goes to Isaac M. Daniel, Professor Emeritus at Northwestern University, and Jonathan Goering, former Divisional Chief Technology Officer at Albany Engineered Composites. 
 
    \begin{appendices}

    \section{Analysis of splitting tensile tests by finite element simulations} \label{Sec:App1}
\begin{figure}[htbp]
		\begin{center}
			\includegraphics[width = 0.8\textwidth]{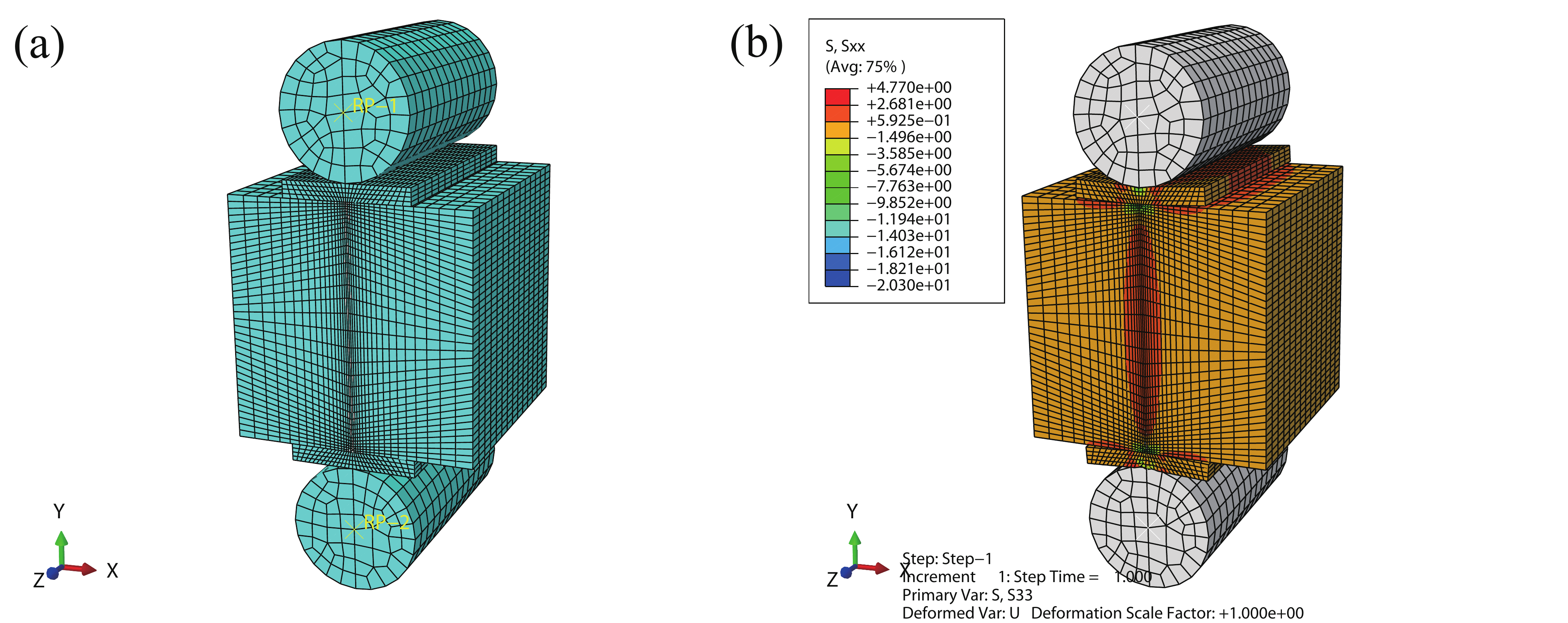}
			\caption[Typical finite element mesh used in the simulation of tensile splitting tests and the horizontal normal stress $\sigma_{xx}$ contour. ]{(a) Typical finite element mesh used in the simulation of tensile splitting tests. (b) Horizontal normal stress $\sigma_{xx}$ contour of a specimen under vertical displacement of -0.5 mm. }
			\label{fig: BTSMesh}
		\end{center}
	\end{figure}

	A 3D finite element model consisted of a cuboid specimen, two strip cushions, and two loading pins was created in Abaqus CAE environment \cite{abaqus}. Eight-node linear brick elements with reduced integration (C3D8R) were adopted for discretization. The general contact algorithm with hard contact as normal behavior was applied for the interaction between the different parts of the model. Mesh refinement with a smallest element size of 0.2 was preformed in the areas where the specimen is in contact with the cushions to ensure the accuracy of the numerical results in the region of high strain/stress gradient. The finite element mesh of the model is shown in Fig. \ref{fig: BTSMesh}a. A linear elastic orthotropic constitutive model was used for the simulation with the material properties given in Table \ref{tab:prop}. The loading pins were defined as rigid bodies to facilitate the application of the boundary conditions with the bottom one fixed vertically and the top one assigned a vertical displacement. Two cases were investigated with case 1 representing specimens loaded in the warp direction whereas case 2 in the weft direction. The simulations were performed in Abaqus Implicit 6.13 \cite{abaqus}. Fig. \ref{fig: BTSMesh}b plotted the deformed shape of the model for case 1 and the horizontal normal stress ($\sigma_{xx}$) contour on the specimen surfaces under the vertical displacement of -0.5 mm. A narrow band of positive $\sigma_{xx}$ (tensile stress, highlighted in red) can be identified. 
	
	\subsection{Stress distribution and and estimation of tensile strength}\label{Sec:App1_1}
\begin{figure}[htbp]
		\begin{center}
			\includegraphics[width = 0.9\textwidth]{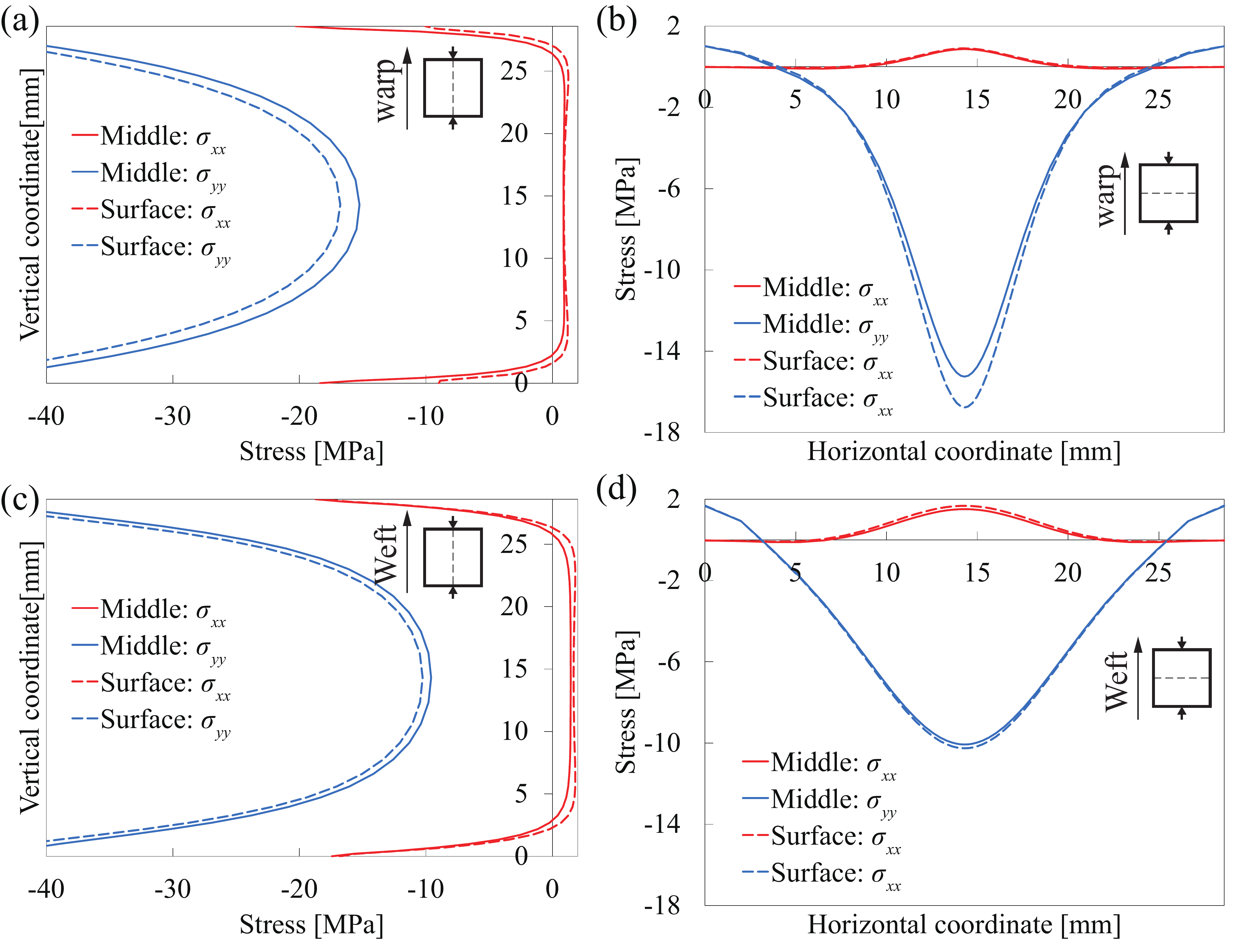}
			\caption[Major stress distribution of a specimen in the simulated tensile splitting test. ]{Major stress distribution of a specimen in simulated tensile splitting test on (a) vertical plane for the case of warp loading, (b) horizontal plane for the case of warp loading, (c) vertical plane for the case of weft loading, (d) horizontal plane for the case of weft loading.}
			\label{fig: BTSCube}
		\end{center}
	\end{figure}

	The numerically calculated stress distribution is shown in Fig. \ref{fig: BTSCube}a and b for case 1, and Fig. \ref{fig: BTSCube}c and d for case 2, respectively. Only the horizontal and vertical normal stress components, $\sigma_{xx}$ and $\sigma_{yy}$, were plotted, whereas the other components were significantly smaller. Fig. \ref{fig: BTSCube}a and c illustrate the stress distribution on the vertical plane of the simulated specimen along the loading line, whereas Fig. \ref{fig: BTSCube}b and d show the stress distribution on the horizontal plane across the center of the specimen. In order to illustrate the 3D effect on the stress distribution, the stresses on the middle and front surfaces of the specimens were plotted together. The 3D effect, however, is not significant for $\sigma_{xx}$ whereas a discrepancy of $\sigma_{yy}$ sampled on the specimen middle and front surfaces due to the 3D effect can be observed. 
	
	As one may notice, the vertical distribution of horizontal tensile stress $\sigma_{xx}$ is almost uniform on both middle and front surfaces, as shown in Fig. \ref{fig: BTSCube}a and c, whereas it attains the maximum value at the center of the specimen, as shown in Fig. \ref{fig: BTSCube}b and d. The horizontal tensile stress is usually associated with the tensile strength of the material when the load reaches its peak if the ultimate failure mode is tensile splitting driven by the maximum horizontal tensile stress. In this way, the relation between the maximum horizontal tensile stress and the applied load can be established, and the parameter $k$ can be obtained correspondingly. For instance, one obtains the maximum $\sigma_{xx}$ of 0.87 MPa (in the middle plane) from Fig. \ref{fig: BTSCube}a and b for case 1, and the corresponding applied load is 3348 N. The parameter $k$ was estimated as 0.213 accordingly. Similarly, one can obtain the parameter $k$ for case 2, being 0.356. 
	
	Although tensile splitting test provides an accessible method to estimate the tensile strength of materials, its accuracy is often questioned. Especially, one may note that along with the uniformly distributed horizontal tensile stress $\sigma_{xx}$ in the middle of the specimen, a large vertical compressive stress $\sigma_{yy}$ can be also observed from Fig. \ref{fig: BTSCube}. As a consequence, the measured tensile strength is obtained from a multiaxial stress state, and could be different from the one obtained from a uniaxial stress state. Another problem which hinders the application of tensile splitting test is noticeable stress concentrations. Especially, as one can observe from Fig. \ref{fig: BTSCube}, the vertical compressive stress $\sigma_{yy}$ in the vicinity of the load application zones could be 50 times larger than the maximum horizontal tensile stress, which could be responsible for the indentation type of failure and the formation of the wedge as one observed in Fig. \ref{fig: BTSLD}b and c. Some researchers proposed that although failure in the tensile splitting test could be sometimes initiated in compression-shear in the vicinity of loading platens due to stress concentrations, the ultimate failure is driven by tension \cite{hobbs1964tensile,steen2005observed}. This may explain the typical failure pattern shown in Fig. \ref{fig: BTSLD} b and c, and it is likely that the formed wedge could further drive crack propagation and eventually lead to the vertical failure path that is the same as splitting failure mode.

	\subsection{Discussion on tensile splitting test improvement}\label{Sec:App1_2}
\begin{figure}[htbp]
		\begin{center}
			\includegraphics[width = 0.5\textwidth]{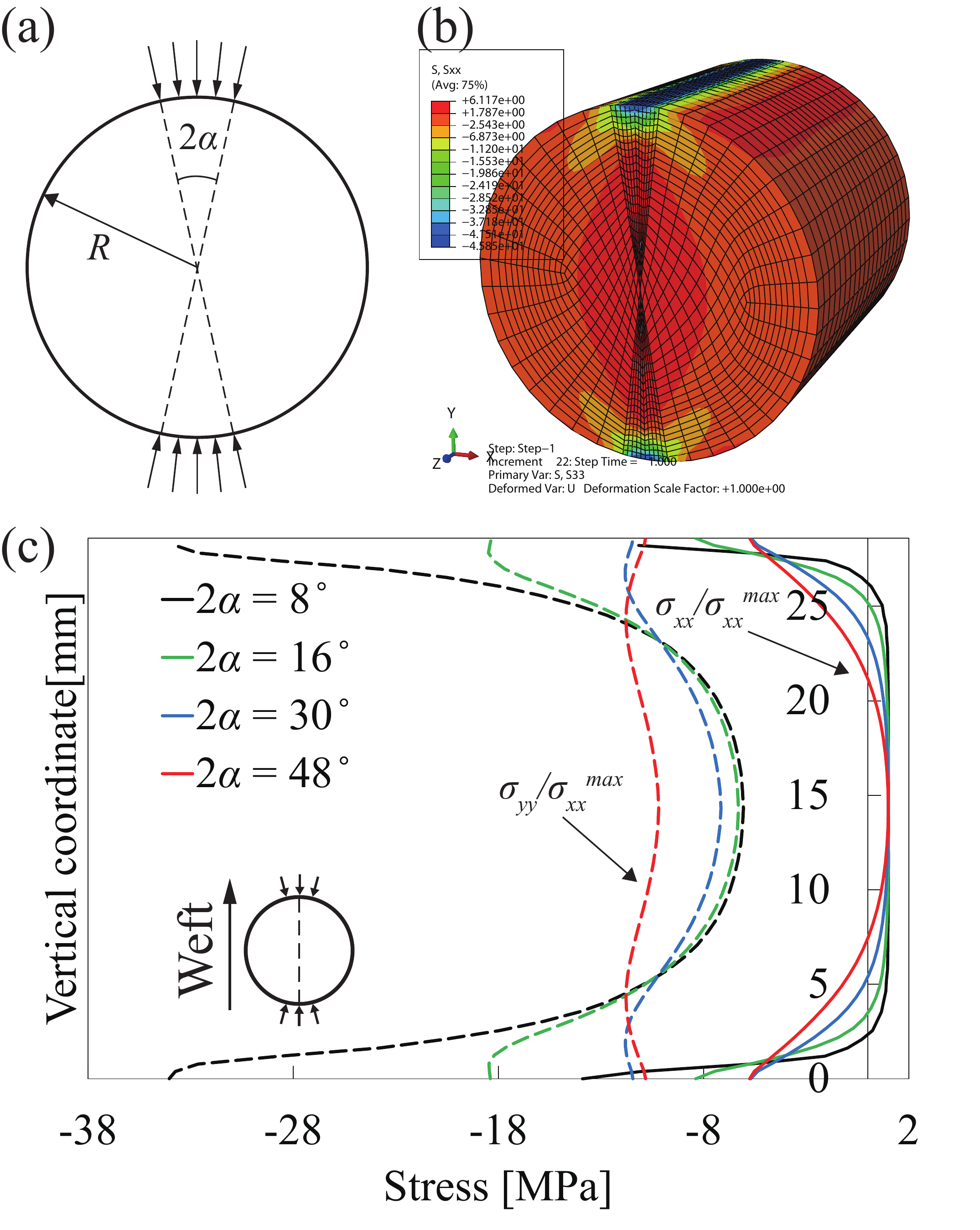}
			\caption[Schematic diagram of the improved tensile splitting test with cylindrical specimen and distributed load and the corresponding stress distribution. ]{(a) Schematic diagram of the improved tensile splitting test with cylindrical specimen and distributed load with a angular width $2\alpha$. (b) Horizontal normal stress $\sigma_{xx}$ contour of a cylindrical specimen under diametrical compression with $2\alpha = \ang{30}$. (c) Distribution of the normalized horizontal and vertical normal stresses on the vertical plane across the center of the specimen. }
			\label{fig: BTSCylinder}
		\end{center}
	\end{figure}

	A remedy to reduce stress concentrations in tensile splitting test is explored numerically herein. Instead of cuboid specimens, disk-shaped or cylindrical specimens are used to facilitate the transmission of load from a load frame. A load of a finite width is applied on the specimen with radius $R$ and thickness $t$. It was found that the loading width, characterized by an angular parameter, $2\alpha$ as depicted in Fig. \ref{fig: BTSCylinder}a, has a substantial influence on the stress distribution within the specimen \cite{perras2014review}. To explore the effect of the loading width and the feasibility of the improved tensile splitting test in determining the out-of-plane tensile strength of the investigated 3D woven composite, a finite element simulation was performed. The simulation setup is similar to the one described above, but instead of applying load through the loading pins, a pressure of 100 MPa is directly applied (radially) on the top and bottom of the specimen with a finite width, as illustrated in Fig. \ref{fig: BTSCylinder}a. Four different angular widths, $2\alpha = \ang{8}$, $\ang{16}$, $\ang{30}$, $\ang{48}$ are considered. The specimens were loaded in the weft direction. The typical horizontal normal stress distribution on the surfaces of the specimen is shown in Fig. \ref{fig: BTSCylinder}b. Similar to the one shown in Fig. \ref{fig: BTSMesh}b, a tensile stress was developed in a region near the center of the specimen. 
	
	The numerically calculated stress distribution on the vertical plane across the center of the specimen is shown in Fig. \ref{fig: BTSCylinder}c. All stresses were normalized by the maximum horizontal tensile stress $\sigma_{xx}^{max}$. It can be found that the stress distribution in all cases is very similar to the one for the cuboid specimen as shown in Fig. \ref{fig: BTSCube}c. However, the stress distribution, especially the concentrations in the vicinity of the load application zones vary with the angular width $2\alpha$. The ratio of the maximum compressive stress, $\max(-\sigma_{ii})$, $i = x$, $y$, to the maximum tensile stress, $\max(\sigma_{xx})$, can serve as an indicator of the degree of stress concentration. One may note that the ratio, and thus the stress concentrations are significantly reduced by increasing $2\alpha$. However, as the angular width $2\alpha$ increases from $\ang{30}$ to $\ang{48}$, the ratio does not decrease anymore. Therefore, the degree of stress concentration is minimized with $2\alpha = \ang{30}$ or $\ang{48}$.  It is also worth noting that the uniformity of the horizontal tensile stress in the vertical line is hampered by an increase of the angular width, which, however, is not desirable. In this case, the angular width $2\alpha = \ang{30}$ is preferred to $2\alpha = \ang{48}$. 
	
	It is promising that the ratio of the maximum compressive to tensile stress can be reduced to around $10$ if the angular width $2\alpha = \ang{30}$. Therefore, as long as the ratio of $F_{2c}$ to $F_{3t}$ is larger than 10, the failure initiation in the improved tensile splitting test could be driven by the maximum horizontal tensile stress rather than compression-shear in the vicinity of the specimen load application zones. As a consequence, the tensile strength of materials can be related to the peak load measured in a tensile splitting test. 
	
	Although conceptually simple, the improved tensile splitting test requires more efforts to manufacture the disk-shaped or cylindrical specimens and to setup the loading system to attain the desired load distribution over a certain width. Thus, the experiments will be left for the future work. In addition, considering the influence of the material inhomogeneity on the stress distribution, one may have reasonable doubt about the results obtained from the finite element simulations with a hidden assumption of material being homogeneous continuum. This doubt needs to be resolved with the assistance of a fine scale numerical model capable of capturing the major heterogeneity of the material, which will be also left for the future work.

	\section{Equivalent elastic crack concept and size effect law}\label{Sec:App_SEL}
	Only Mode I fracture is considered in this study. For an orthotropic structure containing a crack running in the $x$ direction subjected to an external nominal stress $\sigma_N$ in the $y$ direction, the stress intensity factor in the sense of LEFM can be written as \cite{bao1992role}   
	\begin{equation}\label{Eq:CK}
		K_I = \sigma_N \sqrt{D} k(\alpha) = \sigma_N \sqrt{\pi D \alpha} \xi (\alpha, \lambda^{1/4}L/D, \rho)
	\end{equation}
	where $\alpha = a/D$ denotes dimensionless crack length, $k(\alpha) = \sqrt{\pi \alpha} \xi$ is dimensionless stress intensity factor, and $\xi(\alpha, \lambda^{1/4}L/D, \rho)$ presents a dimensionless function accounting for material orthotropy, and $\rho$ and $\lambda$ are dimensionless elastic parameters defined as
	\begin{equation}
		\rho = \frac{\sqrt{E_x E_y}}{2G_{xy}} - \sqrt{\nu_{xy} \nu_{yx}}, \lambda = \frac{E_y}{E_x}
	\end{equation}
	The energy release rate $G(\alpha)$ can be related to the stress intensity factor $K_I$ through the Griffith and Irwin relationship, which reads
	\begin{equation}\label{Eq:CG}
		G(\alpha) = \frac{K_I^2}{E^*} = \frac{\sigma_N^2 D}{E^*}g(\alpha)
	\end{equation}
	where $g(\alpha) = k^2(\alpha)$ is dimensionless energy release rate, and $E^*$ is the effective elastic modulus for orthotropic materials, which reads
	\begin{equation}
		E^* = \sqrt{\frac{2E_x E_y \sqrt{\lambda}}{1+\rho}}
	\end{equation}
	
	According to LEFM, fracture toughness is a critical value of $K_I$ when the peak of $\sigma_N$, i.e. nominal strength $\sigma_{Nu}$, is reached. Identically, fracture energy is a critical value of $G$. These two fracture parameters calculated in this way are hereinafter referred to as the apparent fracture toughness, $K_{IcA}$ and the apparent fracture energy, $G_{fA}$. By substituting the values of $\sigma_{Nu}$ measured in the experiments as reported in Table \ref{tab:SizeEffect} and the elastic constants listed in Table \ref{tab:prop} ($x$ and $y$ axes coincide with 2 and 1 direction in the material system) into Eq. \ref{Eq:CK} and Eq. \ref{Eq:CG}, one can obtain the measures of $K_{IcA}$ and $G_{IcA}$, respectively. The value of $g(\alpha)$ was calculated numerically in the same way as discussed in Refs. \cite{salviato2016experimental,salviato2017graphene,li2019size} by using finite element and the quarter element technique \cite{barsoum1974application}. A uniform remote displacement rather than stress was assumed as a boundary condition, which is more realistic for the actual test setup in this work. The results of apparent fracture toughness and apparent fracture energy are also reported in Table \ref{tab:SizeEffect}. Note that $K_{IcA}$ and $G_{fA}$ are structural characteristics rather than material properties because they depend on specimen size and geometry. 
	
	The equivalent elastic concept provides a simple recipe for the problem of fracture of a quasibrittle material with a finite, non-negligible FPZ. In this case, the equivalent crack length $a = a_0 +c$ at failure must be distinguished from the actual crack (traction free) length $a_0$ where $c = $ elastically equivalent crack extension giving the same compliance according to LEFM as the actual crack growth \cite{bavzant1991identification}. It has been proven by Ba\v zant et al.\cite{bazant1997fracture} that for a sufficiently large structure, the equivalent crack extension $c = c_f$ at maximum load is a material property, and denotes the elastically equivalent length of the FPZ. Correspondingly, the energy required for crack growth by means of the equivalent elastic crack concept can be written as
	\begin{equation}\label{eq:CGa}
		G\left(\alpha_0 + c_f/D \right) = \frac{\sigma_{Nu}^2 D}{E^*}g(\alpha_0 + c_f/D) = G_f
	\end{equation}	
	where $G_f = $ initial fracture energy of the material, and is assumed to be a material property.  Mathematically, $G_f = \lim_{D \to \infty} G_{fA}$ since for $D\to \infty$, one has $c_f/D \to 0$ implying that for an infinitely large specimen, the FPZ occupies an infinitesimal region of the structure, and the equivalent crack extension vanishes.  
	
	The size effect law can be obtained from Eq. \ref{eq:CGa} by expressing the nominal strength $\sigma_{Nu}$ as a function of $D$. The dimensionless function $g(\alpha_0+c_f/D)$ can be approximated with its Taylor series expansion about the value $\alpha_0$ and retained only up to the linear term of the expansion. One obtains:
	\begin{equation}\label{Eq:CSEL1}
		\sigma_{Nu} = \sqrt{\frac{E^*G_f}{Dg(\alpha_0) + c_f g'(\alpha_0)}}
	\end{equation}  
	where the function $g'(\alpha)$ denote the derivative of $g$ with respect to $\alpha$. This equation has the same form as the classic Ba\v zant SEL for isotropic case \cite{bazant1997fracture}, and can be recast in a dimensionless form: 
	\begin{equation}
		\frac{\sigma_{Nu}}{\sigma_0} = \frac{1}{\sqrt{1 + D/D_0}}
	\end{equation}
	where $\sigma_0 = \sqrt{E^*G_f/(c_fg'(\alpha_0))}$ and $D_0 = c_fg'(\alpha_0)/g(\alpha_0)$.
	%The SEL derived from the equivalent elastic crack concept can be applied for the analysis of the size effect fracture data. 	
\end{appendices}

	\bibliographystyle{ieeetr}
	\bibliography{ExpWoven}

\begin{thebibliography}{100}

\bibitem{Tsaibook}
D.~Gay, W.~S. Tsai, and H.~V. Suong, {\em Composite materials: design and
  applications}.
\newblock CRC press, 2014.

\bibitem{Barbero}
E.~J. Barbero, {\em Introduction to composite materials design}.
\newblock CRC press, 2017.

\bibitem{Waas11}
M.~Pankow, A.~Salvi, A.~Waas, C.~Yen, and S.~Ghiorse, ``Resistance to
  delamination of 3{D} woven textile composites evaluated using {E}nd {N}otch
  {F}lexure ({ENF}) tests: Experimental results,'' {\em Composites Part A:
  applied science and manufacturing}, vol.~42, no.~10, pp.~1463--1476, 2011.

\bibitem{Soutis17}
M.~N. Saleh and C.~Soutis, ``Recent advancements in mechanical characterisation
  of 3d woven composites,'' {\em Mechanics of Advanced Materials and Modern
  Processes}, vol.~3, no.~1, p.~12, 2017.

\bibitem{Waas18}
D.~K. Patel, A.~M. Waas, and C.-F. Yen, ``Direct numerical simulation of 3d
  woven textile composites subjected to tensile loading: An experimentally
  validated multiscale approach,'' {\em Composites Part B: Engineering},
  vol.~152, pp.~102--115, 2018.

\bibitem{Waas19}
M.~Pankow, B.~Justusson, M.~Riosbaas, A.~Waas, and C.~Yen, ``Effect of fiber
  architecture on tensile fracture of 3d woven textile composites,'' {\em
  Composite Structures}, vol.~225, p.~111139, 2019.

\bibitem{cox1992mechanisms}
B.~Cox, M.~Dadkhah, R.~Inman, W.~Morris, and J.~Zupon, ``Mechanisms of
  compressive failure in {3D} composites,'' {\em Acta Metallurgica et
  Materialia}, vol.~40, no.~12, pp.~3285--3298, 1992.

\bibitem{cox1994failure}
B.~Cox, M.~Dadkhah, W.~Morris, and J.~Flintoff, ``Failure mechanisms of {3D}
  woven composites in tension, compression, and bending,'' {\em Acta
  Metallurgica et Materialia}, vol.~42, no.~12, pp.~3967--3984, 1994.

\bibitem{cox1995macroscopic}
B.~Cox and M.~Dadkhah, ``The macroscopic elasticity of {3D} woven composites,''
  {\em Journal of Composite Materials}, vol.~29, no.~6, pp.~785--819, 1995.

\bibitem{Castaneda17}
N.~Castaneda, B.~Wisner, J.~Cuadra, S.~Amini, and A.~Kontsos, ``Investigation
  of the z-binder role in progressive damage of 3d woven composites,'' {\em
  Composites Part A: Applied Science and Manufacturing}, vol.~98, pp.~76--89,
  2017.

\bibitem{Lomov09}
S.~V. Lomov, A.~E. Bogdanovich, D.~S. Ivanov, D.~Mungalov, M.~Karahan, and
  I.~Verpoest, ``A comparative study of tensile properties of non-crimp 3{D}
  orthogonal weave and multi-layer plain weave {E}-glass composites. part 1:
  Materials, methods and principal results,'' {\em Composites part a: applied
  science and manufacturing}, vol.~40, no.~8, pp.~1134--1143, 2009.

\bibitem{Warren15a}
K.~C. Warren, R.~A. Lopez-Anido, and J.~Goering, ``Experimental investigation
  of three-dimensional woven composites,'' {\em Composites Part A: Applied
  Science and Manufacturing}, vol.~73, pp.~242--259, 2015.

\bibitem{Warren15b}
K.~C. Warren, R.~A. Lopez-Anido, and J.~Goering, ``Behavior of
  three-dimensional woven carbon composites in single-bolt bearing,'' {\em
  Composite structures}, vol.~127, pp.~175--184, 2015.

\bibitem{Behera15}
B.~Behera, , and B.~Dash, ``Mechanical behavior of 3d woven composites,'' {\em
  Materials \& design}, vol.~67, pp.~261--271, 2015.

\bibitem{callus1999tensile}
P.~Callus, A.~Mouritz, M.~K. Bannister, and K.~Leong, ``Tensile properties and
  failure mechanisms of {3D} woven {GRP} composites,'' {\em Composites Part A:
  Applied Science and Manufacturing}, vol.~30, no.~11, pp.~1277--1287, 1999.

\bibitem{gu2002tensile}
H.~Gu and Z.~Zhili, ``Tensile behavior of 3d woven composites by using
  different fabric structures,'' {\em Materials \& Design}, vol.~23, no.~7,
  pp.~671--674, 2002.

\bibitem{dai2015influence}
S.~Dai, P.~Cunningham, S.~Marshall, and C.~Silva, ``Influence of fibre
  architecture on the tensile, compressive and flexural behaviour of {3D} woven
  composites,'' {\em Composites Part A: Applied Science and Manufacturing},
  vol.~69, pp.~195--207, 2015.

\bibitem{mouritz2010mechanistic}
A.~Mouritz and B.~Cox, ``A mechanistic interpretation of the comparative
  in-plane mechanical properties of {3D} woven, stitched and pinned
  composites,'' {\em Composites Part A: Applied Science and Manufacturing},
  vol.~41, no.~6, pp.~709--728, 2010.

\bibitem{kuo2002failure}
W.-S. Kuo, T.-H. Ko, and T.-S. Lo, ``Failure behavior of three-axis woven
  carbon/carbon composites under compressive and transverse shear loads,'' {\em
  Composites Science and Technology}, vol.~62, no.~7, pp.~989--999, 2002.

\bibitem{cox1996tensile}
B.~N. Cox, M.~S. Dadkhah, and W.~Morris, ``On the tensile failure of 3d woven
  composites,'' {\em Composites Part A: Applied Science and Manufacturing},
  vol.~27, no.~6, pp.~447--458, 1996.

\bibitem{osada2003initial}
T.~Osada, A.~Nakai, and H.~Hamada, ``Initial fracture behavior of satin woven
  fabric composites,'' {\em Composite Structures}, vol.~61, no.~4,
  pp.~333--339, 2003.

\bibitem{bogdanovich2013quasi}
A.~E. Bogdanovich, M.~Karahan, S.~V. Lomov, and I.~Verpoest, ``Quasi-static
  tensile behavior and damage of carbon/epoxy composite reinforced with {3D}
  non-crimp orthogonal woven fabric,'' {\em Mechanics of Materials}, vol.~62,
  pp.~14--31, 2013.

\bibitem{lomov2014monitoring}
S.~V. Lomov, M.~Karahan, A.~Bogdanovich, and I.~Verpoest, ``Monitoring of
  acoustic emission damage during tensile loading of {3D} woven carbon/epoxy
  composites,'' {\em Textile Research Journal}, vol.~84, no.~13,
  pp.~1373--1384, 2014.

\bibitem{yang2000bending}
B.~Yang, V.~Kozey, S.~Adanur, and S.~Kumar, ``Bending, compression, and shear
  behavior of woven glass fiber--epoxy composites,'' {\em Composites Part B:
  Engineering}, vol.~31, no.~8, pp.~715--721, 2000.

\bibitem{kuo2000compressive}
W.-S. Kuo and T.-H. Ko, ``Compressive damage in 3-axis orthogonal fabric
  composites,'' {\em Composites Part A: Applied Science and Manufacturing},
  vol.~31, no.~10, pp.~1091--1105, 2000.

\bibitem{kuo2007effect}
W.-S. Kuo, T.-H. Ko, and C.-P. Chen, ``Effect of weaving processes on
  compressive behavior of {3D} woven composites,'' {\em Composites Part A:
  Applied Science and Manufacturing}, vol.~38, no.~2, pp.~555--565, 2007.

\bibitem{gerlach2012plane}
R.~Gerlach, C.~R. Siviour, J.~Wiegand, and N.~Petrinic, ``In-plane and
  through-thickness properties, failure modes, damage and delamination in {3D}
  woven carbon fibre composites subjected to impact loading,'' {\em Composites
  Science and Technology}, vol.~72, no.~3, pp.~397--411, 2012.

\bibitem{saleh2016characterising}
M.~N. Saleh, A.~Yudhanto, P.~Potluri, G.~Lubineau, and C.~Soutis,
  ``Characterising the loading direction sensitivity of {3D} woven composites:
  Effect of z-binder architecture,'' {\em Composites Part A: Applied Science
  and Manufacturing}, vol.~90, pp.~577--588, 2016.

\bibitem{visrolia2013multiscale}
A.~Visrolia and M.~Meo, ``Multiscale damage modelling of {3D} weave composite
  by asymptotic homogenisation,'' {\em Composite Structures}, vol.~95,
  pp.~105--113, 2013.

\bibitem{pochiraju1999three}
K.~Pochiraju and T.-w. Chou, ``Three-dimensionally woven and braided
  composites. {II}: An experimental characterization,'' {\em Polymer
  Composites}, vol.~20, no.~6, pp.~733--747, 1999.

\bibitem{buchanan2012determination}
S.~Buchanan, E.~Archer, D.~Townsend, S.~Jenkins, A.~T. McIlhagger, and J.~P.
  Quinn, ``Determination of in-plane shear modulus of {3D} woven composites
  with large repeat unit cells,'' {\em Plastics, Rubber and Composites},
  vol.~41, no.~4-5, pp.~194--198, 2012.

\bibitem{daniel2008three}
I.~M. Daniel, J.-J. Luo, and P.~M. Schubel, ``Three-dimensional
  characterization of textile composites,'' {\em Composites Part B:
  Engineering}, vol.~39, no.~1, pp.~13--19, 2008.

\bibitem{abot2004through}
J.~Abot and I.~Daniel, ``Through-thickness mechanical characterization of woven
  fabric composites,'' {\em Journal of Composite Materials}, vol.~38, no.~7,
  pp.~543--553, 2004.

\bibitem{cui1996interlaminar}
W.~Cui, T.~Liu, J.~Len, and R.~Ruo, ``Interlaminar tensile strength ({ILTS})
  measurement of woven glass/polyester laminates using four-point curved beam
  specimen,'' {\em Composites Part A: Applied Science and Manufacturing},
  vol.~27, no.~11, pp.~1097--1105, 1996.

\bibitem{olsson2011survey}
R.~Olsson, ``A survey of test methods for multiaxial and out-of-plane strength
  of composite laminates,'' {\em Composites Science and Technology}, vol.~71,
  no.~6, pp.~773--783, 2011.

\bibitem{hufenbach2011determination}
W.~Hufenbach, A.~Hornig, B.~Zhou, A.~Langkamp, and M.~Gude, ``Determination of
  strain rate dependent through-thickness tensile properties of textile
  reinforced thermoplastic composites using l-shaped beam specimens,'' {\em
  Composites Science and Technology}, vol.~71, no.~8, pp.~1110--1116, 2011.

\bibitem{hufenbach2013characterisation}
W.~Hufenbach, A.~Langkamp, M.~Gude, C.~Ebert, A.~Hornig, S.~Nitschke, and
  H.~B{\"o}hm, ``Characterisation of strain rate dependent material properties
  of textile reinforced thermoplastics for crash and impact analysis,'' {\em
  Procedia Materials Science}, vol.~2, pp.~204--211, 2013.

\bibitem{park2005through}
D.~C. Park, ``Through-thickness compressive strength of carbon--phenolic woven
  composites,'' {\em Composite Structures}, vol.~70, no.~4, pp.~403--412, 2005.

\bibitem{song2014mechanical}
Z.~Song, Z.~Wang, H.~Ma, and H.~Xuan, ``Mechanical behavior and failure mode of
  woven carbon/epoxy laminate composites under dynamic compressive loading,''
  {\em Composites Part B: Engineering}, vol.~60, pp.~531--536, 2014.

\bibitem{whitney1985short}
J.~Whitney and C.~Browning, ``On short-beam shear tests for composite
  materials,'' {\em Experimental Mechanics}, vol.~25, no.~3, pp.~294--300,
  1985.

\bibitem{walter2010monotonic}
T.~Walter, G.~Subhash, B.~Sankar, and C.~Yen, ``Monotonic and cyclic short beam
  shear response of {3D} woven composites,'' {\em Composites Science and
  Technology}, vol.~70, no.~15, pp.~2190--2197, 2010.

\bibitem{gras2013identification}
R.~Gras, H.~Leclerc, S.~Roux, S.~Otin, J.~Schneider, and J.-N. P{\'e}ri{\'e},
  ``Identification of the out-of-plane shear modulus of a {3D} woven
  composite,'' {\em Experimental Mechanics}, vol.~53, no.~5, pp.~719--730,
  2013.

\bibitem{kuo2003failure}
W.-S. Kuo, J.~Fang, and H.-W. Lin, ``Failure behavior of {3D} woven composites
  under transverse shear,'' {\em Composites Part A: Applied Science and
  Manufacturing}, vol.~34, no.~7, pp.~561--575, 2003.

\bibitem{hufenbach2011analysing}
W.~Hufenbach, A.~Langkamp, A.~Hornig, M.~Zscheyge, and R.~Bochynek, ``Analysing
  and modelling the {3D} shear damage behaviour of hybrid yarn
  textile-reinforced thermoplastic composites,'' {\em Composite Structures},
  vol.~94, no.~1, pp.~121--131, 2011.

\bibitem{gude2015modified}
M.~Gude, W.~Hufenbach, M.~Andrich, A.~Mertel, and R.~Schirner, ``Modified
  v-notched rail shear test fixture for shear characterisation of
  textile-reinforced composite materials,'' {\em Polymer Testing}, vol.~43,
  pp.~147--153, 2015.

\bibitem{bascom1980interlaminar}
W.~Bascom, J.~Bitner, R.~Moulton, and A.~Siebert, ``The interlaminar fracture
  of organic-matrix, woven reinforcement composites,'' {\em Composites},
  vol.~11, no.~1, pp.~9--18, 1980.

\bibitem{alif1998effect}
N.~Alif, L.~A. Carlsson, and L.~Boogh, ``The effect of weave pattern and crack
  propagation direction on mode {I} delamination resistance of woven glass and
  carbon composites,'' {\em Composites Part B: Engineering}, vol.~29, no.~5,
  pp.~603--611, 1998.

\bibitem{hufenbach2013influence}
W.~Hufenbach, A.~Hornig, M.~Gude, R.~B{\"o}hm, and F.~Zahneisen, ``Influence of
  interface waviness on delamination characteristics and correlation of
  through-thickness tensile failure with mode {I} energy release rates in
  carbon fibre textile composites,'' {\em Materials \& Design}, vol.~50,
  pp.~839--845, 2013.

\bibitem{koh2011experimental}
T.~Koh, S.~Feih, and A.~Mouritz, ``Experimental determination of the structural
  properties and strengthening mechanisms of z-pinned composite t-joints,''
  {\em Composite Structures}, vol.~93, no.~9, pp.~2222--2230, 2011.

\bibitem{rys2010investigation}
T.~Rys, B.~V. Sankar, and P.~G. Ifju, ``Investigation of fracture toughness of
  laminated stitched composites subjected to mixed mode loading,'' {\em Journal
  of Reinforced Plastics and Composites}, vol.~29, no.~3, pp.~422--430, 2010.

\bibitem{mouritz1999mode}
A.~Mouritz, C.~Baini, and I.~Herszberg, ``Mode {I} interlaminar fracture
  toughness properties of advanced textile fibreglass composites,'' {\em
  Composites Part A: Applied Science and Manufacturing}, vol.~30, no.~7,
  pp.~859--870, 1999.

\bibitem{guenon1989toughness}
V.~A. Gu{\'e}non, T.-W. Chou, and J.~W. Gillespie, ``Toughness properties of a
  three-dimensional carbon-epoxy composite,'' {\em Journal of Materials
  Science}, vol.~24, no.~11, pp.~4168--4175, 1989.

\bibitem{liu2008fracture}
Q.~Liu and M.~Hughes, ``The fracture behaviour and toughness of woven flax
  fibre reinforced epoxy composites,'' {\em Composites Part A: Applied Science
  and Manufacturing}, vol.~39, no.~10, pp.~1644--1652, 2008.

\bibitem{blanco2014intralaminar}
N.~Blanco, D.~Trias, S.~Pinho, and P.~Robinson, ``Intralaminar fracture
  toughness characterisation of woven composite laminates. {Part} {II}:
  Experimental characterisation,'' {\em Engineering Fracture Mechanics},
  vol.~131, pp.~361--370, 2014.

\bibitem{laffan2012translaminar}
M.~Laffan, S.~Pinho, P.~Robinson, and A.~McMillan, ``Translaminar fracture
  toughness testing of composites: A review,'' {\em Polymer Testing}, vol.~31,
  no.~3, pp.~481--489, 2012.

\bibitem{hughes2002fracture}
M.~Hughes, C.~Hill, and J.~Hague, ``The fracture toughness of bast fibre
  reinforced polyester composites {Part} 1 {Evaluation} and analysis,'' {\em
  Journal of Materials Science}, vol.~37, no.~21, pp.~4669--4676, 2002.

\bibitem{green2007experimental}
B.~Green, M.~Wisnom, and S.~Hallett, ``An experimental investigation into the
  tensile strength scaling of notched composites,'' {\em Composites Part A:
  Applied Science and Manufacturing}, vol.~38, no.~3, pp.~867--878, 2007.

\bibitem{li2019size}
W.~Li, Z.~Jin, and G.~Cusatis, ``Size effect analysis for the characterization
  of marcellus shale quasi-brittle fracture properties,'' {\em Rock Mechanics
  and Rock Engineering}, vol.~52, no.~1, pp.~1--18, 2019.

\bibitem{bao1992remarks}
G.~Bao and Z.~Suo, ``Remarks on crack-bridging concepts,'' {\em Applied
  Mechanics Reviews}, vol.~45, no.~8, pp.~355--366, 1992.

\bibitem{salviato2016direct}
M.~Salviato, V.~T. Chau, W.~Li, Z.~P. Ba{\v{z}}ant, and G.~Cusatis, ``Direct
  testing of gradual postpeak softening of fracture specimens of fiber
  composites stabilized by enhanced grip stiffness and mass,'' {\em Journal of
  Applied Mechanics}, vol.~83, no.~11, p.~111003, 2016.

\bibitem{cedolin2008identification}
L.~Cedolin and G.~Cusatis, ``Identification of concrete fracture parameters
  through size effect experiments,'' {\em Cement and Concrete Composites},
  vol.~30, no.~9, pp.~788--797, 2008.

\bibitem{cusatis2009cohesive}
G.~Cusatis and E.~A. Schauffert, ``Cohesive crack analysis of size effect,''
  {\em Engineering Fracture Mechanics}, vol.~76, no.~14, pp.~2163--2173, 2009.

\bibitem{wan2018age}
L.~Wan-Wendner, R.~Wan-Wendner, and G.~Cusatis, ``Age-dependent size effect and
  fracture characteristics of ultra-high performance concrete,'' {\em Cement
  and Concrete Composites}, vol.~85, pp.~67--82, 2018.

\bibitem{cusatis-and-diluzio}
G.~Di~Luzio and G.~Cusatis, ``Cohesive crack analysis of size effect for
  samples with blunt notches and generalized size effect curve for
  quasi-brittle materials,'' {\em Engineering Fracture Mechanics}, vol.~204,
  pp.~15--28, 2018.

\bibitem{bazant1996size}
Z.~P. Ba{\v{z}}ant, I.~M. Daniel, and Z.~Li, ``Size effect and fracture
  characteristics of composite laminates,'' {\em Journal of engineering
  materials and technology}, vol.~118, no.~3, pp.~317--324, 1996.

\bibitem{bavzant1999size}
Z.~P. Ba{\v{z}}ant, J.-J.~H. Kim, I.~M. Daniel, E.~Becq-Giraudon, and G.~Zi,
  ``Size effect on compression strength of fiber composites failing by kink
  band propagation,'' in {\em Fracture scaling}, pp.~103--141, Springer, 1999.

\bibitem{catalanotti2014measurement}
G.~Catalanotti, J.~Xavier, and P.~Camanho, ``Measurement of the compressive
  crack resistance curve of composites using the size effect law,'' {\em
  Composites Part A: Applied Science and Manufacturing}, vol.~56, pp.~300--307,
  2014.

\bibitem{salviato2016experimental}
M.~Salviato, K.~Kirane, S.~E. Ashari, Z.~P. Ba{\v{z}}ant, and G.~Cusatis,
  ``Experimental and numerical investigation of intra-laminar energy
  dissipation and size effect in two-dimensional textile composites,'' {\em
  Composites Science and Technology}, vol.~135, pp.~67--75, 2016.

\bibitem{tube1}
P.~M. Jelf and N.~A. Fleck, ``The failure of composite tubes due to combined
  compression and torsion,'' {\em Journal of Materials Science}, vol.~29,
  pp.~3080--3084, 1994.

\bibitem{tube2}
P.~D. Soden, M.~J. Hinton, and A.~S. Kaddour, ``Biaxial test results for
  strength and deformation of a range of e-glass and carbon fibre reinforced
  composite laminates: failure exercise benchmark data,'' {\em Composites
  Science and Technology}, vol.~62, no.~12, pp.~1489--1514, 2002.

\bibitem{tube3}
S.~R. Swanson, M.~J. Messick, and Z.~Tian, ``Failure of carbon/epoxy lamina
  under combined stress,'' {\em Journal of Composite Materials}, vol.~21,
  no.~7, pp.~619--630, 1987.

\bibitem{tube4}
P.~D. Soden, R.~Kitching, and P.~C. Tse, ``Experimental failure stresses for
  ±55° filament wound glass fibre reinforced plastic tubes under biaxial
  loads,'' {\em Composites}, vol.~20, no.~2, pp.~125--135, 1989.

\bibitem{cruciform1}
A.~Smits, D.~Van~Hemelrijck, T.~P. Philippidis, and A.~Cardon, ``Design of a
  cruciform specimen for biaxial testing of fibre reinforced composite
  laminates,'' {\em Composite Science and Technology}, vol.~66, no.~7-8,
  pp.~964--975, 2006.

\bibitem{cruciform2}
V.~Quaglini, C.~Corazza, and P.~C, ``Experimental characterization of
  orthotropic technical textiles under uniaxial and biaxial loading,'' {\em
  Composites Part A}, vol.~39, no.~8, pp.~1331--1342, 2008.

\bibitem{cruciform3}
A.~Makris, C.~Ramault, D.~Van~Hemelrijck, D.~Zarouchas, E.~Lamkanfi, and
  W.~Van~Paepegem, ``An investigation of the mechanical behavior of carbon
  epoxy cross ply cruciform specimens under biaxial loading,'' {\em Polymer
  Composites}, vol.~31, no.~9, pp.~1554--1561, 2010.

\bibitem{cruciform4}
J.~P. Boehler, S.~Demmerle, and S.~Koss, ``A new direct biaxial testing machine
  for anisotropic materials,'' {\em Experimental Mechanics}, vol.~34, no.~1,
  pp.~1--9, 1994.

\bibitem{arcandevice}
M.~Arcan, Z.~Hashin, and A.~Voloshin, ``A method to produce uniform
  plane-stress states with applications to fiber-reinforced materials,'' {\em
  Experimental Mechanics}, vol.~18, pp.~141--146, 1977.

\bibitem{Yaomulti1}
Y.~Qiao, A.~A. Deleo, and M.~Salviato, ``A study on the multi-axial fatigue
  failure behavior of notched composite laminates,'' {\em Composites Part A},
  vol.~127, p.~105640, 2019.

\bibitem{Pearce}
G.~M. Pearce, C.~Tao, Y.~H.~E. Quek, and N.~T. Chowdhury, ``A modified arcan
  test for mixed-mode loading of bolted joints in composite structures,'' {\em
  Composite Structures}, vol.~187, no.~1, pp.~203--211, 2018.

\bibitem{Tan}
J.~L.~Y. Tan, V.~S. Deshpande, and N.~A. Fleck, ``Failure mechanisms of a
  notched cfrp laminate under multi-axial loading,'' {\em Compos Part A-Appl
  S}, vol.~77, pp.~56--66, 2015.

\bibitem{laux}
T.~Laux, K.~Gan, J.~Dulieu-Barton, and O.~Thomsen, ``A simple nonlinear
  constitutive model based on non-associative plasticity for ud composites:
  Development and calibration using a modified arcan fixture,'' {\em Int J
  Solids Struct}, vol.~162, no.~1, pp.~135--147, 2019.

\bibitem{liechti}
C.~Popelar and K.~Liechti, ``Multiaxial nonlinear viscoelastic characterization
  and modeling of a structural adhesive,'' {\em J Eng Mater Technol}, vol.~119,
  no.~3, pp.~205--210, 1997.

\bibitem{Akhtar}
I.~Ud~Din, P.~Hao, S.~Panier, K.~A. Khan, M.~Aamir, G.~Franz, and K.~Akhtar,
  ``Design of a new arcan fixture for in-plane pure shear and combined
  normal/shear stress characterization of fiber reinforced polymer
  composites,'' {\em Experimental Techniques}, vol.~44, pp.~231--240, 2020.

\bibitem{Tan2}
J.~L.~Y. Tan, V.~S. Deshpande, and N.~A. Fleck, ``The effect of laminate lay-up
  on the multi-axial notched strength of cfrp panels: simulation versus
  experiment,'' {\em Eur J Mech - A/Solids}, vol.~66, pp.~309--321, 2017.

\bibitem{Alfonso}
L.~Alfonso, A.~Uguen, C.~Badulescu, J.~Y. Cognard, T.~Bonnemains, E.~Lolive,
  and N.~Carrere, ``Determination of the 3d failure envelope of a composite
  based on a modified arcan test device,'' {\em Composite Structures},
  vol.~131, pp.~585--593, 2015.

\bibitem{Zeinedini}
A.~Zeinedini, ``A novel fixture for mixed mode i/ii/iii fracture testing of
  brittle materials,'' {\em Fatigue Fract Eng Mater Struct}, vol.~42,
  pp.~838--853, 2019.

\bibitem{cruciformfatigue1}
J.~C. Radon and C.~R. Wchnicki, ``Biaxial fatigue of glass fiber reinforced
  polyester resin,'' {\em ASTM STP 853}, pp.~396--412, 1985.

\bibitem{cruciformfatigue2}
D.~L. Jones, P.~K. Poulose, and H.~Liebowitz, ``Effect of biaxial loads on the
  static and fatigue properties of composite materials,'' {\em ASTM STP 853},
  pp.~413--427, 1985.

\bibitem{cruciformfatigue3}
H.~J. Kwon, P.~Y.~B. Jar, and Z.~Xia, ``Characterization of bi-axial fatigue
  resistance of polymer plates,'' {\em Journal of Materials Science}, vol.~40,
  p.~965–972, 2005.

\bibitem{Yaomulti2}
Y.~Qiao and M.~Salviato, ``Micro-computed tomography analysis of damage in
  notched composite laminates under multi-axial fatigue,'' {\em Composites Part
  B}, vol.~187, p.~107789, 2020.

\bibitem{tubefatigue}
P.~A. Carraro, L.~Maragoni, and M.~Quaresimin, ``Influence of load ratio on the
  biaxial fatigue behaviour and damage evolution in glass/epoxy tubes under
  tension–torsion loading,'' {\em Composites Part A}, vol.~78, pp.~294--302,
  2014.

\bibitem{tubefatigue2}
M.~Quaresimin, P.~A. Carraro, L.~P. Mikkelsen, N.~Lucato, L.~Vivian,
  P.~Brondsted, B.~F. Sørensen, J.~Varna, and R.~Talreja, ``Damage evolution
  under cyclic multi-axial stress state: A comparative analysis between
  glass/epoxy laminates and tubes,'' {\em Composites Part B}, vol.~61,
  pp.~282--290, 2014.

\bibitem{tubefatigue3}
P.~H. Francis, D.~E. Walrath, D.~F. Sims, and D.~N. Weed, ``Biaxial fatigue
  loading of notched composites,'' {\em Journal of Composite Materials},
  vol.~11, pp.~488--501, 1977.

\bibitem{tubefatigue4}
K.~Takemura and T.~Fujii, ``Fatigue strength and damage progression in a
  circular-hole-notched grp composite under combined tension/torsion loading,''
  {\em Composites Science and Technology}, vol.~52, pp.~519--526, 1994.

\bibitem{tubefatigue5}
K.~Takemura and T.~Fujii, ``Fracture mechanics evaluation of progressive
  fatigue damage in a circular-hole-notched grp composite under combined
  tension/torsion loading,'' {\em Composites Science and Technology}, vol.~52,
  pp.~527--534, 1994.

\bibitem{DIC}
Correlated Solutions, Columbia, SC, USA, {\em Vic-2D Reference Manual}, 2009.

\bibitem{warren2015experimental}
K.~C. Warren, R.~A. Lopez-Anido, and J.~Goering, ``Experimental investigation
  of three-dimensional woven composites,'' {\em Composites Part A: Applied
  Science and Manufacturing}, vol.~73, pp.~242--259, 2015.

\bibitem{daniel2006engineering}
I.~M. Daniel and O.~Ishai, {\em Engineering Mechanics of Composite Materials}.
\newblock Oxford University Press, 2006.

\bibitem{hufner2009progressive}
D.~R. Hufner and M.~L. Accorsi, ``A progressive failure theory for woven
  polymer-based composites subjected to dynamic loading,'' {\em Composite
  Structures}, vol.~89, no.~2, pp.~177--185, 2009.

\bibitem{li2013brazilian}
D.~Li and L.~N.~Y. Wong, ``The brazilian disc test for rock mechanics
  applications: review and new insights,'' {\em Rock Mechanics and Rock
  Engineering}, vol.~46, no.~2, pp.~269--287, 2013.

\bibitem{lopez2008meso}
C.~M. L{\'o}pez, I.~Carol, and A.~Aguado, ``Meso-structural study of concrete
  fracture using interface elements. {II}: Compression, biaxial and brazilian
  test,'' {\em Materials and Structures}, vol.~41, no.~3, pp.~601--620, 2008.

\bibitem{bazant1991size}
Z.~P. Ba{\v{z}}ant, M.~T. Kazemi, T.~Hasegawa, and J.~Mazars, ``Size effect in
  brazilian split-cylinder tests: measurements and fracture analysis,'' {\em
  ACI Materials Journal}, vol.~88, no.~3, pp.~325--332, 1991.

\bibitem{hobbs1964tensile}
D.~Hobbs, ``The tensile strength of rocks,'' {\em International Journal of Rock
  Mechanics and Mining Sciences \& Geomechanics Abstracts}, vol.~1, no.~3,
  pp.~385--396, 1964.

\bibitem{steen2005observed}
B.~V.~D. Steen, A.~Vervoort, and J.~Napier, ``Observed and simulated fracture
  pattern in diametrically loaded discs of rock material,'' {\em International
  Journal of Fracture}, vol.~131, no.~1, pp.~35--52, 2005.

\bibitem{bazant1997fracture}
Z.~P. Ba{\v{z}}ant and J.~Planas, {\em Fracture and size effect in concrete and
  other quasibrittle materials}, vol.~16.
\newblock CRC press, 1997.

\bibitem{salviato2019mode}
M.~Salviato, K.~Kirane, Z.~P. Ba{\v{z}}ant, and G.~Cusatis, ``Mode {I} and {II}
  interlaminar fracture in laminated composites: a size effect study,'' {\em
  Journal of Applied Mechanics}, vol.~86, no.~9, p.~091008, 2019.

\bibitem{bazant1992bifurcation}
Z.~P. Ba{\v{z}}ant and M.~R. Tabbara, ``Bifurcation and stability of structures
  with interacting propagating cracks,'' {\em International Journal of
  Fracture}, vol.~53, no.~3, pp.~273--289, 1992.

\bibitem{salviatonotch}
Y.~Qiao and M.~Salviato, ``Strength and cohesive behavior of thermoset polymers
  at the microscale: A size-effect study,'' {\em Engineering Fracture
  Mechanics}, vol.~213, pp.~100--117, 2019.

\bibitem{gomeznotch}
F.~J. Gómez, G.~V. Guinea, and M.~Elices, ``Failure criteria for linear
  elastic materials with u-notches,'' {\em International Journal of Fracture},
  vol.~143, pp.~99--113, 2006.

\bibitem{chou1992effect}
S.~Chou, H.-C. Chen, and H.-E. Chen, ``Effect of weave structure on mechanical
  fracture behavior of three-dimensional carbon fiber fabric reinforced epoxy
  resin composites,'' {\em Composites Science and Technology}, vol.~45, no.~1,
  pp.~23--35, 1992.

\bibitem{D5528}
{ASTM}-D5528-13, ``{Standard Test Method for Mode I Interlaminar Fracture
  Toughness of Unidirectional Fiber-Reinforced Polymer Matrix Composites},''
  standard, ASTM International, West Conshohocken, PA, 2013.

\bibitem{tanzawa1999interlaminar}
Y.~Tanzawa, N.~Watanabe, and T.~Ishikawa, ``Interlaminar fracture toughness of
  3-d orthogonal interlocked fabric composites,'' {\em Composites Science and
  Technology}, vol.~59, no.~8, pp.~1261--1270, 1999.

\bibitem{tamuzs2003delamination}
V.~Tamuzs, S.~Tarasovs, and U.~Vilks, ``Delamination properties of
  translaminar-reinforced composites,'' {\em Composites Science and
  Technology}, vol.~63, no.~10, pp.~1423--1431, 2003.

\bibitem{tamuzs2001progressive}
V.~Tamuzs, S.~Tarasovs, and U.~Vilks, ``Progressive delamination and fiber
  bridging modeling in double cantilever beam composite specimens,'' {\em
  Engineering Fracture Mechanics}, vol.~68, no.~5, pp.~513--525, 2001.

\bibitem{alif1997mode}
N.~Alif, L.~A. Carlsson, and J.~W. Gillespie, ``Mode i, mode ii, and mixed mode
  interlaminar fracture of woven fabric carbon/epoxy,'' in {\em Composite
  Materials: Testing and Design, Thirteenth Volume}, ASTM International, 1997.

\bibitem{bruhwiler1990wedge}
E.~Br{\"u}hwiler and F.~Wittmann, ``The wedge splitting test, a new method of
  performing stable fracture mechanics tests,'' {\em Engineering fracture
  mechanics}, vol.~35, no.~1-3, pp.~117--125, 1990.

\bibitem{brunner2008status}
A.~Brunner, B.~Blackman, and P.~Davies, ``A status report on delamination
  resistance testing of polymer--matrix composites,'' {\em Engineering Fracture
  Mechanics}, vol.~75, no.~9, pp.~2779--2794, 2008.

\bibitem{glessner1989mode}
A.~L. Glessner, M.~T. Takemori, M.~A. Vallance, and S.~K. Gifford, ``Mode i
  interlaminar fracture toughness of unidirectional carbon fiber composites
  using a novel wedge-driven delamination design,'' in {\em Composite
  materials: fatigue and fracture, second volume}, ASTM International, 1989.

\bibitem{brunner2000experimental}
A.~Brunner, ``Experimental aspects of mode i and mode ii fracture toughness
  testing of fibre-reinforced polymer-matrix composites,'' {\em Computer
  methods in applied mechanics and engineering}, vol.~185, no.~2-4,
  pp.~161--172, 2000.

\bibitem{siddique2019finite}
A.~Siddique, B.~Sun, and B.~Gu, ``Finite element modeling on fracture toughness
  of 3d angle-interlock woven carbon/epoxy composites at microstructure
  level,'' {\em Mechanics of Advanced Materials and Structures}, pp.~1--12,
  2019.

\bibitem{Ncorr}
R.~Harilal and M.~Ramji, ``Adaptation of open source 2d dic software ncorr for
  solid mechanics applications,'' {\em In: Proceedings of 9th International
  Symposium on Advanced Science and Technology in Experimental Mechanics, New
  Delhi, November}, 2014.

\bibitem{Blader}
J.~Blaber, B.~Adair, and A.~Antoniou, ``Ncorr: open-source 2d digital image
  correlation matlab software,'' {\em Exp Mech}, vol.~55, no.~6,
  pp.~1106--1122, 2015.

\bibitem{northstar}
``{North Star Imaging, California, USA https://4nsi.com},''

\bibitem{shiyaocompact}
S.~Lin, S.~I. Thorsson, and A.~M. Waas, ``Predicting the low velocity impact
  damage of a quasi-isotropic laminate using est,'' {\em Composite structures},
  vol.~251, p.~112530, 2020.

\bibitem{dye1}
B.~Yu, B.~Bradley, C.~Soutis, P.~Hogg, and P.~Withers, ``2d and 3d imaging of
  fatigue failure mechanisms of 3d woven composites,'' {\em Compos Part A-Appl
  S}, vol.~77, pp.~37--49, 2015.

\bibitem{dye2}
O.~Nexon-Pearson and S.~Hallet, ``An experimental investigation into
  quasi-static and fatigue damage development into bolted hole specimen,'' {\em
  Compos Part B}, vol.~77, pp.~462--473, 2015.

\bibitem{paraview}
J.~Ahrens, B.~Geveci, and C.~Law, {\em ParaView: An End-User Tool for Large
  Data Visualization, Visualization Handbook}.
\newblock Elsevier, 2005.

\bibitem{stressconrete}
H.~Nguyen, M.~Pathirage, M.~Rezaei, M.~Issa, G.~Cusatis, and Z.~P.
  Ba{\v{z}}ant, ``New perspective of fracture mechanics inspired by gap test
  with crack-parallel compression,'' {\em PNAS}, vol.~117, no.~25,
  pp.~14015--14020, 2020.

\bibitem{stresscomposite1}
R.~Talreja, N.~Mitra, and A.~Neogi, ``Cavitation in epoxies under
  composite-like stress states,'' {\em Compos Part A-Appl S}, vol.~106,
  pp.~52--58, 2018.

\bibitem{stresscomposite2}
Y.~Qiao, Q.~Zhang, and M.~Salviato, ``Effects of \emph{in-situ} stress state on
  the plastic deformation, fracture, and size scaling of thermoset polymers and
  related fiber-reinforced composites,'' {\em Proceedings of the American
  Society for Composites—Thirty-five Technical Conference}, 2020.

\bibitem{stressmetal1}
M.~Mirza, D.~Barton, and P.~Church, ``The effect of stress triaxiality and
  strain-rate on the fracture characteristics of ductile metals,'' {\em J Mater
  Sci}, vol.~31, pp.~453--461, 1996.

\bibitem{Yu2016}
B.~Yu, R.~Blanc, C.~Soutis, and P.~Withers, ``Evolution of damage during the
  fatigue of 3d woven glass-fibre reinforced composites subjected to
  tension–tension loading observed by time-lapse x-ray tomography,'' {\em
  Composites Part A- Appl S}, vol.~82, pp.~279--290, 2016.

\bibitem{talrejadamagebook}
R.~Talreja and C.~Singh, {\em Damage and Failure of Composite Materials}.
\newblock Cambridge University Press, 2012.

\bibitem{geubelletransverse}
X.~Zhang, D.~Brandyberry, and P.~Geubelle, ``Igfem-based shape sensitivity
  analysis of the transverse failure of a composite laminate,'' {\em Comput
  Mech}, vol.~64, no.~5, pp.~1455--1472, 2019.

\bibitem{herraez}
M.~Herr\'aez, D.~Mora, F.~Naya, C.~Lopes, C.~Gonz\'alez, and J.~Llorca,
  ``Transverse cracking of cross-ply laminates: A computational micromechanics
  perspective,'' {\em Compos Sci Technol}, vol.~110, pp.~196--204, 2015.

\bibitem{mortell}
D.~Mortell, D.~Tanner, and C.~McCarthy, ``An experimental investigation into
  multi-scale damage progression in laminated composites in bending,'' {\em
  Compos Struct}, vol.~149, pp.~33--40, 2016.

\bibitem{abaqus}
ABAQUS, {\em ABAQUS Users Manual, Ver. 6.13-1}.
\newblock Dassault Systèmes, Providence, RI, USA, 2013.

\bibitem{perras2014review}
M.~A. Perras and M.~S. Diederichs, ``A review of the tensile strength of rock:
  concepts and testing,'' {\em Geotechnical and geological engineering},
  vol.~32, no.~2, pp.~525--546, 2014.

\bibitem{bao1992role}
G.~Bao, S.~Ho, Z.~Suo, and B.~Fan, ``The role of material orthotropy in
  fracture specimens for composites,'' {\em International Journal of Solids and
  Structures}, vol.~29, no.~9, pp.~1105--1116, 1992.

\bibitem{salviato2017graphene}
C.~H. Mefford, Y.~Qiao, and M.~Salviato, ``Failure behavior and scaling of
  graphene nanocomposites,'' {\em Composite Structures}, vol.~176,
  pp.~961--972, 2017.

\bibitem{barsoum1974application}
R.~S. Barsoum, ``Application of quadratic isoparametric finite elements in
  linear fracture mechanics,'' {\em International Journal of Fracture},
  vol.~10, no.~4, pp.~603--605, 1974.

\bibitem{bavzant1991identification}
Z.~P. Ba{\v{z}}ant, R.~Gettu, and M.~Kazemi, ``Identification of nonlinear
  fracture properties from size effect tests and structural analysis based on
  geometry-dependent r-curves,'' {\em International journal of rock mechanics
  and mining sciences \& geomechanics abstracts}, vol.~28, no.~1, pp.~43--51,
  1991.

\end{thebibliography}
	
\end{document}